\documentclass[twocolumn]{aastex701} 
\usepackage{tablefootnote}
\usepackage{hyperref}
\usepackage{amsmath}
\usepackage{color}
\usepackage{caption}

\begin{document}

\title{Spectropolarimetric properties of fast radio bursts co-detected by CHIME/FRB and DSA-110}
\shorttitle{CHIME/FRB and DSA-110 co-detected FRBs}

\author[0000-0002-8897-1973]{Ayush Pandhi}
\email[show]{ayush.pandhi@mcgill.ca}
\affiliation{Department of Physics, McGill University, 3600 rue University, Montr\'eal, QC H3A 2T8, Canada}
\affiliation{Trottier Space Institute, McGill University, 3550 rue University, Montr\'eal, QC H3A 2A7, Canada}
\correspondingauthor{Ayush Pandhi}

\author[0000-0002-6573-7316]{Myles Sherman}
\email{mylesbsherman@gmail.com}
\affiliation{Cahill Center for Astronomy and Astrophysics, MC 249-17 California Institute of Technology, Pasadena CA 91125, USA}

\author[0000-0001-9855-5781]{Jakob Faber}
\email{jfaber@caltech.edu}
\affiliation{Cahill Center for Astronomy and Astrophysics, MC 249-17 California Institute of Technology, Pasadena CA 91125, USA}

\author[0000-0002-4795-697X]{Ziggy Pleunis}
\email{z.pleunis@uva.nl}
\affiliation{Anton Pannekoek Institute for Astronomy, University of Amsterdam, Science Park 904, 1098 XH Amsterdam, The Netherlands}
\affiliation{ASTRON, Netherlands Institute for Radio Astronomy, Oude Hoogeveensedijk 4, 7991 PD Dwingeloo, The Netherlands}

\author[0000-0003-0510-0740]{Kenzie Nimmo}
\email{kenzie.nimmo@cfa.harvard.edu}
\altaffiliation{NHFP Einstein Fellow}
\affiliation{Center for Astrophysics | Harvard \& Smithsonian, 60 Garden St, Cambridge, MA 02138, USA}

\author[0000-0002-4941-5333]{Stella Ocker}
\email{socker@caltech.edu}
\affiliation{Cahill Center for Astronomy and Astrophysics, MC 249-17 California Institute of Technology, Pasadena CA 91125, USA}
\affiliation{Observatories of the Carnegie Institution for Science, Pasadena, CA 91101, USA}

\author[0000-0003-3236-8769]{Dylan Jow}
\email{dylanjow@stanford.edu}
\affiliation{Kavli Institute for Particle Astrophysics \& Cosmology, P.O. Box 2450, Stanford University, Stanford, CA 94305, USA}

\author[0000-0002-7587-6352]{Liam Connor}
\email{liam.connor@cfa.harvard.edu}
\affiliation{Center for Astrophysics | Harvard \& Smithsonian, 60 Garden St, Cambridge, MA 02138, USA}

\author[0000-0002-8376-1563]{Alice P.~Curtin}
\email{alice.curtin@mcgill.ca}
\affiliation{Department of Physics, McGill University, 3600 rue University, Montr\'eal, QC H3A 2T8, Canada}
\affiliation{Trottier Space Institute, McGill University, 3550 rue University, Montr\'eal, QC H3A 2A7, Canada}
\affiliation{Anton Pannekoek Institute for Astronomy, University of Amsterdam, Science Park 904, 1098 XH Amsterdam, The Netherlands}

\author[0000-0002-3382-9558]{B.~M.~Gaensler}
\email{gaensler@ucsc.edu}
\affiliation{Department of Astronomy and Astrophysics, University of California, Santa Cruz, 1156 High Street, Santa Cruz, CA 95060, USA}
\affiliation{Dunlap Institute for Astronomy and Astrophysics, 50 St. George Street, University of Toronto, ON M5S 3H4, Canada}
\affiliation{David A.\ Dunlap Department of Astronomy and Astrophysics, 50 St. George Street, University of Toronto, ON M5S 3H4, Canada}

\author[0000-0003-2317-1446]{Jason W.~T.~Hessels}
\email{jason.hessels@mcgill.ca}
\affiliation{Department of Physics, McGill University, 3600 rue University, Montr\'eal, QC H3A 2T8, Canada}
\affiliation{Trottier Space Institute, McGill University, 3550 rue University, Montr\'eal, QC H3A 2A7, Canada}
\affiliation{Anton Pannekoek Institute for Astronomy, University of Amsterdam, Science Park 904, 1098 XH Amsterdam, The Netherlands}
\affiliation{ASTRON, Netherlands Institute for Radio Astronomy, Oude Hoogeveensedijk 4, 7991 PD Dwingeloo, The Netherlands}

\author[0000-0001-9345-0307]{Victoria Kaspi}
\email{victoria.kaspi@mcgill.ca}
\affiliation{Department of Physics, McGill University, 3600 rue University, Montr\'eal, QC H3A 2T8, Canada}
\affiliation{Trottier Space Institute, McGill University, 3550 rue University, Montr\'eal, QC H3A 2A7, Canada}

\author[0009-0004-4176-0062]{Afrokk Khan}
\email{afrasiyab.khan@mcgill.ca}
\affiliation{Department of Physics, McGill University, 3600 rue University, Montr\'eal, QC H3A 2T8, Canada}
\affiliation{Trottier Space Institute, McGill University, 3550 rue University, Montr\'eal, QC H3A 2A7, Canada}

\author[0000-0002-4119-9963]{Casey J.~Law}
\email{caseylaw@caltech.edu}
\affiliation{Cahill Center for Astronomy and Astrophysics, MC 249-17 California Institute of Technology, Pasadena CA 91125, USA}
\affiliation{Owens Valley Radio Observatory, California Institute of Technology, Big Pine CA 93513, USA}

\author[0000-0002-7164-9507]{Robert Main}
\email{robert.main@nrc-cnrc.gc.ca}
\affiliation{Dominion Radio Astrophysical Observatory, Herzberg Astronomy \& Astrophysics Research Centre, National Research Council Canada, 717 White Lake Rd., Kaleden, BC, V0H 1K0 Canada}

\author[0000-0001-7348-6900]{Ryan Mckinven}
\email{ryan.mckinven@mcgill.ca}
\affiliation{Department of Physics, McGill University, 3600 rue University, Montr\'eal, QC H3A 2T8, Canada}
\affiliation{Trottier Space Institute, McGill University, 3550 rue University, Montr\'eal, QC H3A 2A7, Canada}

\author[0000-0002-0940-6563]{Mason Ng}
\email{mason.ng@mcgill.ca}
\affiliation{Department of Physics, McGill University, 3600 rue University, Montr\'eal, QC H3A 2T8, Canada}
\affiliation{Trottier Space Institute, McGill University, 3550 rue University, Montr\'eal, QC H3A 2A7, Canada}

\author[0000-0002-8912-0732]{Aaron B.~Pearlman}
\email{aaron.b.pearlman@mit.edu}
\altaffiliation{NASA Hubble Fellow}
\affiliation{MIT Kavli Institute for Astrophysics and Space Research, Massachusetts Institute of Technology, 77 Massachusetts Ave, Cambridge, MA 02139, USA}
\affiliation{Department of Physics, McGill University, 3600 rue University, Montr\'eal, QC H3A 2T8, Canada}
\affiliation{Trottier Space Institute, McGill University, 3550 rue University, Montr\'eal, QC H3A 2A7, Canada}

\author[0000-0002-7252-5485]{Vikram Ravi}
\email{vikram@caltech.edu}
\affiliation{Cahill Center for Astronomy and Astrophysics, MC 249-17 California Institute of Technology, Pasadena CA 91125, USA}
\affiliation{Owens Valley Radio Observatory, California Institute of Technology, Big Pine CA 93513, USA}

\author[0000-0002-4623-5329]{Mawson Sammons}
\email{mawson.sammons@lam.fr}
\affiliation{Laboratoire d'Astrophysique de Marseille, Aix-Marseille Univ., CNRS, CNES, Marseille, France}

\author[0000-0002-7374-7119]{Paul Scholz}
\email{pscholz@yorku.ca}
\affiliation{Department of Physics and Astronomy, York University, 4700 Keele Street, Toronto, ON M3J 1P3, Canada}

\author[0000-0002-4477-3625]{Kritti Sharma}
\email{kritti@caltech.edu}
\affiliation{Cahill Center for Astronomy and Astrophysics, MC 249-17 California Institute of Technology, Pasadena CA 91125, USA}

\author[0000-0002-6823-2073]{Kaitlyn Shin}
\email{kaitshin@caltech.edu}
\affiliation{Cahill Center for Astronomy and Astrophysics, MC 249-17 California Institute of Technology, Pasadena CA 91125, USA}

\author[0000-0002-2088-3125]{Kendrick Smith}
\email{kmsmith@perimeterinstitute.ca}
\affiliation{Perimeter Institute for Theoretical Physics, 31 Caroline Street N, Waterloo, ON N2L 2Y5, Canada}

\author[0000-0001-9784-8670]{Ingrid Stairs}
\email{stairs@astro.ubc.ca}
\affiliation{Department of Physics and Astronomy, University of British Columbia, 6224 Agricultural Road, Vancouver, BC V6T 1Z1 Canada}

\begin{abstract}
We present spectropolarimetric analysis of twelve fast radio bursts (FRBs) co-detected by CHIME/FRB and DSA-110, eleven of which are apparently non-repeating. We obtain broadband spectral coverage (400$-$1530~MHz) for these FRBs and constrain propagation effects in their polarized spectra. We observe two notable spectral features: (i) in three FRBs, we find decreasing linear polarization fraction with increasing frequency and, (ii) in seven FRBs, we find that the linear polarization fraction decreases with decreasing frequency. The former spectral signature is commonly observed in radio pulsars and is suggestive of magnetospheric emission, where either the mixing of orthogonally polarized modes or incoherent summing of unresolved burst components may cause depolarization at high frequencies. The latter is consistent with depolarization due to multipath propagation. We report the first statistically significant correlation ($3.5-3.9\sigma$) between the extent of depolarization and amplitude of Faraday rotation arising from FRB host galaxies, implying that the two effects likely arise in the same turbulent, magnetoionic environments around FRB sources. The prevalence of depolarization in our sample suggests that turbulent environments that have, thus far, been predominantly associated with some repeating FRBs, may also be commonly present around a subset of apparently non-repeating FRBs. We highlight a systematic bias for detecting a disproportionately high fraction of constant polarization angle (PA) profiles in low-frequency FRB surveys due to amplified scatter broadening, which we also observe in two of our co-detected FRBs.
\end{abstract}

\keywords{Radio bursts (1339) --- Radio transient sources (2008) --- Polarimetry (1278)}

\section{Introduction} \label{sec:intro}
The physical origins and emission mechanisms of fast radio bursts \citep[FRBs;][]{2007Sci...318..777L}, coherent radio transients lasting only microseconds to milliseconds, and of extragalactic origin, remain elusive. The discovery of a radio burst associated with Galactic magnetar SGR~1935$+$2154 that had comparable energies to some FRBs provided a direct connection between FRB sources and magnetars \citep{2020Natur.587...54C, 2020Natur.587...59B}. Since then, spectropolarimetric properties of some FRBs have pointed towards their emission sites being located in the magnetospheres of neutron stars \citep[e.g., FRB~20221022A;][]{2025Natur.637...43M, 2025Natur.637...48N}. However, the range of burst morphologies \citep[e.g.,][]{2021ApJ...923....1P, 2023NatAs...7.1486S, 2024ApJ...974..274F, 2025ApJ...992..206C}, polarization properties \citep[e.g.,][]{2024ApJ...968...50P, 2024ApJ...964..131S, 2025PASA...42..133S, 2026arXiv260921177D}, and host environments \citep[e.g.,][]{2022Natur.602..585K, 2023ApJ...954...80G, 2025ApJ...979L..21S, 2025ApJ...979L..22E} observed in the larger FRB population \citep[$\sim 4500$ published sources to date;][]{FRBCollaboration_2026_arXiv}, make it challenging to reconcile all FRBs with a single, common origin.

Spectropolarimetric properties of FRBs provide key insights into both the FRB emission process and their source environments \citep[e.g.,][]{2018Natur.553..182M, 2020Natur.586..693L, 2023Sci...380..599A, 2023ApJ...957L...8S}. The intrinsic linear and circular polarization fractions ($L/I$ and $|V|/I$, respectively) of FRB signals help to constrain the emission mechanism responsible for their production \citep[e.g.,][]{2023MNRAS.522.2448Q}. Temporal variability of the polarization angle (PA) across the burst duration may trace our viewing geometry of the FRB emission region over $\lesssim$~ms timescales, which can be parameterized using the ``rotating vector model'' that has been successfully applied to many radio pulsars \citep[e.g.,][]{1969ApL.....3..225R, 2023MNRAS.520.4801J} and has been proposed for some FRBs \citep[e.g.,][]{2025Natur.637...43M}. 

As FRB signals traverse between their emitting source and the observer, they accrue frequency-dependent propagation effects from intervening plasma. The radio signals are dispersed by free electrons along the line of sight (LOS), causing lower frequencies to arrive at the telescope at later times; this is characterized by the dispersion measure (DM), which traces the total electron column towards the FRB source. As FRB signals are typically linearly polarized, they also undergo Faraday rotation: their polarization angles rotate as a function of frequency from propagation through magnetoionic plasma. This is characterized by the rotation measure (RM) that encodes information about the product of the LOS electron density and magnetic field:
\begin{equation}
{\rm{DM}} = \int_{0}^{z} \frac{n_e(z')}{(1+z')} \frac{\mathrm{d}l(z')}{\mathrm{d}z'}\mathrm{d}z'~\mathrm{pc}~\mathrm{cm}^{-3} \label{eq:DM} \,;
\end{equation}
\begin{equation}
{\rm{RM}} = 0.812 \int_{0}^{z} \frac{n_e(z') B_\parallel(z')}{(1 + z')^2}\frac{\mathrm{d}l(z')}{\mathrm{d}z'}\mathrm{d}z'~\mathrm{rad}~\mathrm{m}^{-2} \label{eq:RM} \,.
\end{equation}
Here, $z$ refers to the redshift of the emitting source. The electron density, $n_e(z)$, is in units of cm$^{-3}$, the LOS magnetic field strength, $B_\parallel(z)$, is in units of $\mu$G, and $\mathrm{d}l(z)$ is the LOS line element at $z$ in units of pc. In Equation \ref{eq:RM}, the RM is positive when the average LOS magnetic field is pointed towards the observer and negative when it is pointed away. Spectral depolarization has also been observed in a handful of (mostly repeating) FRBs and it has been attributed to multipath propagation through inhomogeneous magnetoionic plasma \citep{2022Sci...375.1266F, 2025MNRAS.tmp.1894U, 2026ApJ..1000L..53P, 2026arXiv260918086B, 2026arXiv260921177D}.

Some frequency-dependent polarimetric effects have been studied in radio pulsars but are rarely observed in FRBs. For example, scattering broadening of some pulsars at $\lesssim 1$~GHz is accompanied by flattening of their PA profiles, presumably due to pulse-phase mixing in the scattering tail \citep[e.g.,][]{1972AuJPh..25..759K, 2003A&A...410..253L, 2009MNRAS.392L..60K, 2009MNRAS.396.1559N, 2023RAA....23j4002W}. This effect has been observed in the uniquely bright FRB~20250316A, where PA variability of amplitude $\sim 30~\mathrm{deg}$ at $800$~MHz was suppressed by scattering at $<500$~MHz \citep{2025ApJ...989L..48C}.

Many pulsars also show a decreasing linear polarization fraction with increasing frequency \citep[e.g.,][]{1973ApJ...179L...7M, 1996A&A...309..481X, 1997ApJ...475..763M, 2008MNRAS.388..261J, 2015A&A...576A..62N}. While there is no consensus on the specific mechanism responsible for this phenomenon, most of the proposed models rely on the birefringence of plasma in the extreme magnetospheres of neutron stars \citep{1975ApJ...196...51R, 1986ApJ...302..138B, 1997ApJ...475..763M, 1998A&A...336..209V}. Other explanations include superimposed orthogonal polarization modes with different spectral indices \citep{2005MNRAS.359..481K} and incoherent summing of short duration bursts with higher frequency emission arising from deeper in the magnetosphere of a neutron star \citep{2008MNRAS.388..261J}. All of these scenarios necessitate an emission site located near the surface of a neutron star. One FRB shows a tentative decrease in $L/I$ with increasing frequency over 400$-$800~MHz \citep{2024ApJ...968...50P}, but there are no definitive detections of this effect in FRBs.

While current FRB experiments reliably measure DM and RM for most detected bursts, constraining the aforementioned spectropolarimetric effects requires a large lever arm in observing frequency ($\nu$) afforded by low-frequency and broadband observations. The problem is exacerbated as most FRB sources do not repeat \citep{2023ApJ...947...83C, 2026arXiv260508410C}, requiring instantaneous broadband observations for a single burst. Uniquely, two FRB experiments in the Northern hemisphere, the Canadian Hydrogen Intensity Mapping Experiment FRB Project \citep[CHIME/FRB, observing at 400$-$800~MHz;][]{2018ApJ...863...48C} and the Deep Synoptic Array 110 \citep[DSA-110, observing at 1280$-$1530~MHz;][]{2024ApJ...967...29L}, have overlapping survey footprints, enabling co-detections of FRBs.

In a companion paper (S.~K.~Ocker \& K.~Nimmo et.~al.~submitted; hereafter ``Paper I''), we introduce a sample of twelve FRBs (11 apparent non-repeaters and one repeater)\footnote{FRB~20230814C/B is identified as a repeater by DSA-110, which detected one other burst from the same source. The source has not been observed to repeat by CHIME/FRB. See Paper I for further details.} co-detected by CHIME/FRB and DSA-110 and with recorded channelized voltage data at both observatories. Here, we explore the spectropolarimetric properties of these co-detected FRBs over 400$-$1530~MHz. In Sections \ref{sec:obs} and \ref{sec:obs_codetections}, we describe our polarimetric data and summarize the process of identifying co-detected FRBs. We detail our modeling of the $L/I$ and PA profile as a function of frequency in Section \ref{sec:models}. In Section \ref{sec:results}, we describe our resulting model fits to the co-detected FRB data, finding complex $L/I$ spectra for eight FRBs, and PA profiles that flatten at low frequencies for two sources. We interpret our measurements in Section \ref{sec:discussion}, showcasing that some of the FRBs may be emitted in the magnetosphere of neutron stars, that many FRBs may be associated with turbulent and magnetized local environments, and we discuss the implications of low-frequency flattening in PA profiles as a systematic bias for current polarimetric FRB population studies. We conclude by summarizing our findings and discussing potential applications of our work in upcoming, broadband FRB surveys.

\section{Observations and Data Reduction} \label{sec:obs}
\subsection{CHIME/FRB} \label{sec:obs_chime}
CHIME, located at the Dominion Radio Astrophysical Observatory in British Columbia, Canada, operates as a transit telescope over 400$-$800~MHz, observing the sky above a Declination of $-10~\mathrm{deg}$ on a daily cadence \citep{2022ApJS..261...29C}. The CHIME/FRB system conducts a real-time FRB search \citep[for details, see][]{2018ApJ...863...48C} and saves full polarization channelized voltage data, or ``baseband data'', at a frequency resolution of $390.625$~kHz and a time resolution of $2.56~\mu\mathrm{s}$, for FRBs exceeding a real-time detection signal-to-noise $\mathrm{S/N} > 12$.\footnote{The threshold is reduced to $\mathrm{S/N} > 10$ for subsequent bursts from a known repeating FRB source.} The localizations for FRBs with recorded baseband data are later refined by a S/N-maximizing search over a dense grid of formed beams, achieving median sky position uncertainties of $\sim20''$ \citep{Michilli2024}. Twelve of the fifteen co-detected FRBs identified in Paper I have recorded baseband data at both observatories, enabling spectropolarimetric analysis.

To derive polarization data products at 400$-$800~MHz, we employ the CHIME/FRB polarization pipeline \citep[for details, see][]{2021ApJ...920..138M}. The FRB burst envelope temporal limits are identified as the region in which the total intensity is $>20$\% of the peak burst intensity, or if the burst has multi-component structure, we manually set the limits to ensure all components are encompassed within them. Stokes $I$, $Q$, $U$, and $V$ spectra are produced by integrating over this burst envelope with uniform weights in time. We apply RM-synthesis \citep{2005A&A...441.1217B} to the FRB Stokes spectra, obtaining a robust RM detection if the peak polarized intensity in Faraday depth space is $>6\sigma$ after applying {\tt RM-CLEAN} \citep{2009A&A...503..409H}. We then apply the QU-fitting algorithm \citep[implemented with a modified version of V1.2 of the {\tt RM-tools} package;][]{2020ascl.soft05003P, 2026arXiv260120092V}, using the default model described by \cite{2021ApJ...920..138M}. This QU-fitting step provides an independent measurement of the FRB RM, as well as an estimate of the physical (cable) delay between the two linear polarizations, which can act as a source of contamination between Stokes $U$ and $V$.\footnote{Instrumental polarization can cause the sign of the RM derived via QU-fitting to be incorrect \citep[see Appendix A by][]{2021ApJ...920..138M}. Thus, we opt to use the RM estimate from RM-synthesis throughout this work. A detailed analysis to mitigate this QU-fitting sign error in CHIME/FRB data will be presented in upcoming work by M.~Ng et al.~(in prep.).} We then return to the Stokes dynamic spectra, derotate them given the FRB RM, apply a correction based on the fitted cable delay term, and integrate over the burst envelope temporal limits to obtain a set of ``corrected'' Stokes $I$, $Q$, $U$, and $V$ spectra that we use for any further analysis.

For any bursts that do not satisfy the $6\sigma$ linearly polarized signal-to-noise criterion, we only measure an upper limit on their $L/I$ averaged across 400$-$800~MHz of $6/(\mathrm{S/N}(I))$, where $\mathrm{S/N}(I)$ is the signal-to-noise of the Stokes $I$ signal \citep{2024ApJ...968...50P}.

\subsection{DSA-110} \label{sec:obs_dsa}
The DSA-110 is a radio interferometer located at the Owens Valley Radio Observatory that operates over 1280$-$1530~MHz \citep{2023ApJ...949L...3R, 2024ApJ...967...29L}. During science commissioning, 63 antennas across $2.5$~km were pointed at a Declination of $71.6~\mathrm{deg}$, with a circular $10~\mathrm{deg}^2$ effective field of view, and used to conduct a real-time FRB search. Full-polarization channelized voltage data are saved for FRB candidates that exceed a real-time detection threshold of $\mathrm{S/N} > 8.5$, are not associated with terrestrial radio-frequency interference, and have DMs greater than 75\% of the Galactic DM estimate along the LOS by the {\tt NE2001} model \citep{2002astro.ph..7156C}. The channelized voltage data is saved at frequency and time resolutions of $30.5$~kHz and $32.7$~$\mu$s; FRB candidate positions are refined offline by forming interferometric visibilities and calibrating them using nearby calibrator observations and a National Radio Astronomical Observatory Very Large Array Sky Survey sky model \citep{1998AJ....115.1693C}, and imaging the burst with the long-baseline data \citep{2023ApJ...949L...3R}, yielding sky position uncertainties of $\lesssim 2''$. We analyze the DSA-110 channelized voltage data for the twelve co-detected FRBs that also have CHIME/FRB baseband data.

We use a modified version of the DSA-110 FRB polarization pipeline \citep[described in detail by][]{2024ApJ...964..131S} to derive polarization data products at 1280$-$1530~MHz. Two bright radio sources, one with known polarization (3C286) and one with $\lesssim 5\%$ linear polarization level at L-band (3C48) are used to calibrate gain and phase differences between the two linear feeds, mitigating instrumental polarization effects. Though DSA-110 applies absolute PA calibration \citep[for further details, see Appendices D$-$G by][]{2024ApJ...964..131S}, CHIME/FRB does not; thus, we normalize the PA profiles from both data sets to be centered at $0~\mathrm{deg}$, and only compare their relative PA variability.

For a given FRB, Stokes spectra are generated by averaging over the temporal extent of the burst, which is determined manually to encompass all components in the case of multi-component bursts. To remain consistent with the CHIME/FRB data products, we conduct an RM search using RM-synthesis \citep[implemented with {\tt RM-tools};][]{2005A&A...441.1217B, 2020ascl.soft05003P, 2026arXiv260120092V} with a detection threshold of linearly-polarized $\mathrm{S/N} > 6$. The Stokes dynamic spectra are derotated to correct for the measured RM. Integrating these derotated Stokes spectra across the temporal extent of the FRB then provides the ``corrected'' Stokes $I$, $Q$, $U$, and $V$ spectra at 1280$-$1530~MHz that we use for further analysis.

\section{Co-detected FRB sample} \label{sec:obs_codetections}
Paper I identifies fifteen FRBs jointly detected by CHIME/FRB and DSA-110 between 2022 February and 2024 February, twelve of which have channelized voltage data with both instruments. The DMs of candidate pairs estimated at each facility broadly agree; their arrival times have a median offset $\sim 2$~ms (1$\sigma$ $\pm 2.5$~ms), and all but one pair have consistent sky positions at the $3\sigma$ level.\footnote{The CHIME/FRB and DSA-110 $3\sigma$ localization regions for FRB~20220207A/C are in disagreement, with an offset in best-fit positions of $\sim 100''$. Paper I discusses possible systematics impacting these localizations and argues that FRB~20220207A/C is still a co-detected FRB given the overwhelming evidence from its consistent arrival time, DM, RM, and burst morphology.} The resulting sample comprises twelve broadband FRBs spanning $262$--$960$~pc~cm$^{-3}$ in DM; spectroscopic host-galaxy redshifts are available for nine bursts (ranging from $0.0430 \leq z \leq 0.5535$), while three co-detected FRBs lack a confirmed host redshift. Details of the co-detection identification process, dynamic spectra of the bursts, analyses of their scattering timescales, scintillation bandwidths, host-galaxy associations, and energetics are reported in Paper~I. In Figure \ref{fig:pol_profiles}, we present the total intensity, linearly and circularly polarized intensity, and PA profiles (for more details regarding the PA profiles see Section \ref{sec:models_pa}) measured by both CHIME/FRB and DSA-110 for the co-detected FRBs.

These FRBs are assigned two Transient Name Server (TNS) names, one for the detection by each instrument. Thus, we present FRB TNS names in the following format: e.g., FRB~20220207A/C, where FRB~20220207A is the TNS name assigned to the CHIME/FRB detection and FRB~20220207C refers to the DSA-110 detection.

\begin{figure*}
    \centering 
    \includegraphics[width=0.49\textwidth]{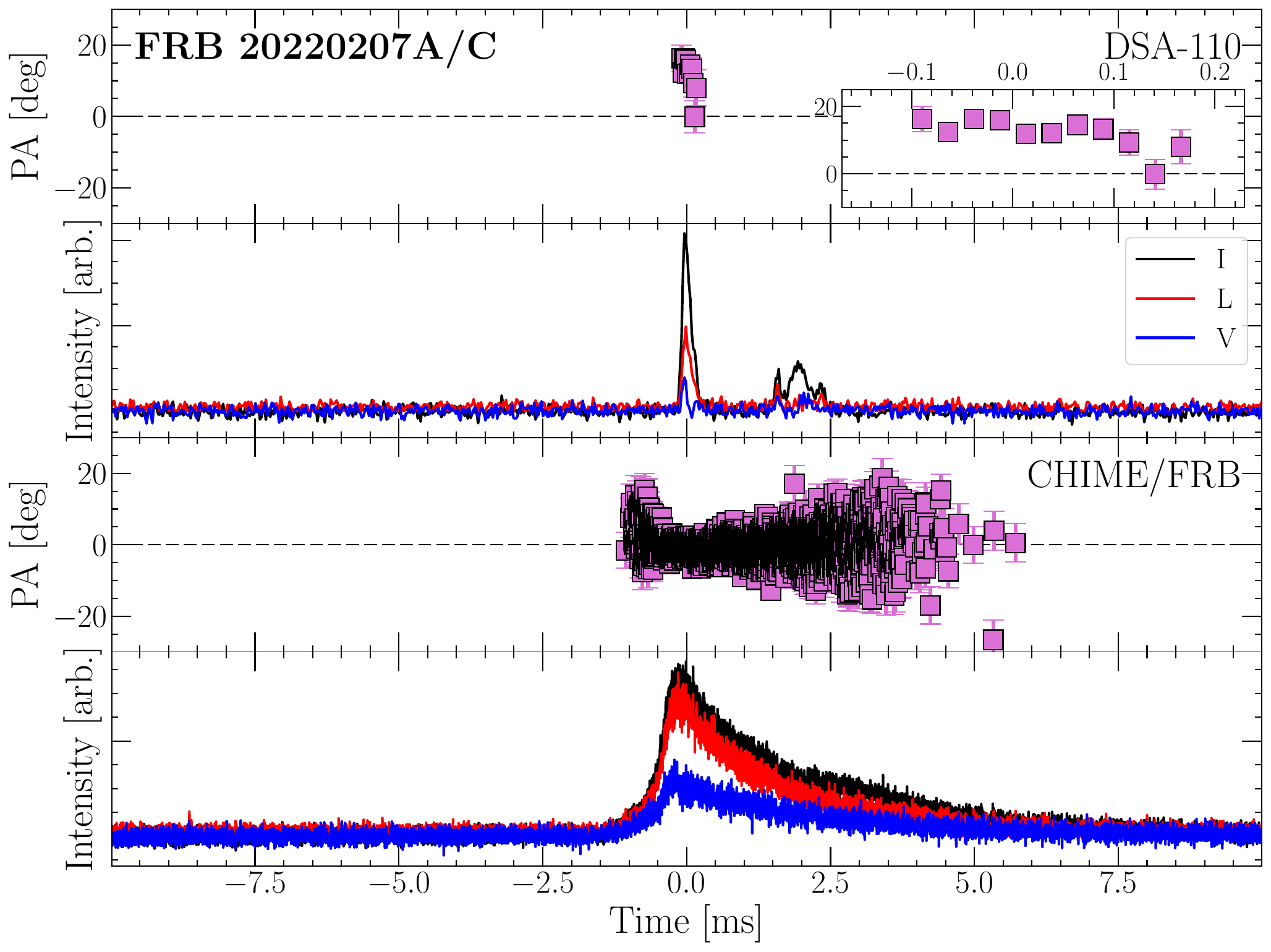}
    \includegraphics[width=0.49\textwidth]{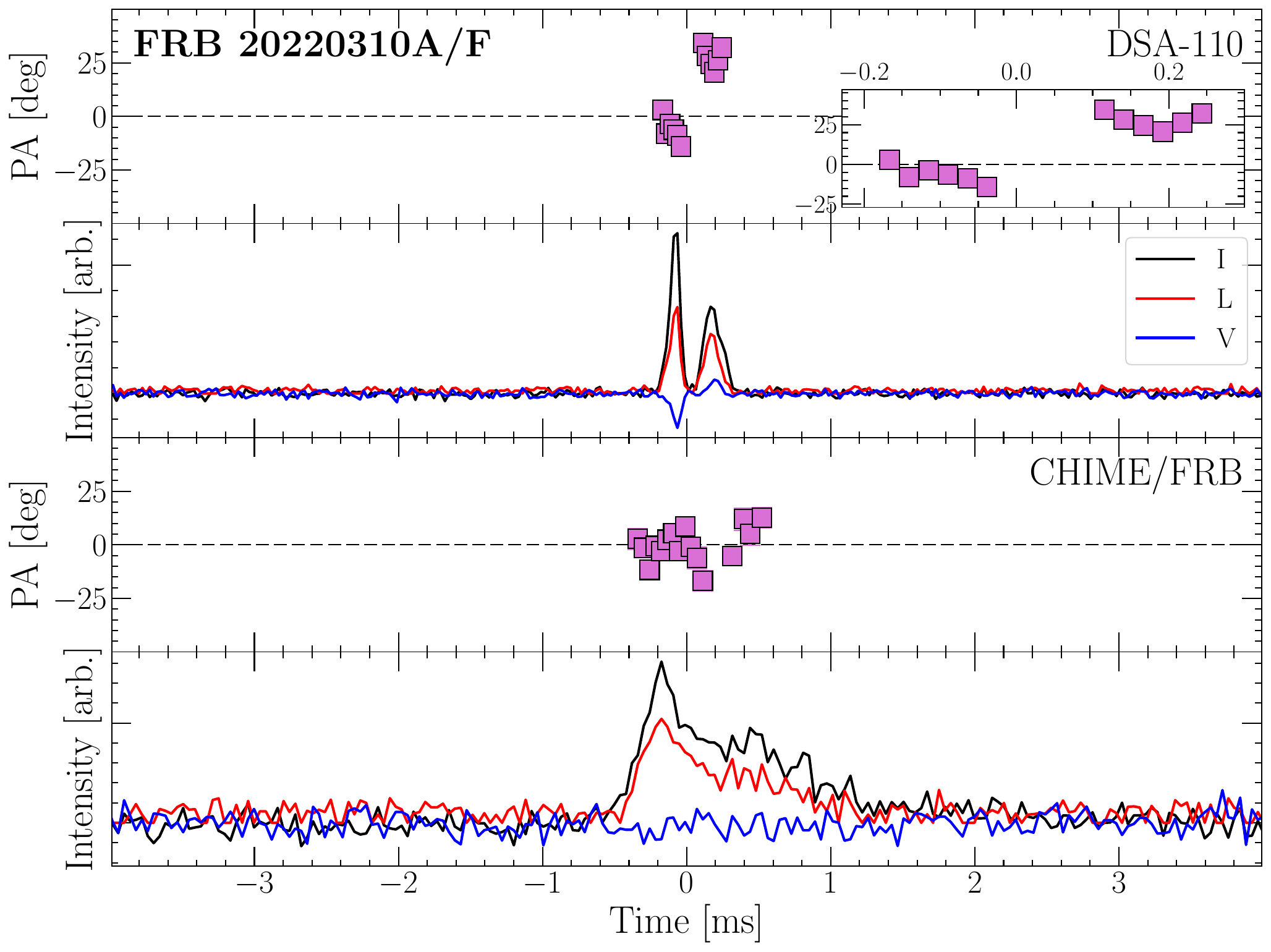}
    \includegraphics[width=0.49\textwidth]{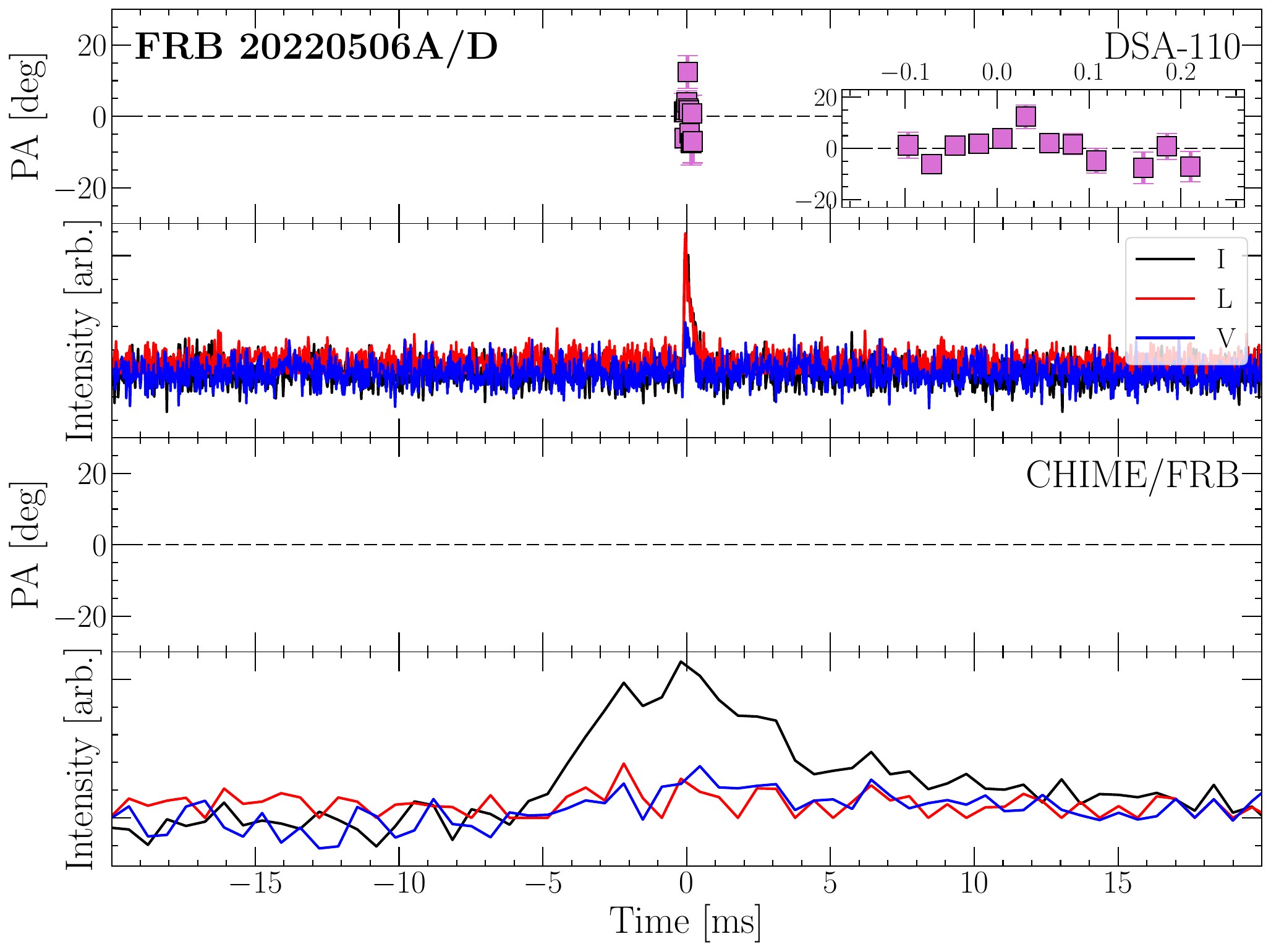}
    \includegraphics[width=0.49\textwidth]{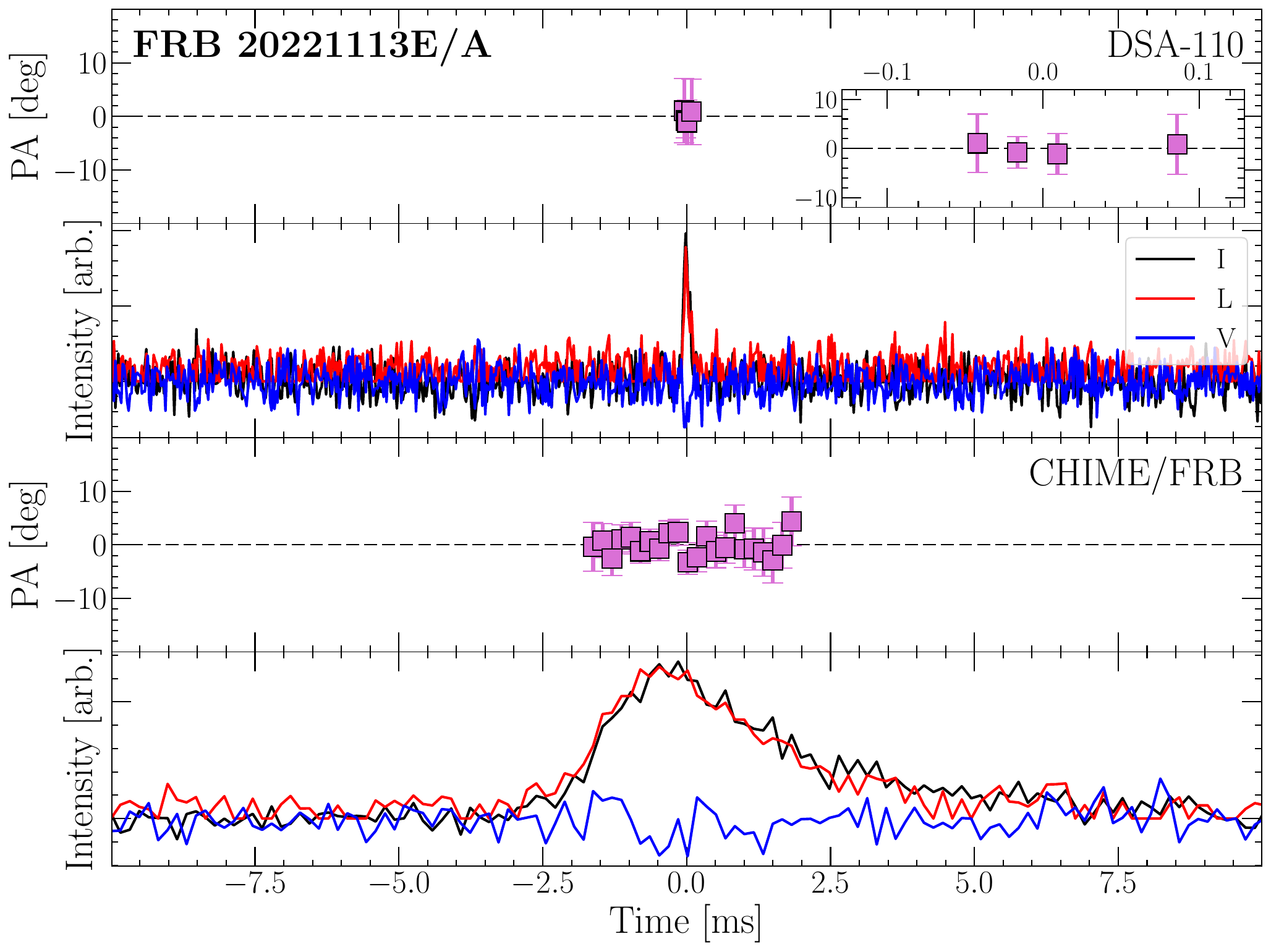}
    \includegraphics[width=0.49\textwidth]{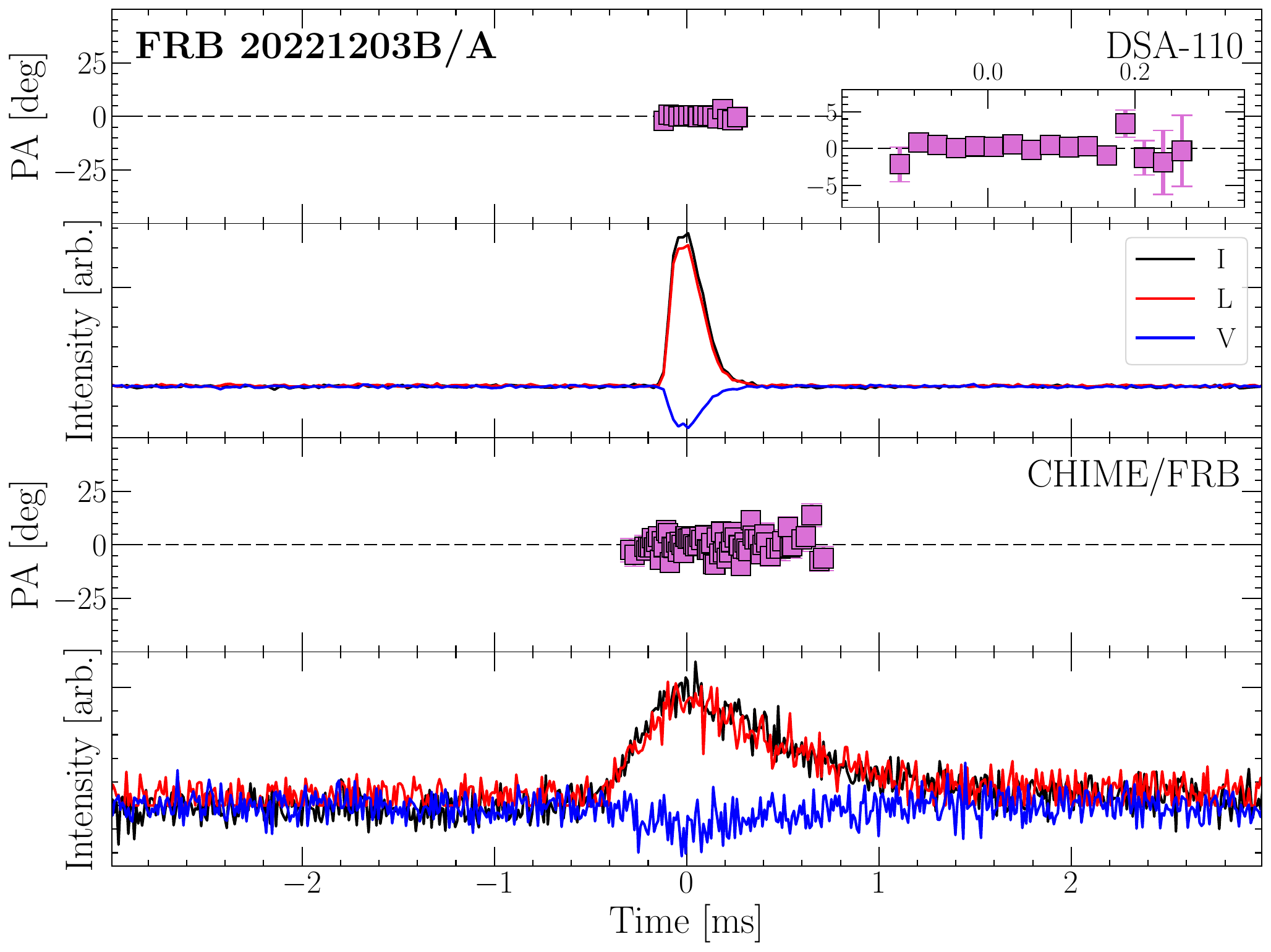}
    \includegraphics[width=0.49\textwidth]{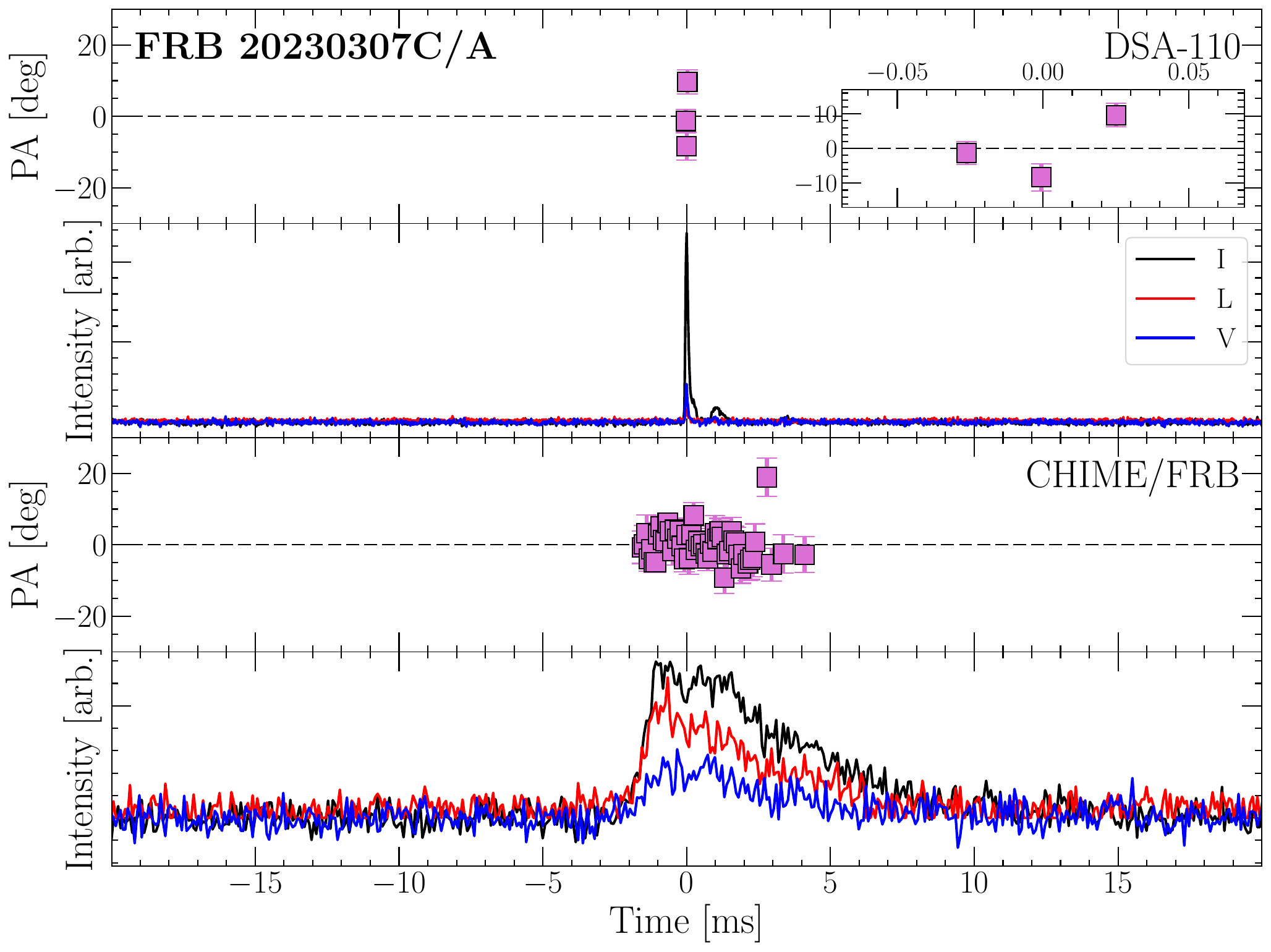}
    \caption{Temporal profiles of Stokes $I$ (black line), $L = \sqrt{Q^2 + U^2}$ (red line), $V$ (blue line), and PA (purple squares) for the co-detected FRBs summed over frequency. Insets of zoomed in DSA-110 PA profiles are provided for clarity. For each burst, DSA-110 data are plotted above the CHIME/FRB data. FRB TNS names are provided in the top-left corner of each panel and $\mathrm{PA=0~\mathrm{deg}}$ is indicated by a dashed black line.} 
    \label{fig:pol_profiles}
\end{figure*}

\begin{figure*}
    \ContinuedFloat
    \centering 
    \includegraphics[width=0.49\textwidth]{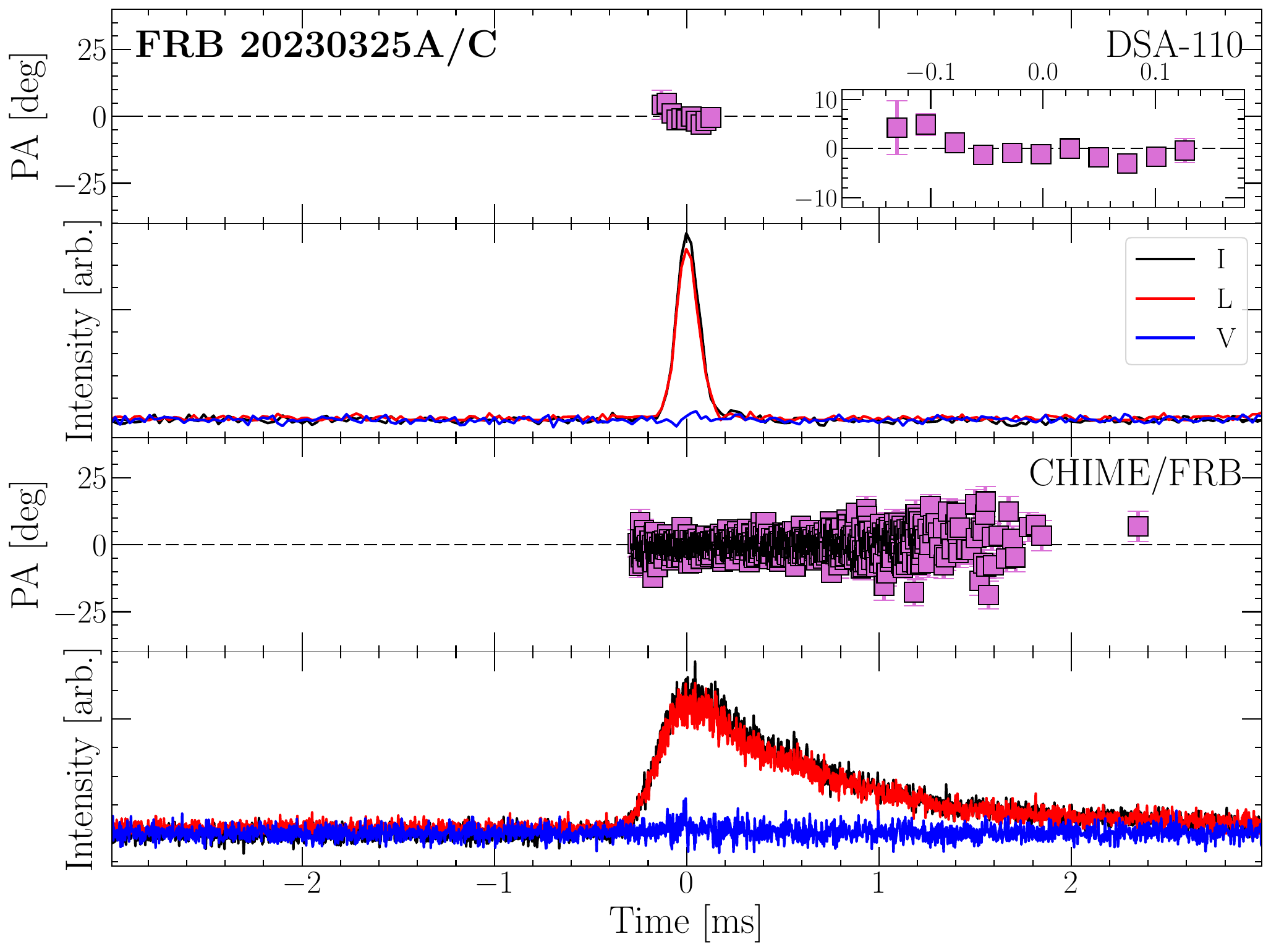}
    \includegraphics[width=0.49\textwidth]{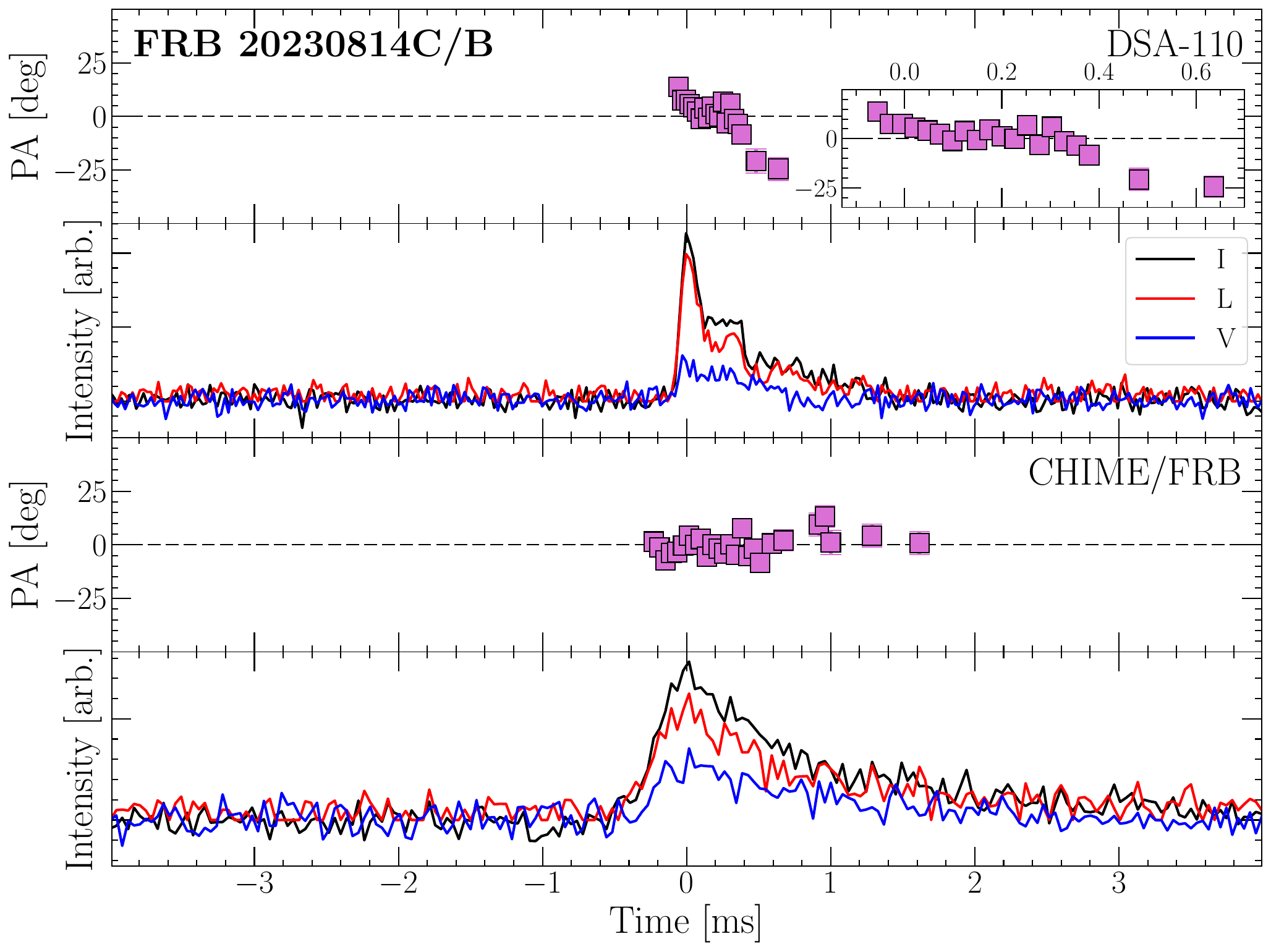}
    \includegraphics[width=0.49\textwidth]{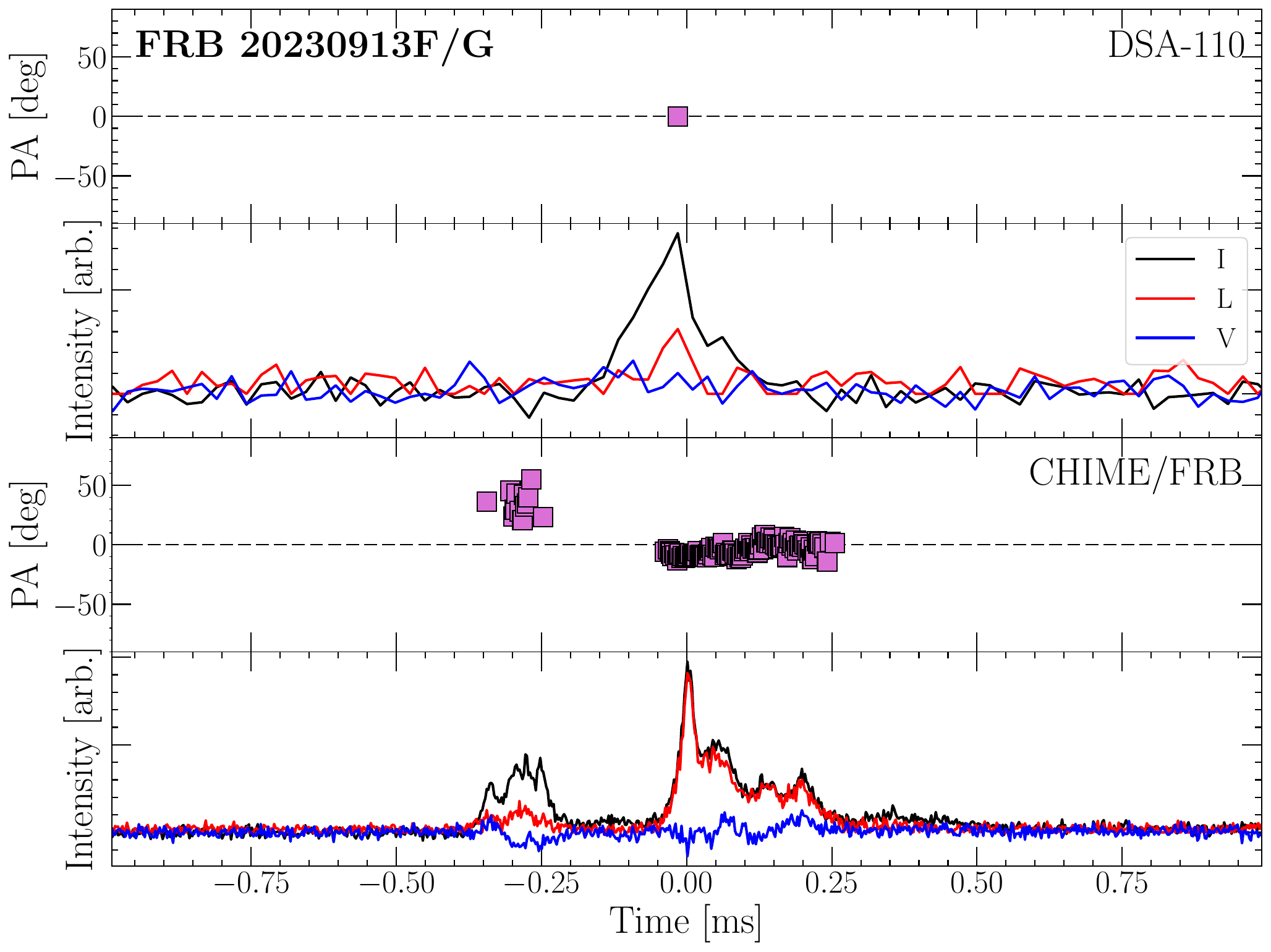}
    \includegraphics[width=0.49\textwidth]{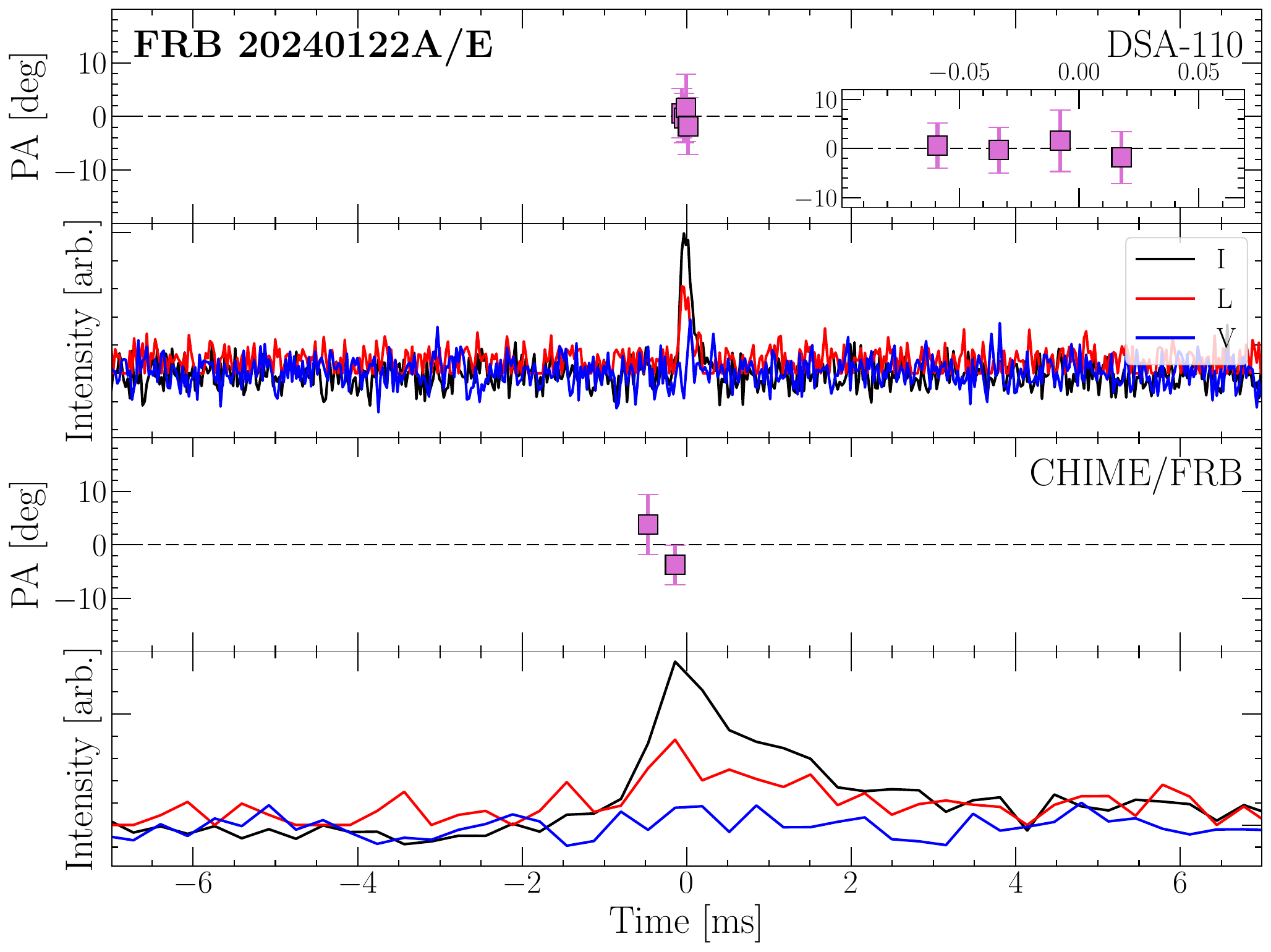}
    \includegraphics[width=0.49\textwidth]{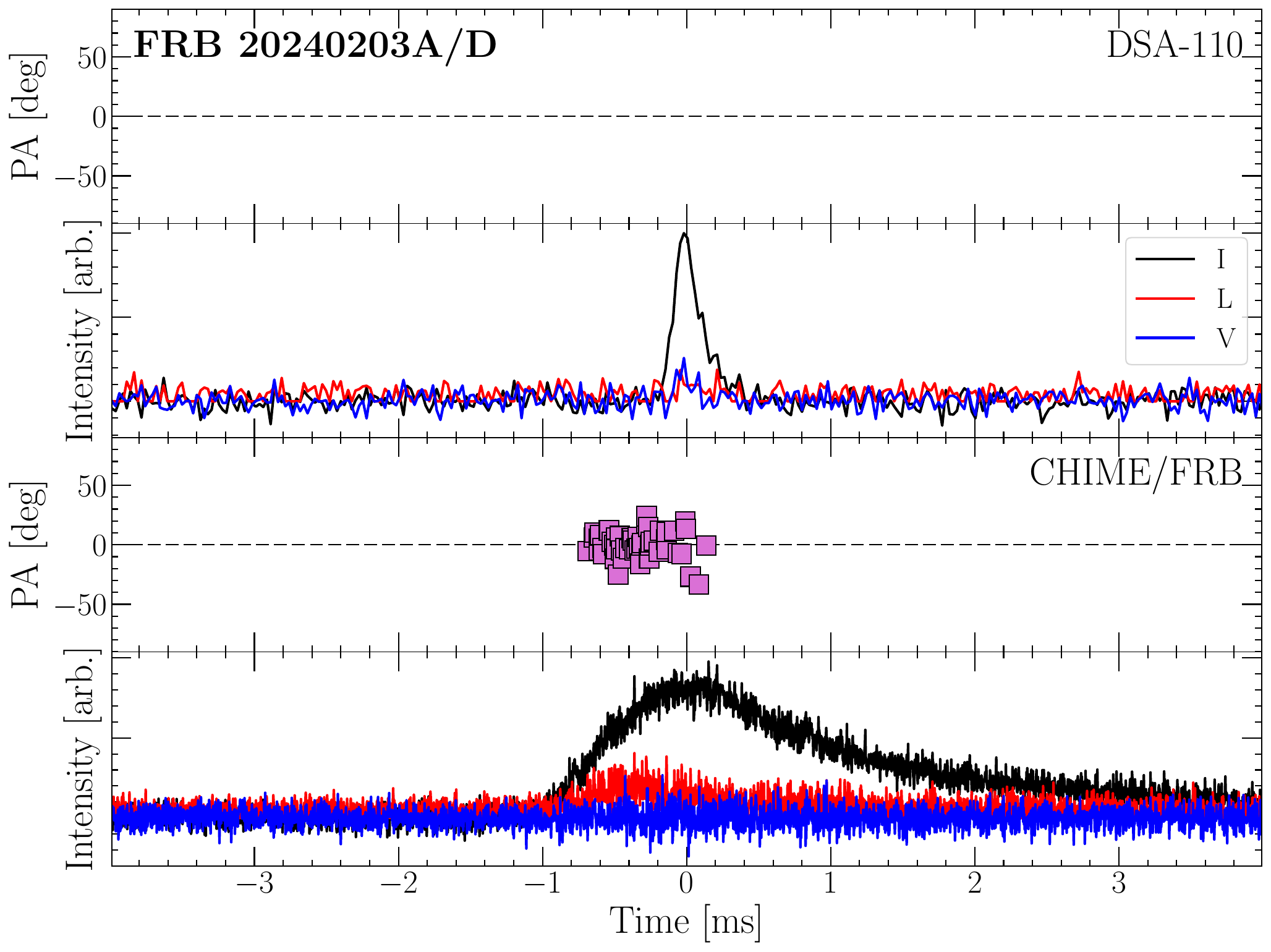}
    \includegraphics[width=0.49\textwidth]{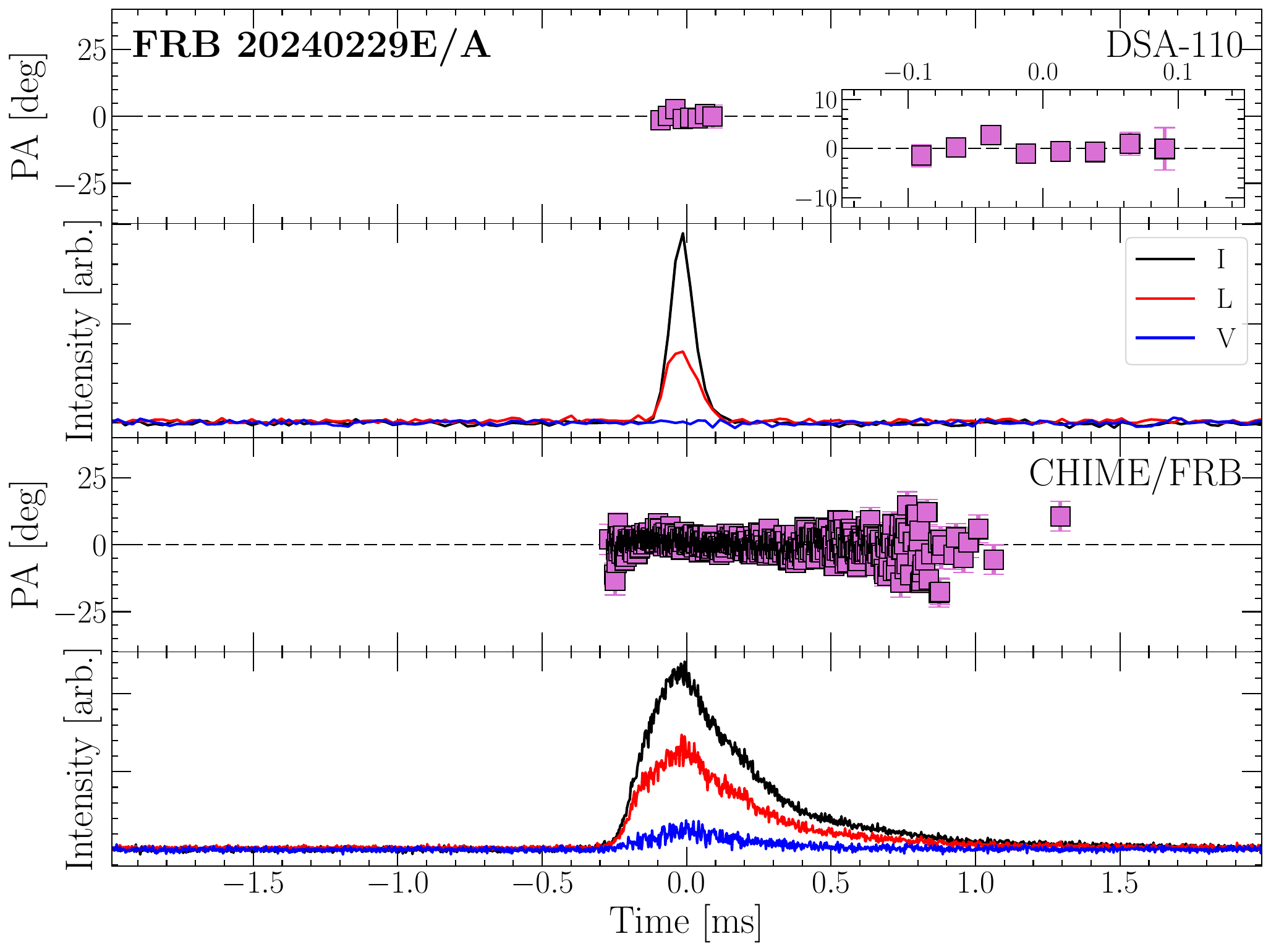}
    \caption{\emph{Continued.}} 
\end{figure*}

\section{Spectropolarimetric modeling} \label{sec:models}
We analyze the $L/I$ and PAs of the co-detected FRBs as a function of frequency, modeling their spectral properties based on frequency-dependent polarization effects that have been previously observed in FRBs and radio pulsars. We do not consider spectral evolution of the circular polarization fraction, $|V|/I$, in our analysis as CHIME/FRB data suffer from unmodeled instrumental effects that can lead to $|V|/I \sim 30\%$ uncertainties. In general, the co-detected FRBs show circular polarization fractions $0.0 \lesssim |V|/I \lesssim 0.4$ (across 1280$-$1530~MHz), with only $2/12$ FRBs having $|V|/I > 0.3$, and at 400$-$800~MHz, all bursts are consistent with the level of possible instrumental polarization, $|V|/I \lesssim 0.3$.

\subsection{Linear polarization} \label{sec:models_li}
We model the observed linear polarization fraction as a function of frequency, $(L/I)(\nu)$, using three different frameworks, each adding a layer of complexity to describe potential propagation effects arising in foreground magnetoionic plasma. These models are described below and fiducial $(L/I)(\nu)$ spectra between 400$-$1530~MHz for each of the three scenarios are presented in Figure \ref{fig:li_models}. The models are meant to be representative but they are not the only possible models to describe these physical phenomena. Qualitatively, the fits provide an adequate description of the data, suggesting that more complex modeling is not justified in the present work.

\subsubsection{Model 1: Constant L/I} \label{subsec:model_1}
\begin{figure*}
    \centering
    \includegraphics[width=0.329\textwidth]{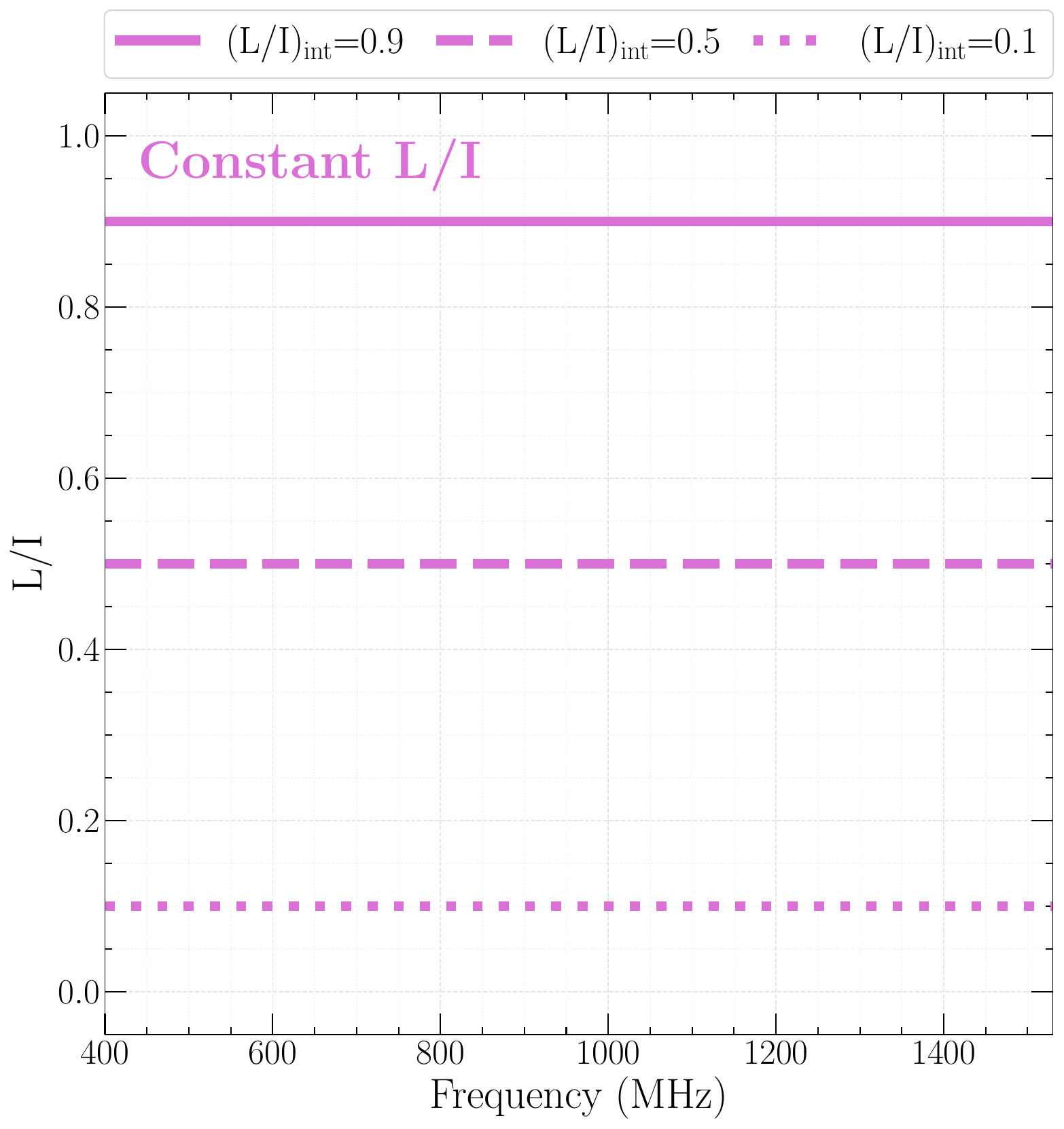}
    \includegraphics[width=0.329\textwidth]{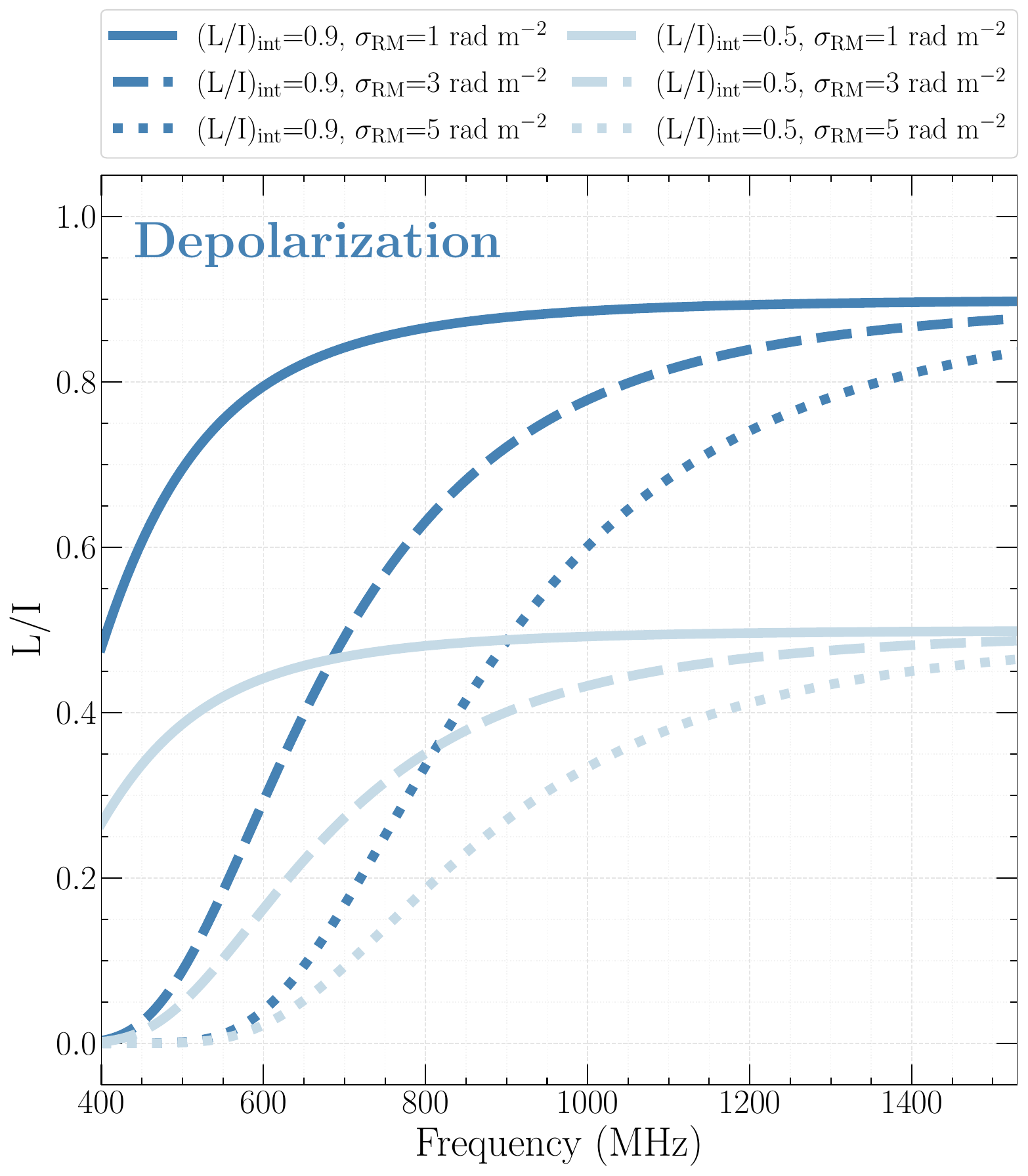}
    \includegraphics[width=0.329\textwidth]{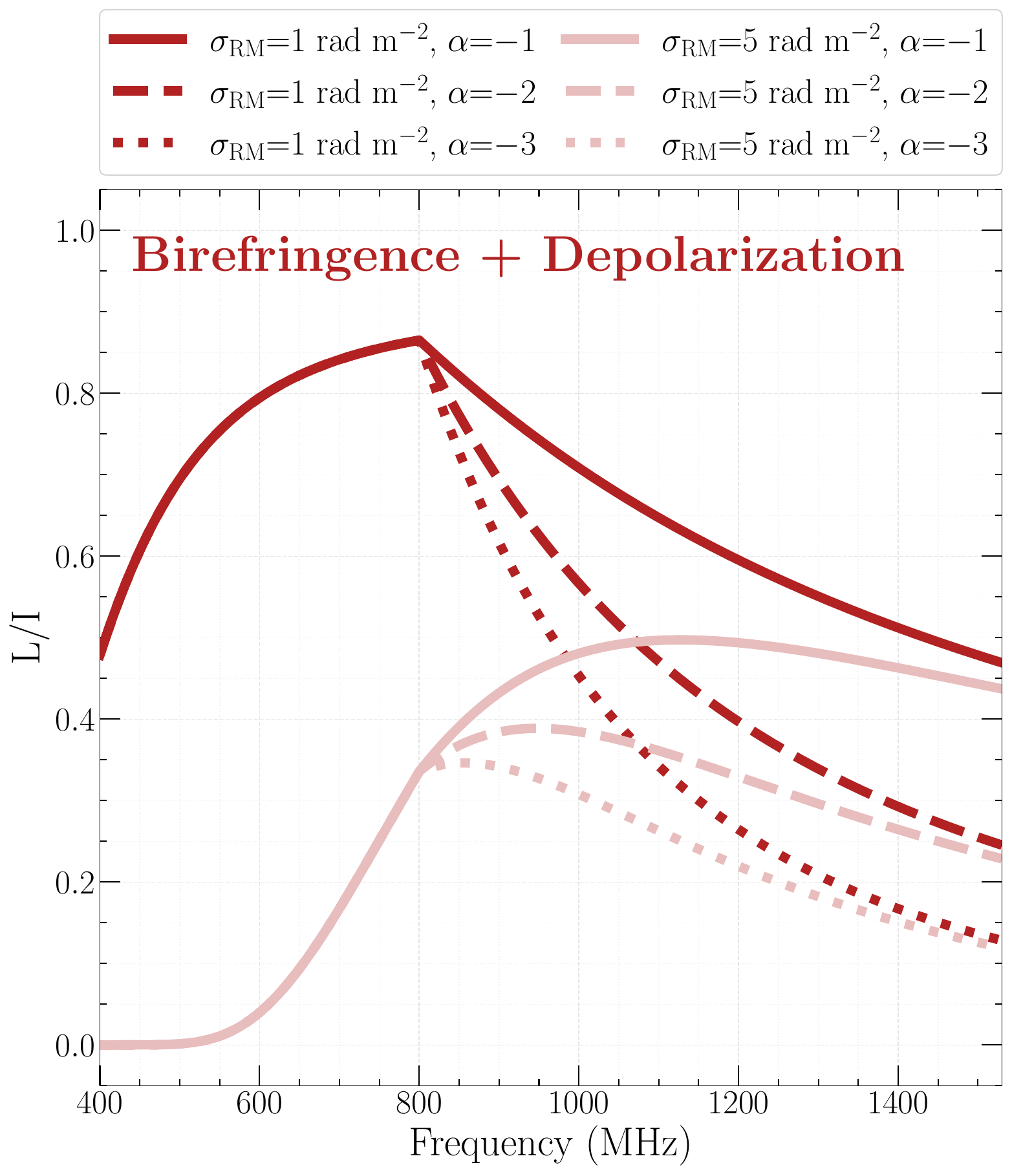}
    \caption{(Left) Toy model of a constant $L/I$ versus frequency for different values of the intrinsic linear polarization fraction, $(L/I)_\mathrm{int}$. (Middle) Toy model of depolarization following Equation \ref{eq:model_2} for various input $\sigma_\mathrm{RM}$ values with $(L/I)_\mathrm{int}=0.9$ (dark lines) and $(L/I)_\mathrm{int}=0.5$ (faint lines). (Right) Toy model combining birefringence and depolarization effects (Equation \ref{eq:model_3}). Dark lines show the model behavior for various $\alpha$ indices with $\sigma_\mathrm{RM} = 1~\mathrm{rad}~\mathrm{m}^{-2}$; faint lines show curves with $\sigma_\mathrm{RM} = 5~\mathrm{rad}~\mathrm{m}^{-2}$ and the same range of $\alpha$. All shown curves for the birefringence and depolarization model assume $(L/I)_\mathrm{int}=0.9$ and $\nu_c = 800$~MHz. The frequency range in all panels spans 400$-$1530~MHz.} 
    \label{fig:li_models}
\end{figure*}

In the most straightforward model, the FRB is emitted at a constant intrinsic linear polarization fraction, $(L/I)_\mathrm{int}$, across its emitting bandwidth, and the signal is unaffected by any propagation effects:
\begin{equation}
(L/I)(\nu) = (L/I)_\mathrm{int}\,. \label{eq:model_1}
\end{equation}
Examples of $(L/I) (\nu)$ following Equation \ref{eq:model_1} for a range of $(L/I)_\mathrm{int}$ are presented in the left panel of Figure \ref{fig:li_models}.

\subsubsection{Model 2: Depolarization} \label{subsec:model_2}
Decreasing $(L/I) (\nu)$ towards lower frequencies has been reported in a number of different radio sources, including in radio galaxies \citep[e.g.,][]{2012AJ....144..105H, 2014ApJS..212...15F}, pulsars \citep[e.g.,][]{2015A&A...576A..62N, 2019PASA...36...25X, 2021MNRAS.504..228S}, and FRBs \citep[][]{2022Sci...375.1266F, 2025MNRAS.tmp.1894U, 2026ApJ..1000L..53P, 2026arXiv260918086B, 2026arXiv260921177D}, and it has been attributed to depolarization from intervening, inhomogeneous Faraday screens.

Consider now that the FRB emitted with $(L/I)_\mathrm{int}$ initially passes through an inhomogeneous plasma screen, causing the signal to be scattered to a finite angular size. If this scattered signal then propagates through inhomogeneous, magnetized plasma, it undergoes differential Faraday rotation due to intervening spatial fluctuations in electron density ($\delta n_e$) and/or LOS magnetic field strength ($\delta B_\parallel$) in this media. When the FRB is detected, these spatial RM variations are averaged across the angularly separated paths, leading to a decrease in $(L/I) (\nu)$ relative to $(L/I)_\mathrm{int}$ (referenced to zero wavelength):
\begin{equation}
(L/I) (\nu) = (L/I)_\mathrm{int}~\mathrm{exp} \left(-2 c^4 \nu^{-4} \sigma_\mathrm{RM}^2 \right)\,. \label{eq:model_2}
\end{equation}
Here, $c$ is the speed of light and $\sigma_\mathrm{RM}$ is the rms RM variations across scattered paths through the inhomogeneous magnetoionic medium \citep{1966MNRAS.133...67B}. The middle panel of Figure \ref{fig:li_models} shows the expected $(L/I) (\nu)$ for a range of fiducial $\sigma_\mathrm{RM}$ and $(L/I)_\mathrm{int}$.

Equation \ref{eq:model_2} assumes a Gaussian distribution of RM fluctuations across the resolved source. Under other physical assumptions, e.g., if the RM structure function follows a power-law or if the RM fluctuations trace anisotropic turbulence, $(L/I) (\nu)$ may take on a power-law, or more complex, spectral shape \citep{1991MNRAS.250..726T, 2016ApJ...818..178L, 2016ApJ...825..154Z}, though the overall trend would still have linear polarization levels decreasing at lower frequencies. We choose to use the depolarization formulation described in Equation \ref{eq:model_2} for ease of comparison to previous FRB studies \citep[e.g.,][]{2022Sci...375.1266F, 2025MNRAS.tmp.1894U, 2026ApJ..1000L..53P, 2026arXiv260918086B, 2026arXiv260921177D}. However, understanding that this is a simplified case, we refrain from deriving the magnetoionic properties of the media giving rise to the depolarization (i.e., the density fluctuations in electron density, $\delta n_e$, or in the LOS magnetic field, $\delta B_\parallel$). Instead, we treat $\sigma_\mathrm{RM}$ as an analog for the level of inhomogeneity in intervening magnetoionic plasma that we can compare between FRB sources.

\subsubsection{Model 3: Birefringence and depolarization} \label{subsec:model_3}
Many complex polarization signatures in coherent radio transients, as a function of frequency and pulse phase, are thought to arise from propagation through birefringent plasma in a compact object magnetosphere, such as in pulsars \citep[e.g.,][]{1981A&A...100..107M, 1998MNRAS.301..235G, 2008MNRAS.388..261J, 2015A&A...576A..62N} and in long period radio transients \citep[e.g.,][]{2024MNRAS.535..909D, 2026arXiv260307857P}. A key spectral signature of this phenomena is decreasing linear polarization fraction with increasing frequency, which has been observed in many radio pulsars across several octaves in frequency \citep[e.g.,][]{2015A&A...576A..62N}.

Standard, dipolar radio emission from pulsars is thought to originate in the magnetosphere of the neutron star above their polar caps,\footnote{This is not necessarily true for all modes of transient radio emission from pulsars, e.g., giant pulses or sporadic magnetar bursts likely originate farther from the neutron star surface.} with the emitting frequency relating to the height above the neutron star surface where the emission is generated \citep[sometimes referred to as ``radius-to-frequency mapping'';][]{1975ApJ...196...51R}. Their polarized emission is thought to be comprised of two orthogonal propagation modes, the ordinary and extraordinary modes \citep[e.g.,][]{1975ApJ...196...83M}. The differing, frequency-dependent refractive indices of these two modes cause their beaming angles to diverge as they propagate through a birefringent medium, such as a neutron star magnetosphere \citep{1986ApJ...302..138B}. The angular separation between the two modes increases at lower frequencies (higher emission heights) where refraction is stronger such that the observer may only see emission from one of the two modes, which are each highly polarized. At higher frequencies (lower emission heights), the beaming angle between the two modes decreases and the observed emission becomes depolarized from the mixing of the orthogonal polarization modes. Empirically, \cite{1973ApJ...179L...7M} describe the $(L/I) (\nu)$ of these radio pulsars as a power-law decrease above a characteristic frequency, $\nu_c$, where the orthogonal modes become increasingly mixed, and the $(L/I) (\nu)$ is constant below $\nu_c$, when the observed emission is dominated by only one mode. 

If FRB emission is produced in the magnetosphere of a highly magnetized neutron star, it is plausible that the FRB spectra may have signatures of depolarization at both high (due to the birefringent magnetosphere) and low (due to propagation through inhomogeneous magnetoionic plasma) frequencies. We combine these two effects in Model 3, such that the observed $(L/I) (\nu)$ of the FRB emission is:
\begin{multline}
(L/I) (\nu) = ~\mathrm{exp} \left(-2 c^4 \nu^{-4} \sigma_\mathrm{RM}^2 \right)\\
\times
\begin{cases}
    (L/I)_\mathrm{int}, & \nu \leq \nu_c \\
    (L/I)_\mathrm{int} \left(\frac{\nu}{\nu_c}\right)^{\alpha}, & \nu > \nu_c 
\end{cases}
\,. \label{eq:model_3}
\end{multline}
Example $(L/I) (\nu)$ spectra for fiducial $(L/I)_\mathrm{int}$, $\nu_c$, $\sigma_\mathrm{RM}$, and $\alpha$ are plotted in the rightmost panel of Figure \ref{fig:li_models}.

\subsection{Polarization angle} \label{sec:models_pa}
From the derotated Stokes $Q$ and $U$ dynamic spectra (denoted by the subscript ``derot'' for clarity) we can derive the observed PA as a function of time, $t$, across the burst envelope:
\begin{equation}
\mathrm{PA}(t) = \frac{1}{2} \mathrm{arctan} \left( \frac{U_\mathrm{derot}(t)}{Q_\mathrm{derot}(t)} \right)\,. \label{eq:pa}
\end{equation}
PA uncertainties are calculated by propagating uncertainties in the derotated Stokes $Q$ and $U$ spectra, following the procedure detailed by \cite{2019ApJ...878...92V}. We mask points in the PA profile below a linearly-polarized $\mathrm{S/N} < 5$, a conservative threshold above which the PA uncertainties have been shown to be robust \citep[see Appendix B by][]{2024ApJ...968...50P}. Since the CHIME/FRB polarization pipeline does not calibrate for an absolute PA due to unknown beam phase effects off of CHIME's meridian axis, we focus our analysis on the relative PA evolution across the burst. Hence, we center both the 400$-$800~MHz and 1280$-$1530~MHz PA profiles to have a mean of $0~\mathrm{deg}$. The PA profiles for all co-detected FRBs are shown in Figure \ref{fig:pol_profiles}. We quantify the magnitude of PA variability in a given PA$(t)$ timeseries by evaluating the goodness of fit between the observed PA profile, centered at $0~\mathrm{deg}$, and a constant PA$(t) = 0~\mathrm{deg}$ model using a reduced $\chi_\nu^2$ test. We use a threshold of $\chi_\nu^2 \geq 5$, above which we consider the PA profile to be variable, as prescribed by \cite{2024ApJ...968...50P}.

We aim to characterize whether PA profiles of scattered FRBs become flattened at lower frequencies due to phase-mixing of signal in the scattering tail. In the FRB literature, this effect is most clearly seen in FRB~20250316A, in which a clear PA swing of $30~\mathrm{deg}$ at $800$~MHz becomes suppressed by scatter broadening at $<500$~MHz, where the PA is almost entirely constant across the burst envelope \citep[see Figure 5 by][]{2025ApJ...989L..48C}. We search for such signals in our co-detected FRBs by applying the following criteria: (i) $\chi_\nu^2 \geq 5$ at 1280$-$1530~MHz while $\chi_\nu^2 < 5$ at 400$-$800~MHz and (ii) the amplitude of the maximum PA variation across the profile decreases from 1280$-$1530 to 400$-$800~MHz.

\section{Results} \label{sec:results}
\subsection{FRB magnetoionic environments} \label{sec:results_host_env}
The observed RMs, $\mathrm{RM}_\mathrm{obs}$, are summarized along with subsequent results in Table \ref{tb:host_props}. Performing RM-synthesis on joint CHIME/FRB$-$DSA-110 spectra only marginally improves the theoretical RM uncertainty while generating artifacts in the FDF due to differences in the relative polarized intensity scalings between instruments and irregular frequency sampling. Thus, we opt to report the $\mathrm{RM}_\mathrm{obs}$ from the CHIME/FRB data as the lower frequency and larger bandwidth observations provide greater accuracy in the RM determination. We confirm that the FDF peaks of the CHIME/FRB and DSA-110 data agree to within their respective full width at half maximum when robust RM detections are possible in both data sets. Independent polarization properties of many of these FRBs are provided in catalogs by the CHIME/FRB (M.~Ng et al.~in prep.) and DSA-110 teams \citep{2026arXiv260930616V}, respectively. The observed DM (denoted $\mathrm{DM}_\mathrm{obs}$) and spectroscopic host galaxy redshift, $z$, of these FRBs are reported in Paper I and adopted for this work. 

$\mathrm{DM}_\mathrm{obs}$ and $\mathrm{RM}_\mathrm{obs}$ can be written as the sum of individual contributions from plasma along the LOS:
\begin{multline}
\mathrm{DM}_\mathrm{obs} = \mathrm{DM}_{\mathrm{ion}} + \mathrm{DM}_{\mathrm{disk}} + \mathrm{DM}_{\mathrm{halo}} \\
+ \int_{0}^{z}\mathrm{DM}_{\mathrm{IGM}}(z')~\mathrm{d}z' + \frac{\mathrm{DM}_{\mathrm{host}}}{(1 + z)}~\mathrm{pc}~\mathrm{cm}^{-3}\,; \label{eq:dm_comps}
\end{multline}
\begin{multline}
\mathrm{RM}_\mathrm{obs} = \mathrm{RM}_{\mathrm{ion}} + \mathrm{RM}_{\mathrm{disk}} + \mathrm{RM}_{\mathrm{halo}} \\
+ \int_{0}^{z}\mathrm{RM}_{\mathrm{IGM}}(z')~\mathrm{d}z' + \frac{\mathrm{RM}_{\mathrm{host}}}{(1+z)^2}~\mathrm{rad}~\mathrm{m}^{-2}\,, \label{eq:rm_comps}
\end{multline}
where ``ion" refers to the Earth's ionosphere, ``disk" and ``halo", are the Milky Way disk and halo, respectively, ``IGM" is the intergalactic medium and ``host" is the host galaxy. We can calculate the electron density-weighted average LOS magnetic field strength in the FRB host galaxy using the $\mathrm{DM}_{\mathrm{host}}$ and $\mathrm{RM}_{\mathrm{host}}$:
\begin{equation}
\left<B_{\parallel,\mathrm{host}}\right> = 1.232 \frac{\mathrm{RM}_\mathrm{host}}{\mathrm{DM}_\mathrm{host}}~\mu\text{G}\,. \label{eq:b_host}
\end{equation}

We employ Monte Carlo random sampling to determine $\mathrm{DM}_\mathrm{host}$, $\mathrm{RM}_\mathrm{host}$, and $\left<B_{\parallel,\mathrm{host}}\right>$, following the procedure described by \cite{2025ApJ...982..146P}. For each FRB, we perform 10$^4$ trials, drawing foreground DM and RM contributions from assumed distributions and using the host galaxy redshift to convert to rest-frame quantities. The median of the resulting $\mathrm{DM}_\mathrm{host}$, $\mathrm{RM}_\mathrm{host}$, and $\left<B_{\parallel,\mathrm{host}}\right>$ distributions are taken as the maximum likelihood estimates with 68\% confidence intervals representing the associated uncertainties. $\mathrm{DM}_\mathrm{obs}$ and $\mathrm{RM}_\mathrm{obs}$ are drawn from Gaussian distributions centered on their measured values and with standard deviation equal to their uncertainties. We assume that the ionospheric contribution is negligible \citep[since, typically, $\mathrm{DM}_\mathrm{ion} \lesssim 10^{-5}~\mathrm{pc}~\mathrm{cm}^{-3}$ and $\mathrm{RM}_\mathrm{ion} \lesssim 1~\mathrm{rad}~\mathrm{m}^{-2}$;][]{2016ApJ...821...66L, 2019MNRAS.484.3646S}. $\mathrm{DM}_{\mathrm{disk}}$ is drawn from a uniform distribution centered on the DM estimated by the NE2025 model towards the FRB position \citep{2026arXiv260211838O} and assuming a $20$\% uncertainty. $\mathrm{DM}_\mathrm{halo}$ is drawn from a Lognormal distribution with mean $\mathrm{log}_{10}(30~\mathrm{pc}~\mathrm{cm}^{-3})$ and a 0.2~dex standard deviation \citep{2015MNRAS.451.4277D,2020ApJ...888..105Y,2023ApJ...946...58C}. $\mathrm{RM}_{\mathrm{disk}} + \mathrm{RM}_{\mathrm{halo}}$ follow a Gaussian distribution based on the reconstructed Galactic mean RM and variance at the FRB position \citep{2022A&A...657A..43H}. $\mathrm{DM}_\mathrm{IGM}$ is sampled from the probability distribution function presented in Equation 1 of \cite{2024ApJ...965...57B}, based on the empirical $\mathrm{DM}_\mathrm{IGM} - z$ relation. $\mathrm{RM}_\mathrm{IGM}$ follows a Gaussian distribution centered at $0~\mathrm{rad}~\mathrm{m}^{-2}$ with standard deviation $6~\mathrm{rad}~\mathrm{m}^{-2}$ \citep{2010MNRAS.409L..99S}. 

The best-fit $\mathrm{DM}_\mathrm{host}$, $\mathrm{RM}_\mathrm{host}$, and $\left<B_{\parallel,\mathrm{host}}\right>$ for each co-detected FRB associated with a host galaxy that has a measured spectroscopic redshift is presented in Table \ref{tb:host_props}. For these FRBs, we find mean $\mathrm{DM}_\mathrm{host} = 230 \pm 110~\mathrm{pc}~\mathrm{cm}^{-3}$, $|\mathrm{RM}_\mathrm{host}| = 170 \pm 220~\mathrm{rad}~\mathrm{m}^{-2}$, and $\left|\left<B_{\parallel,\mathrm{host}}\right>\right| = 0.85 \pm 0.74~\mu\mathrm{G}$. Notably, FRB~20230307C/A has both the largest $\mathrm{DM}_\mathrm{host}$ and $|\mathrm{RM}_\mathrm{host}|$ across the sample; meanwhile, FRB~20240229E/A has a large $\mathrm{DM}_\mathrm{host} = 305^{+56}_{-112}~\mathrm{pc}~\mathrm{cm}^{-3}$ but its $|\mathrm{RM}_\mathrm{host}|$ is consistent with $0~\mathrm{rad}~\mathrm{m}^{-2}$. For FRB~20220207A/C, we assume that the host galaxy association and redshift reported previously by \cite{2024ApJ...967...29L} are robust; however, we note that the rest-frame results for this specific FRB should be interpreted with some caution in light of the discrepancy in its localization between observatories (see Paper I for details).

\setlength{\tabcolsep}{26pt}
\setlength\extrarowheight{2pt}
\setlength{\LTcapwidth}{1.0\textwidth}
\begin{table*}[ht!]
\begin{center}
\caption{Summary of magnetoionic host environment properties for the 12 co-detected FRBs.} \label{tb:host_props}
\begin{tabular}{ccccc}
\hline
\hline
TNS Name & $\mathrm{RM}_\mathrm{obs}$ & $\mathrm{RM}_\mathrm{host}$ & $\mathrm{DM}_\mathrm{host}$ & $\left<B_{\parallel,\mathrm{host}}\right>$\\
 & (rad m$^{-2}$) & (rad m$^{-2}$) & (pc cm$^{-3}$) & ($\mu$G)\\
\hline
FRB~20220207A/C & $-161.54(1)$ & $-170(19)$ & $140^{+19}_{-27}$ & $-1.5^{+0.3}_{-0.4}$\\
FRB~20220310A/F & $-5.3(2)$ & $+20(21)$ & $137^{+68}_{-79}$ & $+0.18^{+0.34}_{-0.19}$\\
FRB~20220506A/D & $+32(4)\tablenotemark{a}$ & $+70(28)$ & $133^{+50}_{-70}$ & $+0.68^{+0.74}_{-0.31}$\\
FRB~20221113E/A & $-13.65(9)$ & $+50(31)$ & $173^{+48}_{-76}$ & $+0.38^{+0.38}_{-0.23}$\\
FRB~20221203B/A$\tablenotemark{b}$ & $+21.0(2)$ & $-$ & $-$ & $-$\\
FRB~20230307C/A & $-473.49(9)$ & $-756(16)$ & $463^{+57}_{-127}$ & $-2.0^{+0.2}_{-0.8}$\\
FRB~20230325A/C$\tablenotemark{b}$ & $-10.48(2)$ & $-$ & $-$  & $-$\\
FRB~20230814C/B & $+23.3(1)$ & $+39(47)$ & $310^{+103}_{-154}$ & $+0.2^{+0.3}_{-0.2}$\\
FRB~20230913F/G & $-81.39(3)$ & $-163(26)$ & $286^{+59}_{-111}$ & $-0.72^{+0.17}_{-0.45}$\\
FRB~20240122A/E$\tablenotemark{b}$ & $-2.4(3)$ & $-$ & $-$ & $-$\\
FRB~20240203A/D & $+178.18(9)$ & $+221(19)$ & $134^{+23}_{-34}$ & $+2.0^{+0.7}_{-0.3}$\\
FRB~20240229E/A & $-11.21(1)$ & $+3(13)$ & $305^{+56}_{-112}$ & $+0.01^{+0.06}_{-0.06}$\\
\hline
\hline
\end{tabular}
\end{center}
\tablenotetext{$a$}{Obtained from DSA-110 data as the CHIME/FRB linearly polarized intensity was too low to robustly measure RM.}
\tablenotetext{$b$}{Does not have an associated host galaxy with a measured spectroscopic redshift.}
\end{table*}

\subsection{Linear polarization} \label{sec:results_li}
We utilize an ensemble Markov Chain Monte Carlo (MCMC) sampling routine, implemented with the {\tt emcee} package \citep{2013PASP..125..306F}, to fit the observed $L/I$ spectra of each co-detected FRB to the three models presented in Section \ref{sec:models_li} (Equations \ref{eq:model_1}$-$\ref{eq:model_3}). Assuming Gaussian uncertainties, the log-likelihood is defined as:
\begin{multline}
\mathrm{ln}~\mathcal{L} = -\frac{1}{2}\sum_i \bigg[ \left( \frac{(L/I)_i - (L/I)_{\mathrm{model,}i}}{\sigma_{(L/I),i}} \right)^2 \\
+ \mathrm{ln} \left(2 \pi \sigma_{(L/I),i}^2 \right) \bigg], \label{eq:log_likelihood}
\end{multline}
where $(L/I)_i$ is the observed linear polarization fraction and $\sigma_{(L/I),i}$ is the associated measurement uncertainty; $(L/I)_{\mathrm{model,}i}$ is the expected model $L/I$ following one of Equations \ref{eq:model_1}, \ref{eq:model_2}, or \ref{eq:model_3}. We adopt uniform priors over the following ranges: $0 \leq(L/I)_\mathrm{int} \leq 1$, $0~\mathrm{rad}~\mathrm{m}^{-2} \leq \sigma_\mathrm{RM} \leq 100~\mathrm{rad}~\mathrm{m}^{-2}$, $0~\mathrm{MHz} < \nu_c <10^4~\mathrm{MHz}$, and $-10 < \alpha < 0$. Maximum-likelihood estimates for each parameter are determined as the medians of the resulting posterior distributions, with uncertainties quoted as the $68$\% confidence intervals. We conduct model comparison in two steps: (i) we compute the Bayesian Information Criterion \citep[BIC;][]{1978AnSta...6..461S} and determine which model minimizes the BIC for a given FRB $L/I$ spectrum, and (ii) if the BIC between two or more models are $\Delta\mathrm{BIC} < 6$ \citep[following the guidelines by][]{bayesfactor1995}, and their model spectra are visually similar, we default to the model with fewer degrees of freedom. 

\begin{figure*}
    \centering \hspace{14pt}
    \includegraphics[width=0.79\textwidth]{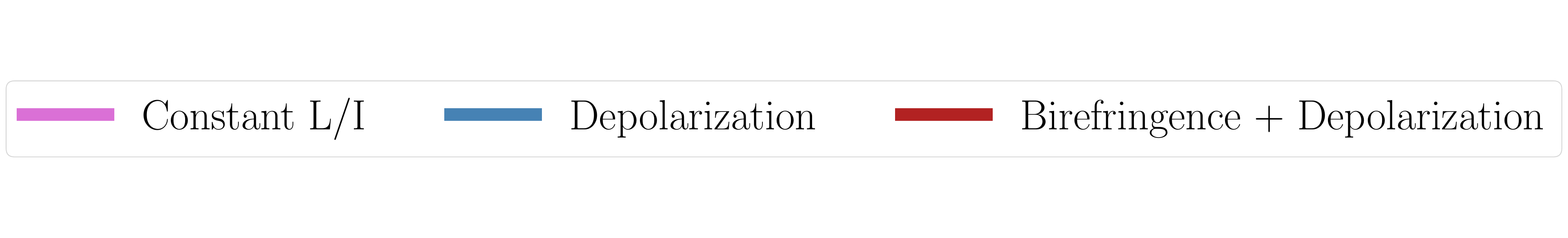}\\ \vspace{-20pt}
    \includegraphics[width=0.27\textwidth]{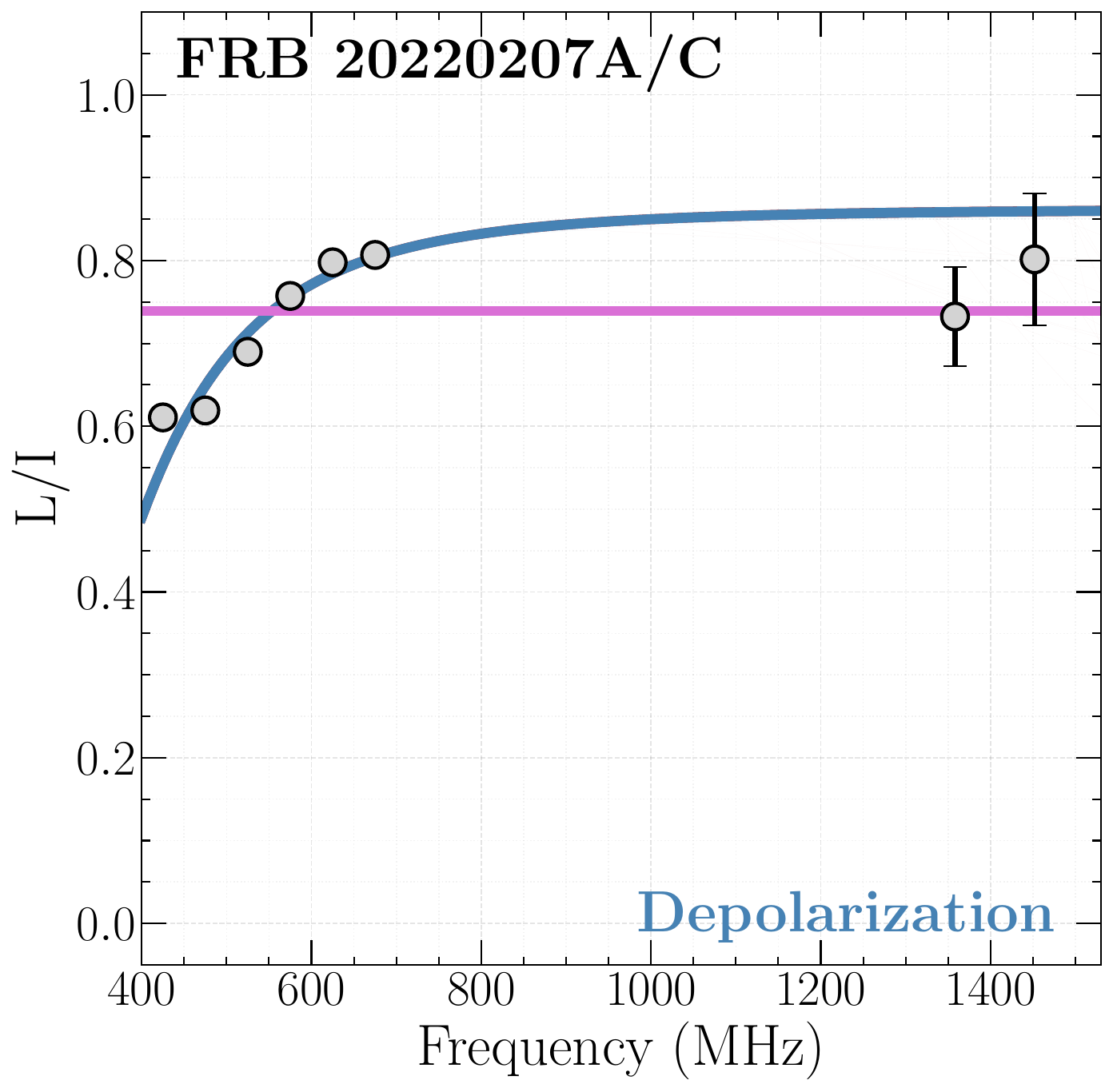}
    \includegraphics[width=0.27\textwidth]{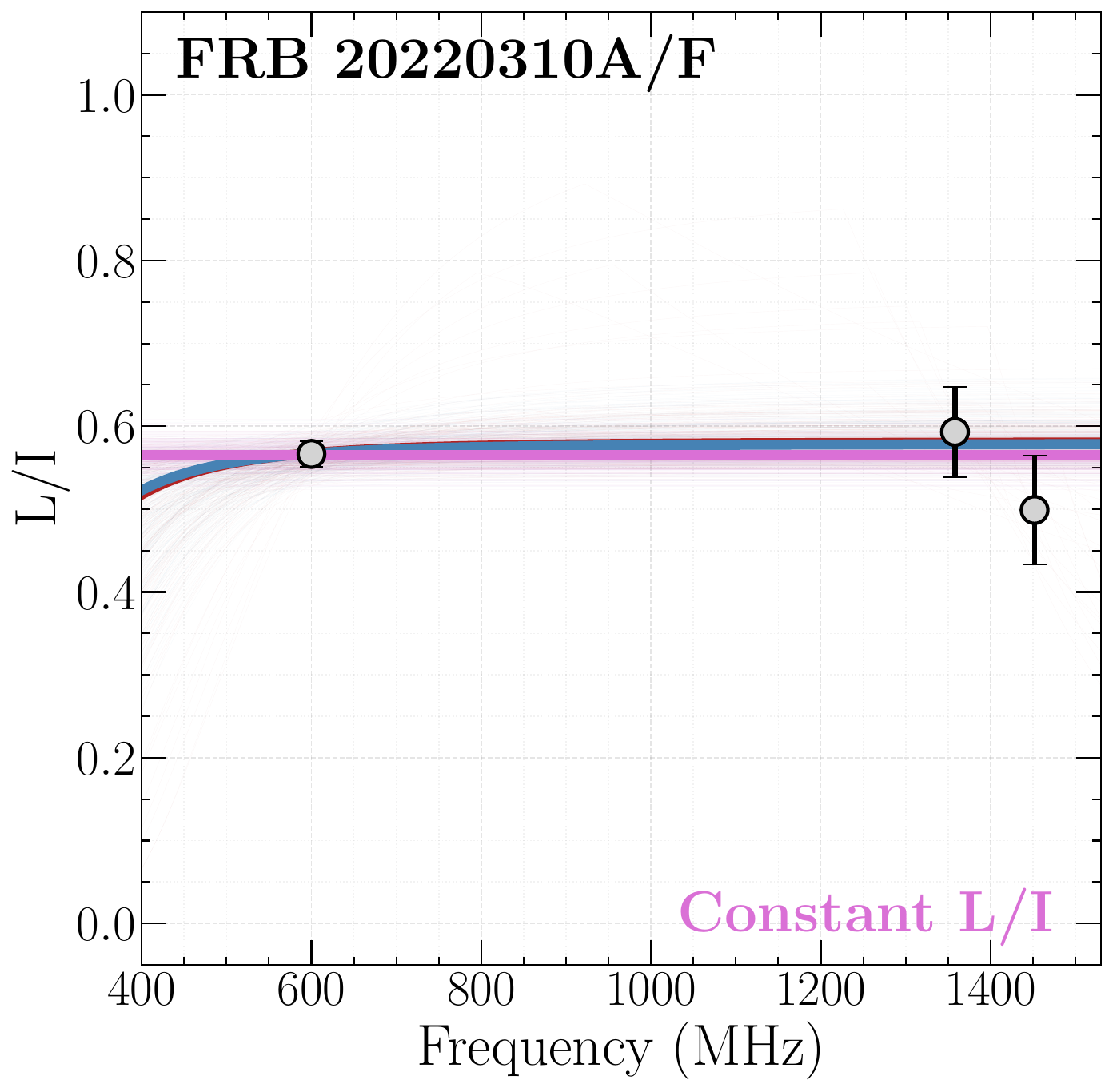}
    \includegraphics[width=0.27\textwidth]{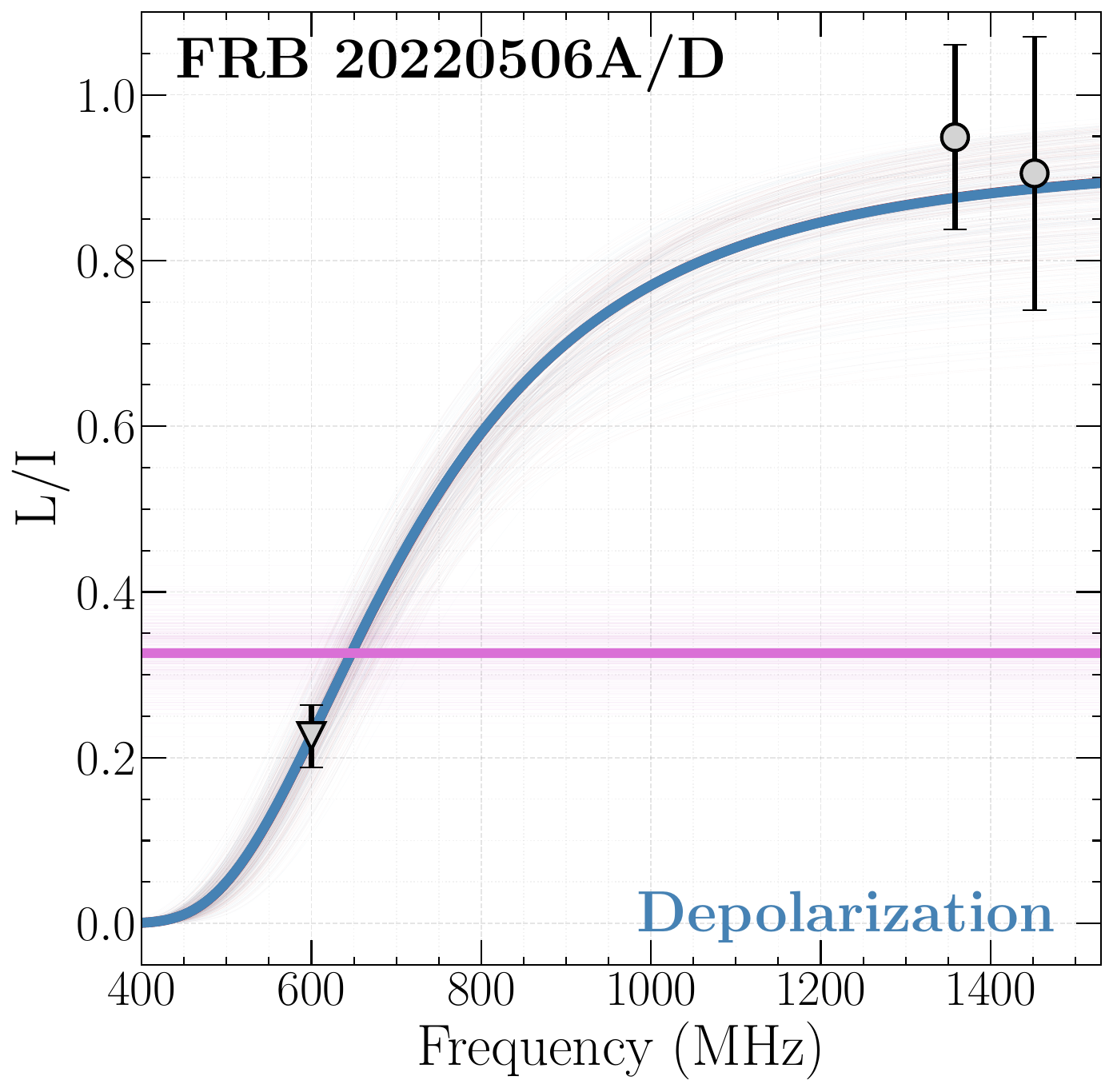}
    \includegraphics[width=0.27\textwidth]{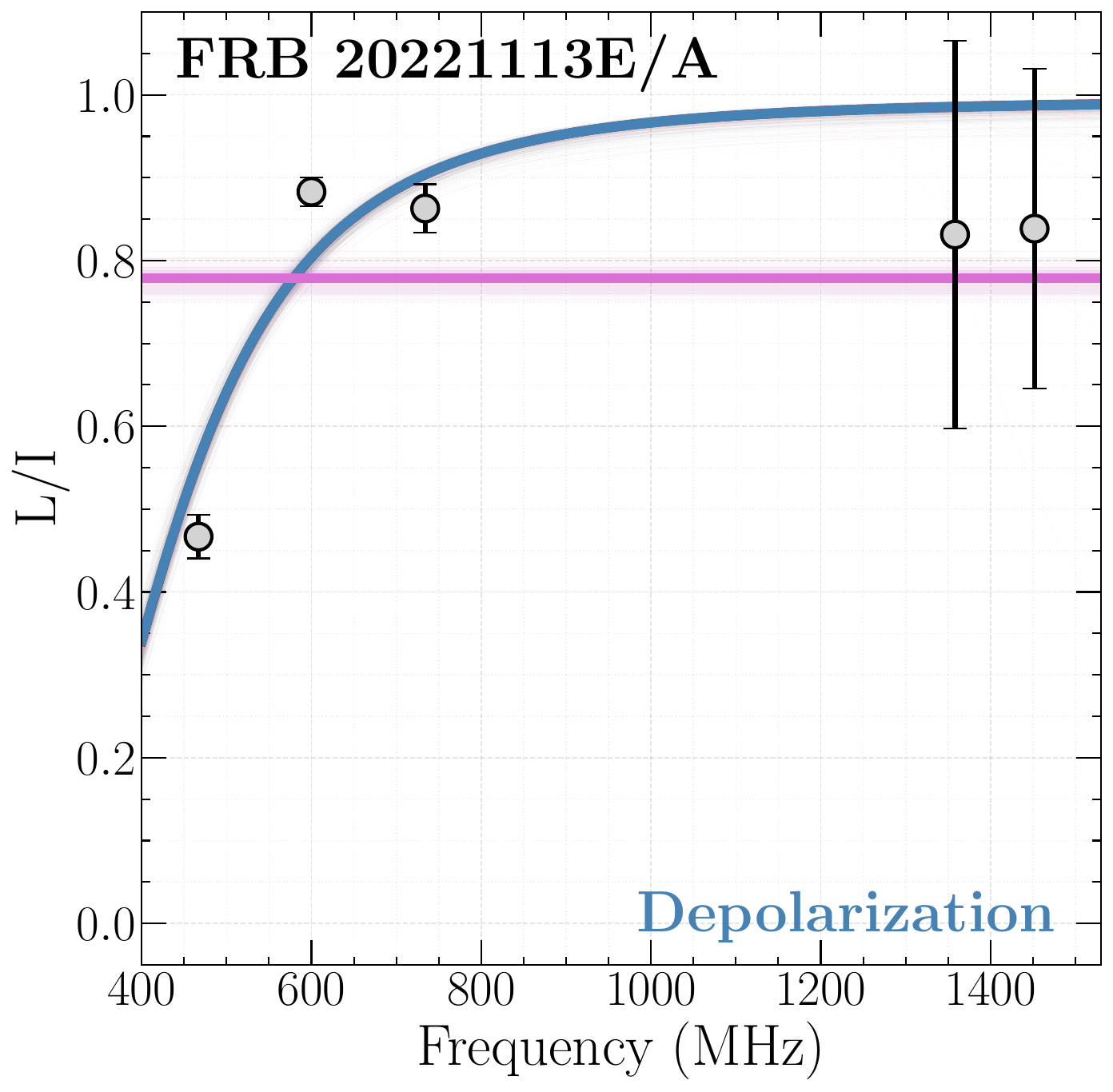}
    \includegraphics[width=0.27\textwidth]{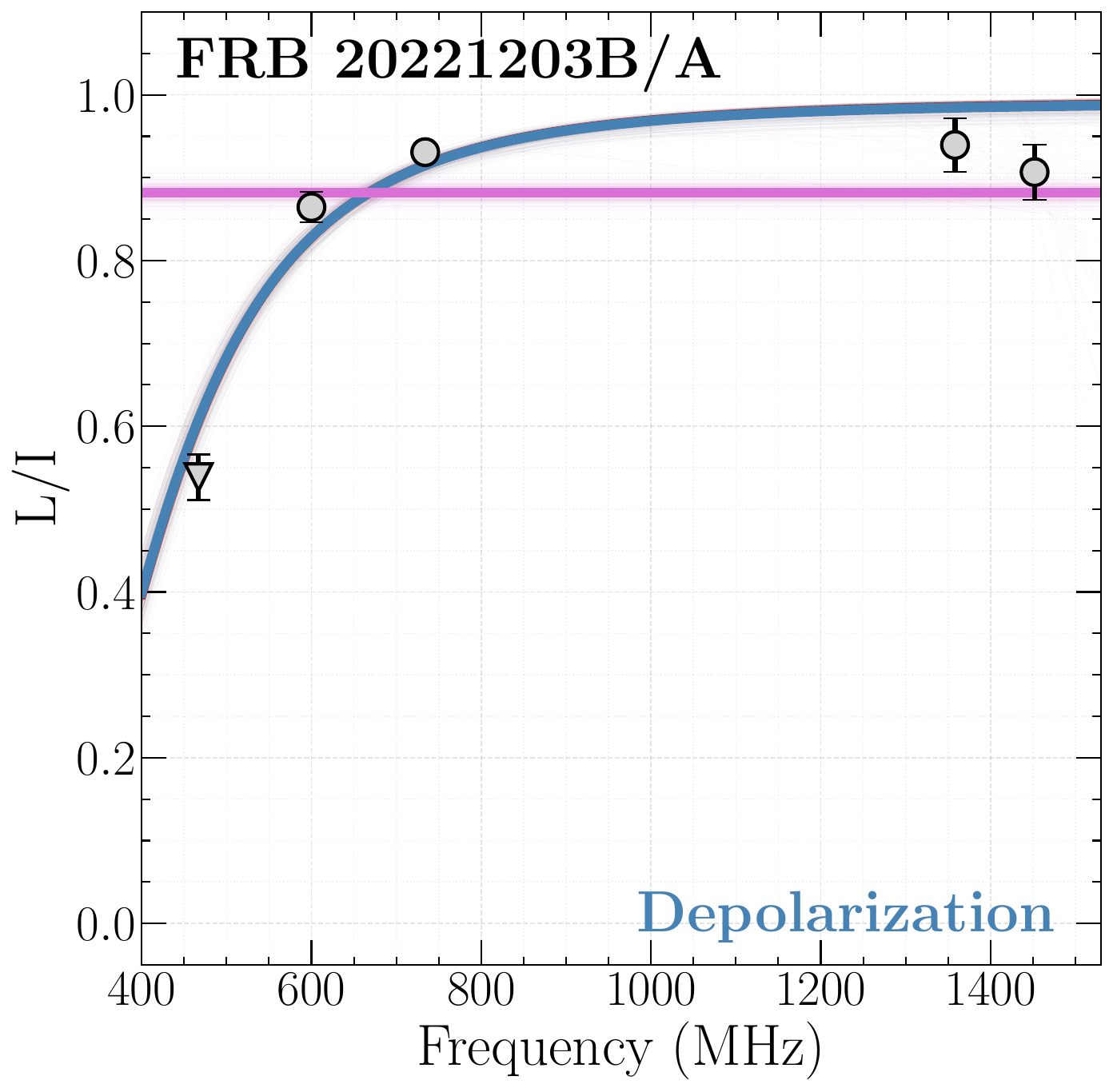}
    \includegraphics[width=0.27\textwidth]{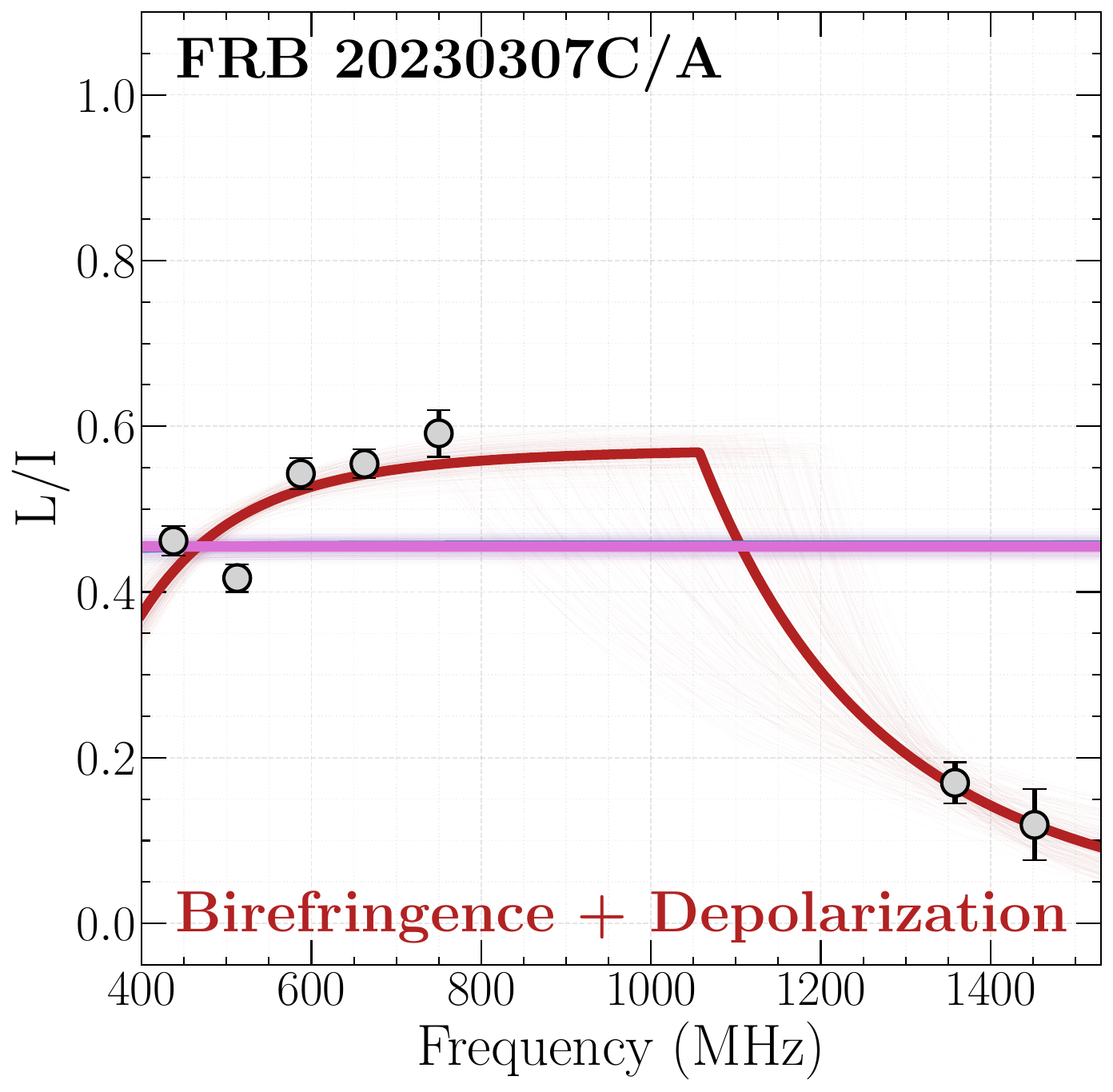}
    \includegraphics[width=0.27\textwidth]{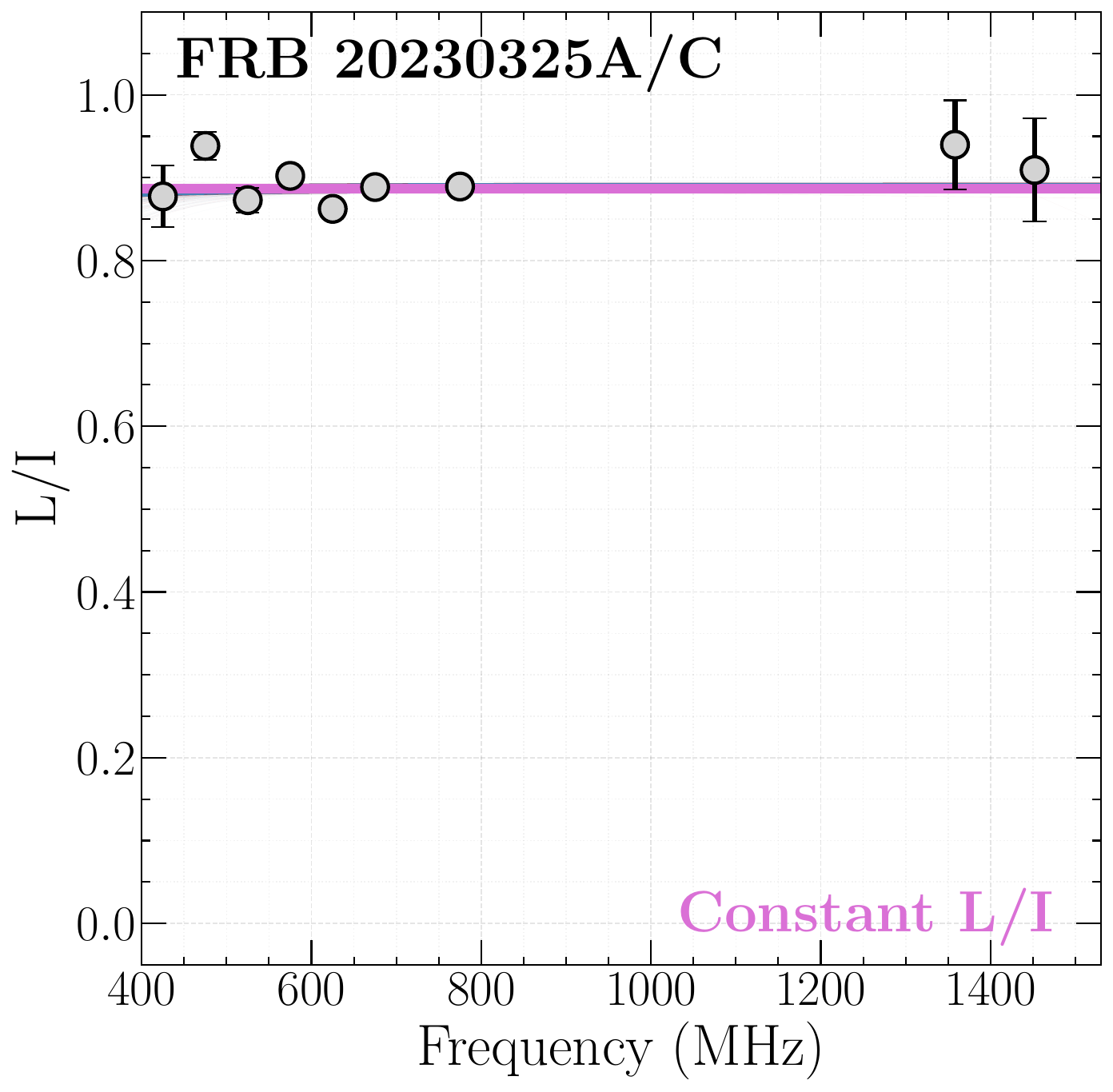}
    \includegraphics[width=0.27\textwidth]{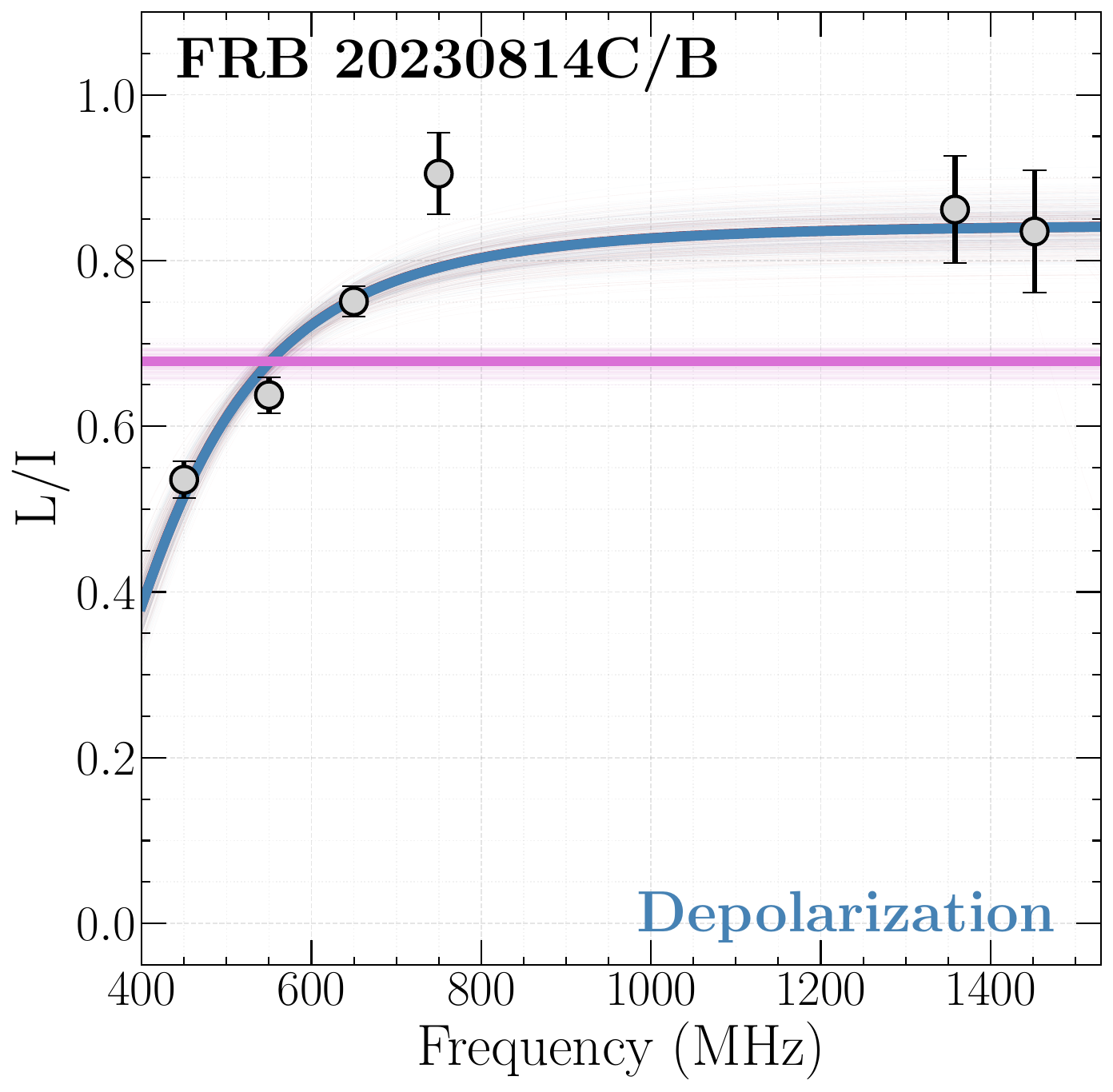}
    \includegraphics[width=0.27\textwidth]{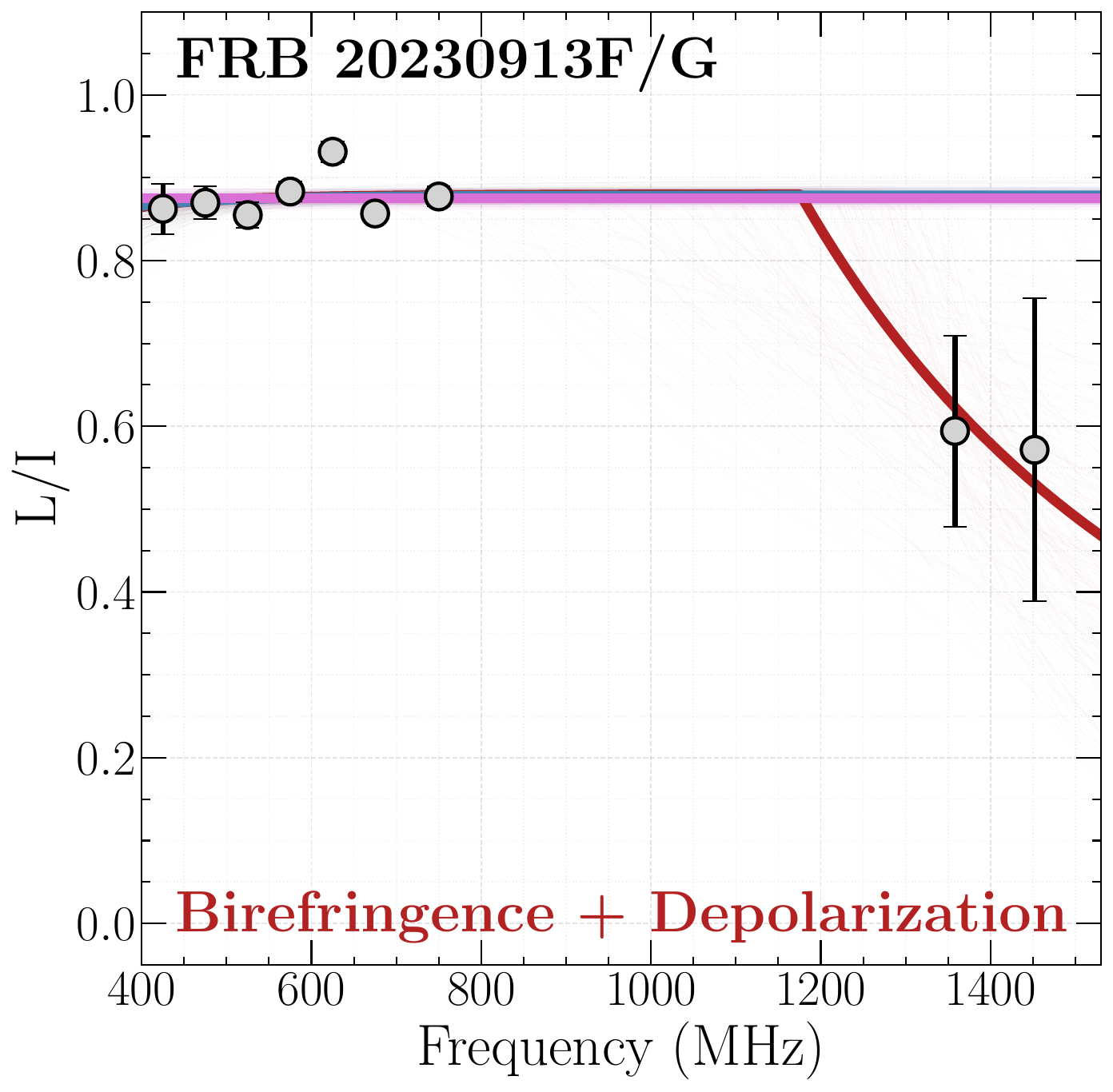}
    \includegraphics[width=0.27\textwidth]{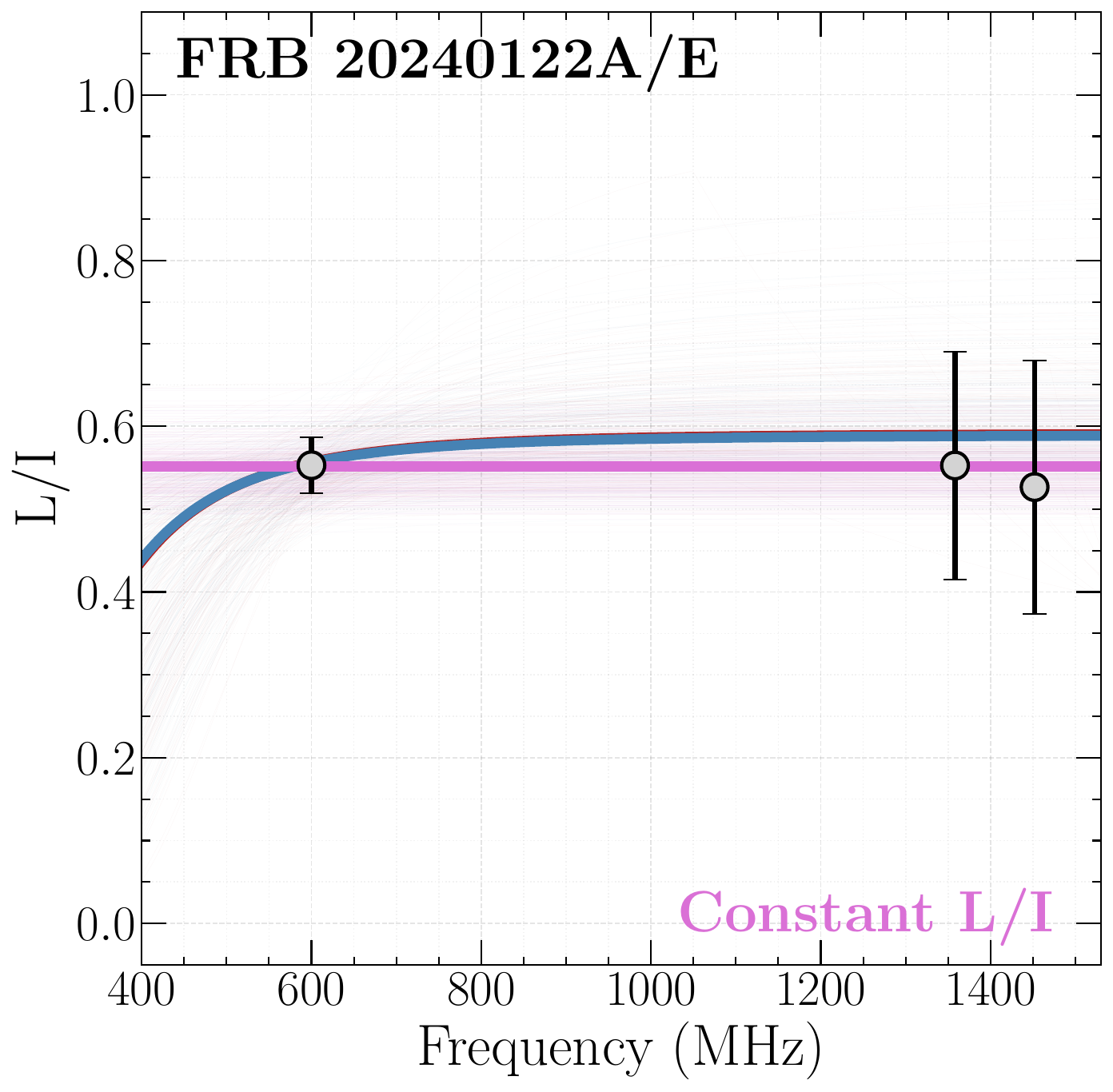}
    \includegraphics[width=0.27\textwidth]{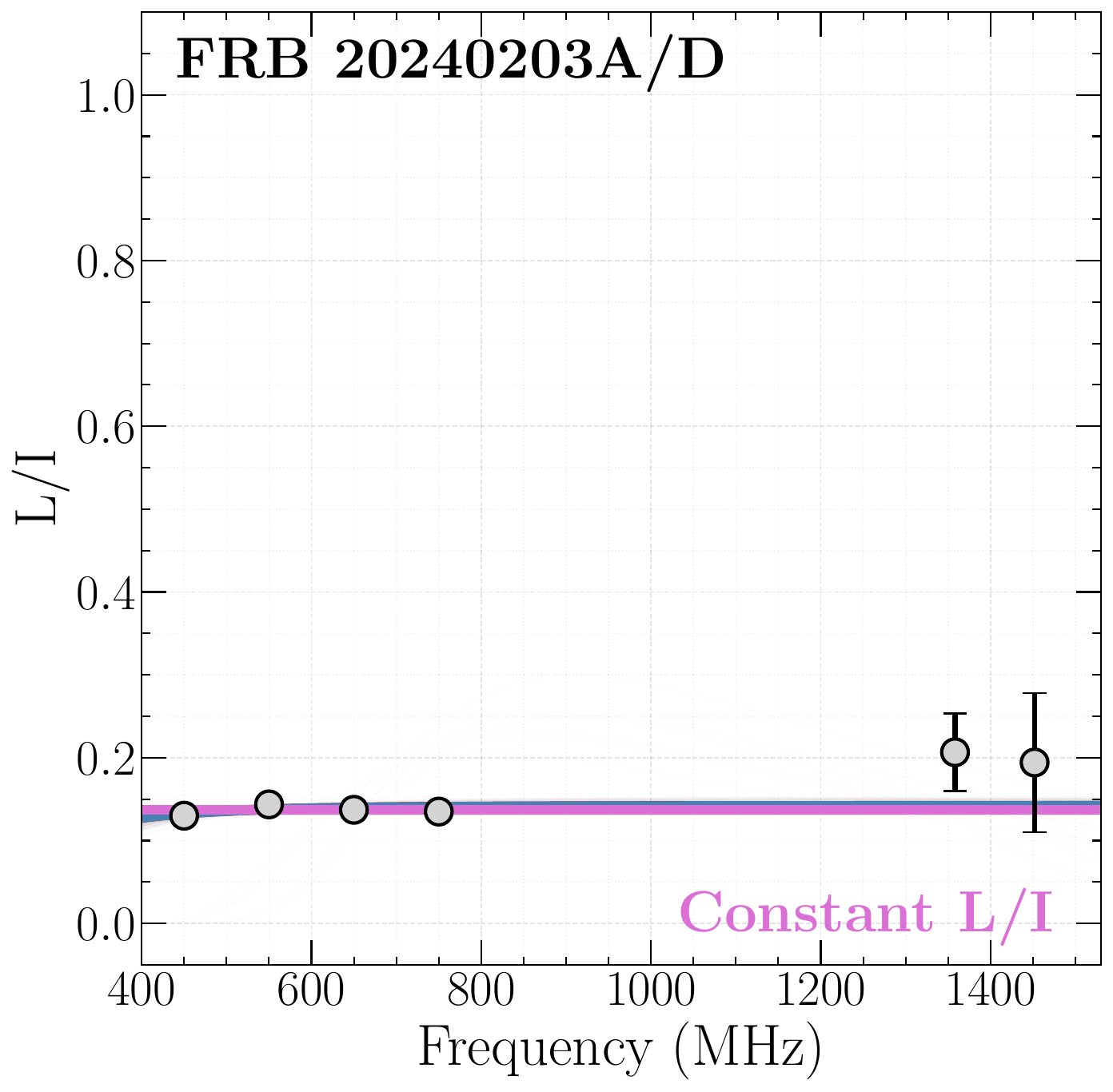}
    \includegraphics[width=0.27\textwidth]{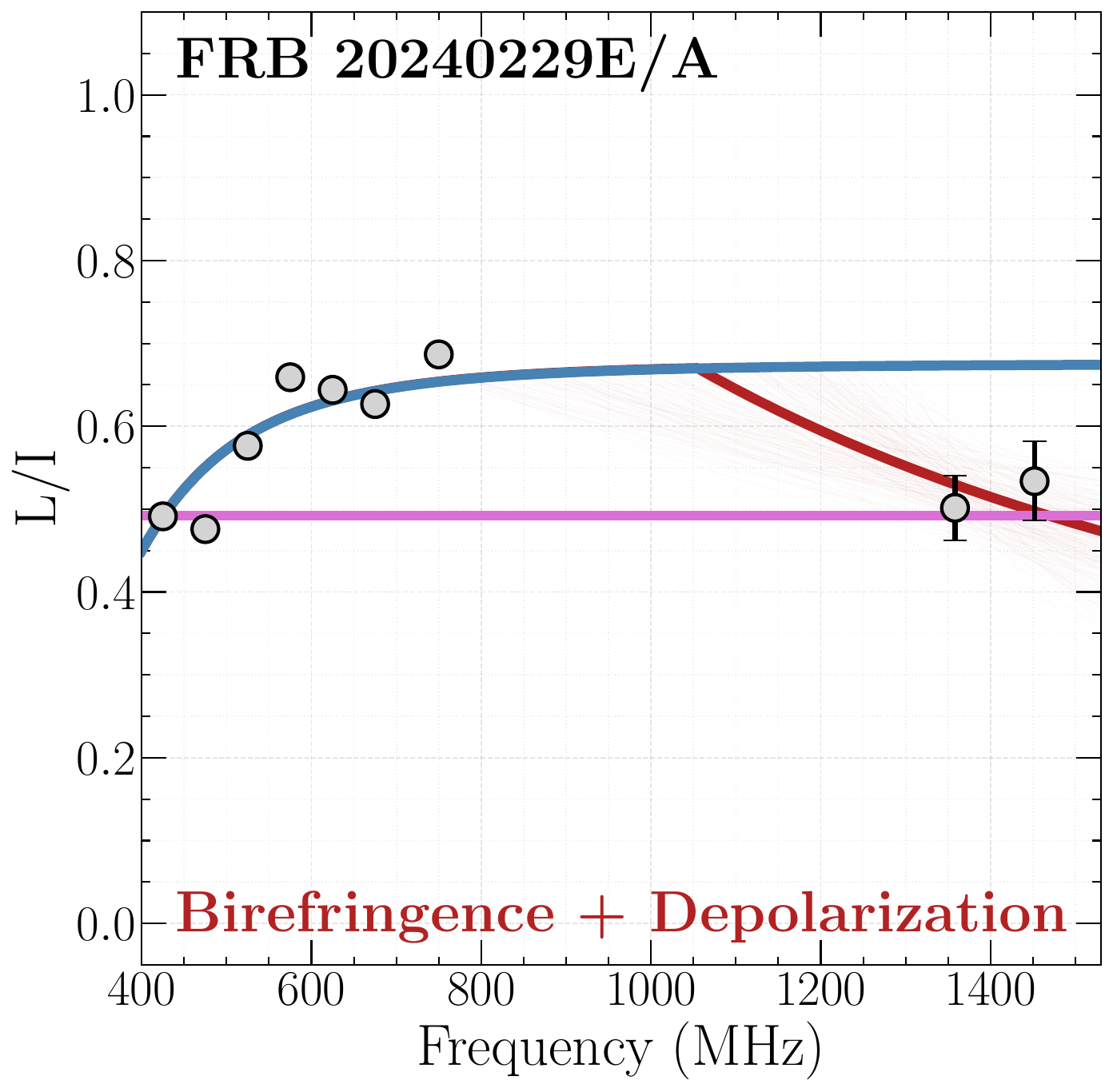}
    \caption{$(L/I) (\nu)$ across 400$-$1530~MHz for the 12 co-detected FRBs are plotted as grey circles. Models following Equations \ref{eq:model_1}, \ref{eq:model_2}, and \ref{eq:model_3} fit to each spectra are plotted as purple, blue, and red solid lines, respectively. Note that all three models are plotted in every panel, but their best-fit lines often overlap, making it hard to visually distinguish them. Individual MCMC trial fits are plotted as thinner, translucent lines in the background. The best-fit model is labeled in the bottom right of each panel.} 
    \label{fig:li_spectra}
\end{figure*}

For each co-detected FRB, we plot $(L/I) (\nu)$ and the MCMC model fits for our three models in Figure \ref{fig:li_spectra}. We enforce a minimum frequency-bin size of $50$~MHz to ensure sufficient linearly polarized signal in each bin with clear signal in Stokes $I$. We increase the bin sizes until we maximize the number of bins with robust $L/I$ measurements (i.e. bins in which we can independently derive an RM using the CHIME/FRB polarization pipeline that is consistent with that of the full burst across 400$-$800~MHz). The best-fit model is identified in the bottom right of each panel, and the maximum-likelihood estimates for those model parameters are provided in Table \ref{tb:model_params}. Note that the lowest frequency bin for both FRBs~20220506A/D and 20221203B/A are upper limits in $L/I$ but, for the purposes of our model fitting, they are treated as data points. This is specifically because each of the upper limits show a large downturn in $L/I$ from the next-highest frequency bin. By treating the upper limit as a measured $L/I$, we are effectively able to constrain a lower limit on $\sigma_\mathrm{RM}$ in these two cases, rather than making a direct measurement. For the FRBs best fit by Equation \ref{eq:model_3}, $\nu_c$ has large uncertainties due to the observing-frequency gap between the top of the CHIME/FRB and bottom of the DSA-110 bands (800$-$1280~MHz).

\setlength{\tabcolsep}{11.9pt}
\setlength\extrarowheight{2pt}
\setlength{\LTcapwidth}{1.0\textwidth}
\begin{table*}[h!]
\begin{center}
\caption{Summary of best-fit parameters from $L/I$ spectra modeling.} \label{tb:model_params}
\begin{tabular}{cccccc}
\hline
\hline
TNS Name & Best-fit model & $(L/I)_\mathrm{int}$ & $\sigma_\mathrm{RM}$ & $\nu_c$ & $\alpha$\\
 &  &  & (rad m$^{-2}$) & (MHz) & \\
\hline
FRB~20220207A/C & Depolarization & $0.86^{+0.03}_{-0.03}$ & $0.946^{+0.008}_{-0.008}$ & $-$ & $-$\\
FRB~20220310A/F & Constant $L/I$ & $0.56^{+0.01}_{-0.01}$ & $-$ & $-$ & $-$\\
FRB~20220506A/D & Depolarization & $0.92^{+0.05}_{-0.08}$ & $>3.4^{+0.2}_{-0.2}\tablenotemark{a}$ & $-$ & $-$\\
FRB~20221113E/A & Depolarization & $0.993^{+0.005}_{-0.010}$ & $1.30^{+0.04}_{-0.04}$ & $-$ & $-$\\
FRB~20221203B/A & Depolarization & $0.991^{+0.006}_{-0.010}$ & $>1.20^{+0.04}_{-0.05}\tablenotemark{a}$ & $-$ & $-$\\
FRB~20230307C/A & Birefringence + Depolarization & $0.57^{+0.02}_{-0.02}$ & $0.82^{+0.08}_{-0.08}$ & $1057^{+108}_{-177}$ & $-5^{+2}_{-3}$\\
FRB~20230325A/C & Constant $L/I$ & $0.887^{+0.002}_{-0.002}$ & $-$ & $-$ & $-$\\
FRB~20230814C/B & Depolarization & $0.844^{+0.02}_{-0.02}$ & $1.12^{+0.06}_{-0.07}$ & $-$ & $-$\\
FRB~20230913F/G & Birefringence + Depolarization & $0.880^{+0.007}_{-0.006}$ & $0.2^{+0.1}_{-0.1}$\tablenotemark{b} & $1176^{+125}_{-271}$ & $-2^{+2}_{-3}$\\
FRB~20240122A/E & Constant $L/I$ & $0.55^{+0.03}_{-0.03}$ & $-$ & $-$ & $-$\\
FRB~20240203A/D & Constant $L/I$ & $0.137^{+0.002}_{-0.002}$ & $-$ & $-$ & $-$\\
FRB~20240229E/A & Birefringence + Depolarization & $0.676^{+0.002}_{-0.002}$ & $0.803^{+0.03}_{-0.03}$ & $1054^{+150}_{-191}$ & $-1.0^{+0.4}_{-0.9}$\\
\hline
\hline
\end{tabular}
\end{center}
\tablenotetext{$a$}{The lowest frequency bin, in which most of the depolarization occurs, is an upper limit on $L/I$, thus only a $\sigma_\mathrm{RM}$ lower limit is provided by the model fitting.}
\tablenotetext{$b$}{While FRB~20230913F/G is best fit by Equation \ref{eq:model_3}, its $L/I$ is approximately constant between 400$-$800~MHz (hence, its estimated $\sigma_\mathrm{RM}$ is small). We do not consider this to be a robust detection of depolarization.}
\end{table*}

Overall, we find that $4/12$ FRBs favor the constant $L/I$ model, $5/12$ prefer the depolarization model, and $3/12$ are best described by the combination of magnetospheric birefringence effects and depolarization. Two of the three FRBs in the latter category (FRBs~20230307C/A and 20240229E/A) show a clear downturn in $L/I$ towards $400$~MHz, while the $L/I$ of FRB~20230913F/G increases from DSA-110 to CHIME/FRB frequencies, and it remains approximately constant across 400$-$800~MHz. Thus, depolarization with decreasing frequency is observed in $7/12$ FRBs in our sample. 

\subsection{Polarization angle} \label{sec:results_pa} 
We find that FRBs~20220310A/F and 20230814C/B exhibit flattening PA profiles towards lower frequencies. Both of their PA profiles, integrated over 400$-$800~MHz and 1280$-$1530~MHz, are presented in Figure \ref{fig:pol_profiles}. FRB~20220310A/F shows a $60~\mathrm{deg}$ PA jump over $0.1$~ms at 1280$-$1530~MHz (PA $\chi_\nu^2 = 226$), which is reduced to a maximum PA variation of only $30~\mathrm{deg}$ over the $1$~ms burst duration at 400$-$800~MHz (PA $\chi_\nu^2 = 3$). At 1280$-$1530~MHz, FRB~20230814C/B shows a $40~\mathrm{deg}$ PA swing over $0.1$~ms (PA $\chi_\nu^2 = 7$), while at 400$-$800~MHz the PA variability is $< 20~\mathrm{deg}$ in amplitude (PA $\chi_\nu^2 = 2$). Most of the remaining $10/12$ FRBs in our sample show flat PA profiles at both CHIME/FRB and DSA-110 frequencies, and a few have such a low $S/N$ in one of the two datasets that we cannot construct a well-sampled PA profile over multiple frequency bands (see Figure \ref{fig:pol_profiles}).

\subsection{Correlation between depolarization, Faraday rotation, and scattering} \label{sec:results_depol}
\begin{figure*}
    \centering
    \includegraphics[width=0.325\textwidth]{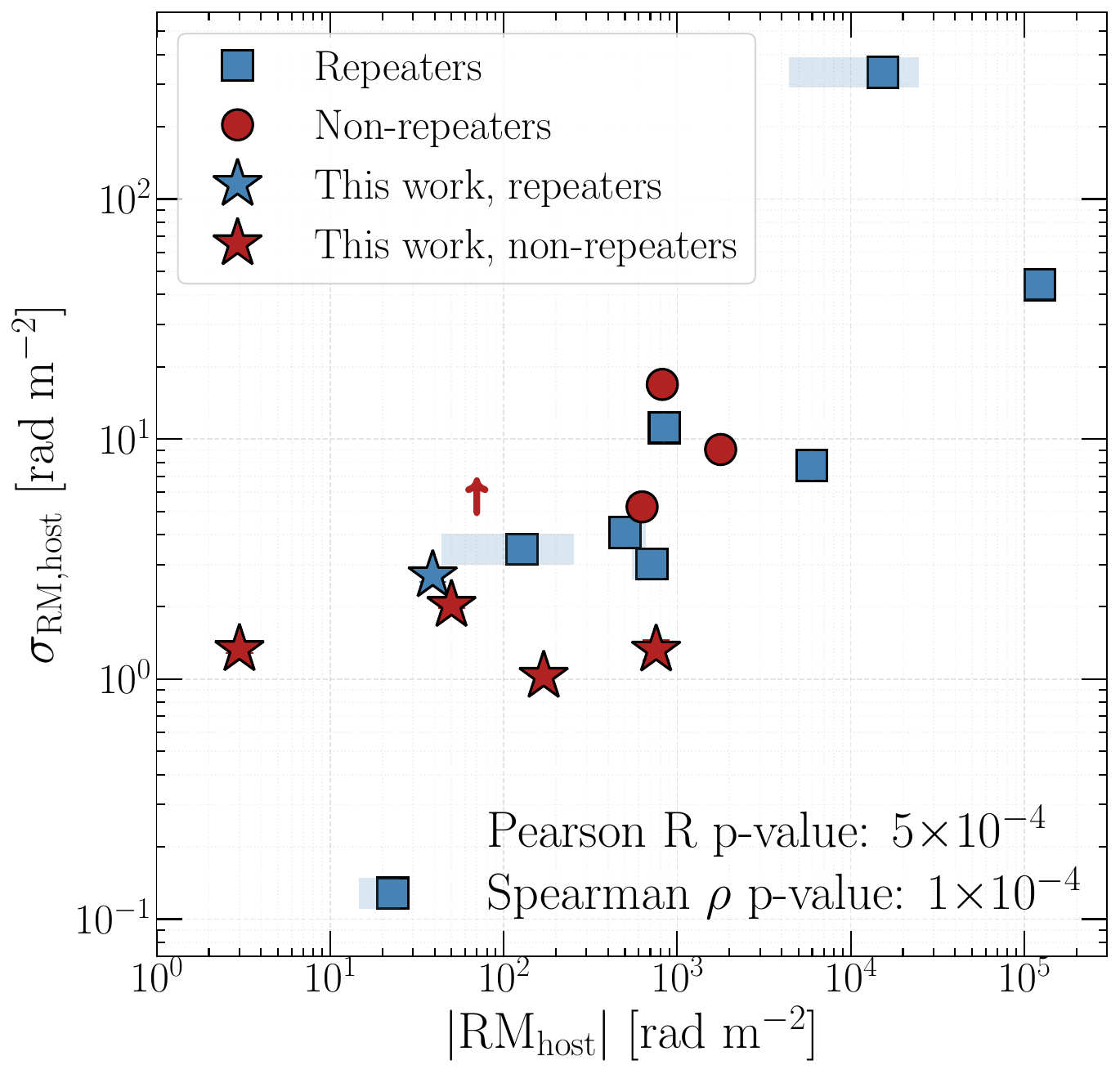}
    \includegraphics[width=0.325\textwidth]{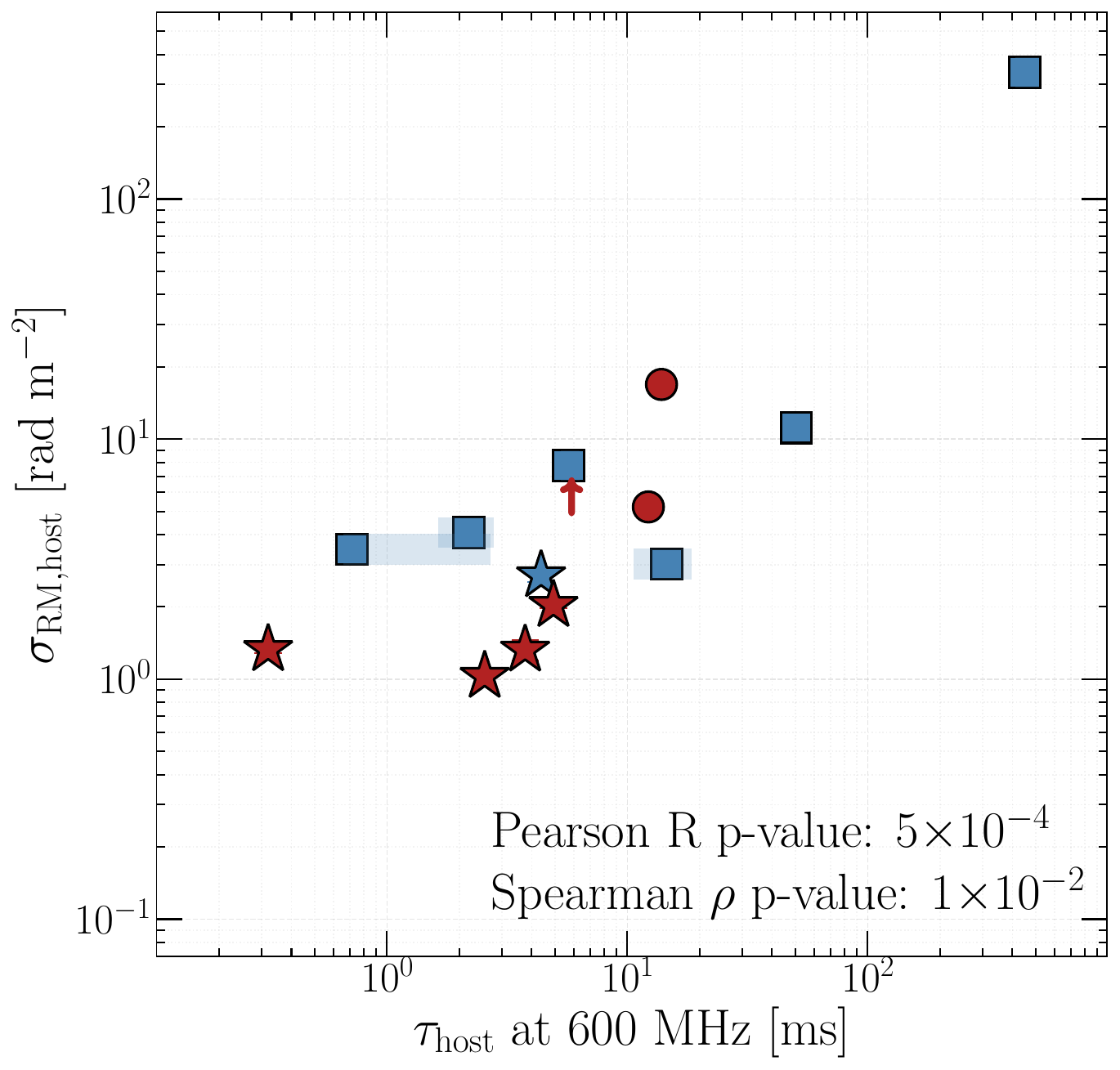}
    \includegraphics[width=0.318\textwidth]{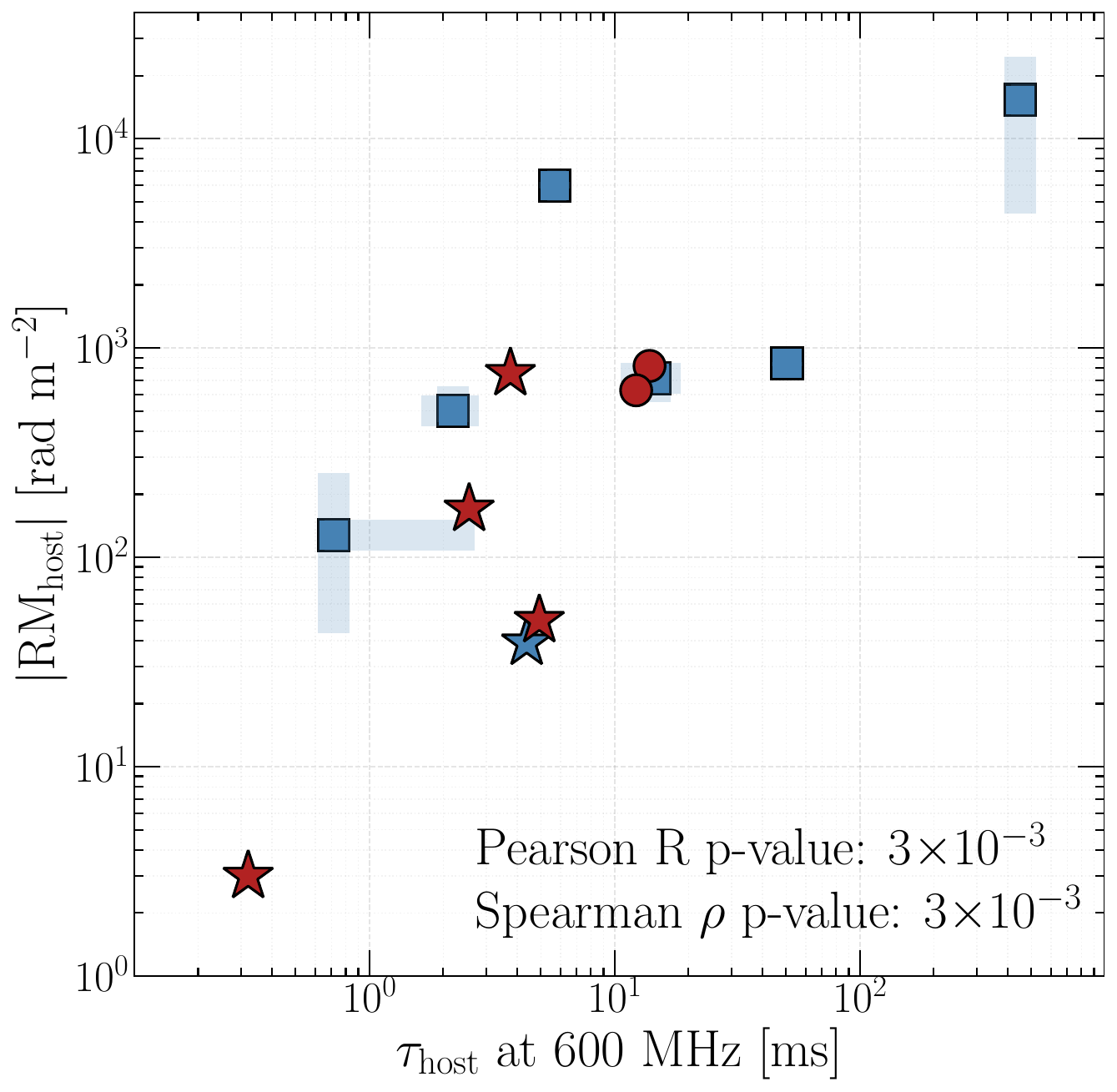}
    \caption{(Left) Rest-frame RM scatter, $\sigma_\mathrm{RM,host}$, versus $|\mathrm{RM}_\mathrm{host}|$ for the five co-detected, depolarizing FRBs and that have a host galaxy with a spectroscopic redshift (stars). The $\sigma_\mathrm{RM,host}$ lower limit for FRB~20220506A/D is plotted as a red arrow; we do not include FRB~20221203B/A as it does not have a reliable spectroscopic redshift. Depolarizing FRBs from the literature are overplotted as blue squares \citep[repeaters;][]{2022Sci...375.1266F, 2026ApJ..1000L..53P} and red circles \citep[apparent non-repeaters;][]{2025MNRAS.tmp.1894U, 2026arXiv260918086B, 2026arXiv260921177D}. (Middle) $\sigma_\mathrm{RM,host}$ versus rest-frame scattering timescale, $\tau_\mathrm{host}$, referenced to 600~MHz, where available, for the same sample of FRBs. (Right) $|\mathrm{RM}_\mathrm{host}|$ versus $\tau_\mathrm{host}$ referenced to 600~MHz for the same sample as the middle panel. Blue shaded regions in all panels represent the 68\% confidence interval around the median $|\mathrm{RM}_\mathrm{host}|$ and $\tau$ for repeaters. Uncertainties on $\sigma_\mathrm{RM,host}$ are smaller than the markers. The p-values from Pearson R and Spearman $\rho$ tests in all cases are presented in the bottom right of each panel.} 
    \label{fig:sigmarm_correlation}
\end{figure*}

Correlations between $\sigma_\mathrm{RM}$, $|\mathrm{RM}|$, and scattering time scale ($\tau$) have been suggested as indicators for FRBs traversing dense, magnetized, and inhomogeneous plasma in their host galaxies. However, the significance of these correlations has only been marginal thus far, chiefly owing to the small sample size of depolarizing FRBs \citep{2022Sci...375.1266F, 2025MNRAS.tmp.1894U, 2026ApJ..1000L..53P, 2026arXiv260918086B, 2026arXiv260921177D}. Here, we revisit the $\sigma_\mathrm{RM}-|\mathrm{RM}|$, $\sigma_\mathrm{RM}-\tau$, and $|\mathrm{RM}|-\tau$ relationships, using depolarizing FRBs from the literature and adding our results for our five depolarizing FRBs that also have secure host galaxy associations and spectroscopic redshifts \citep[FRBs~20220207A/C, 20221113E/A, 20230307C/A, 20230814C/B, and 20240229E/A; for measurements of their scattering timescales see Paper I, and for their host galaxy associations see][]{2024ApJ...967...29L, 2026arXiv260930616V}. We also plot the $\sigma_\mathrm{RM}$ upper limit for FRB~20220506A/D in Figure \ref{fig:sigmarm_correlation}, but we do not include it in any statistical tests. Again, we echo that the rest-frame values of FRBs~20220207A/C should be interpreted with some caution.

We evaluate these correlations in the FRB rest-frame and we account for any RM variability in repeating FRBs. First, we convert to the rest-frame RM scatter, $\sigma_\mathrm{RM,host}$, given the cosmological redshift, $z$, of the FRB host galaxies \citep{2024MNRAS.528.2511T},
\begin{equation}
\sigma_\mathrm{RM,host} = \sigma_\mathrm{RM} (1+z)^2\,. \label{eq:sigma_rm_host}
\end{equation}
Following Equation \ref{eq:rm_comps}, we assume $\mathrm{RM}_\mathrm{ion} = \mathrm{RM}_\mathrm{IGM} = 0~\mathrm{rad}~\mathrm{m}^{-2}$, subtract the Galactic RM as measured by \cite{2022A&A...657A..43H}, convert to the rest frame by applying a $(1+z)^2$ correction factor, and take the absolute value to get $|\mathrm{RM}_\mathrm{host}|$. 

Figure \ref{fig:sigmarm_correlation} (left panel) presents the $\sigma_\mathrm{RM,host} - |\mathrm{RM}_\mathrm{host}|$ correlation for all depolarizing FRBs discovered thus far \citep[those found in this work as well as those discovered by][]{2022Sci...375.1266F, 2025MNRAS.tmp.1894U, 2026ApJ..1000L..53P, 2026arXiv260918086B, 2026arXiv260921177D}. We use both Pearson R and Spearman $\rho$ tests on the $\mathrm{log}_{10}(\sigma_\mathrm{RM,host})-\mathrm{log}_{10}(|\mathrm{RM}_\mathrm{host}|)$ relationship, finding test statistics equal to $0.767$ and $0.812$ (positive correlation) with respective p-values of $5 \times 10^{-4}$ ($3.5 \sigma$) and $1 \times 10^{-4}$ ($3.9 \sigma$). We consider this correlation to be statistically significant (i.e., $\geq 3\sigma$).

Next, we restrict our depolarizing FRB sample to only those with measured scattering timescales exceeding the contribution predicted by the NE2025 model \citep[scaled to the same reference frequency assuming a frequency dependence of $\nu^{-4}$;][]{2026arXiv260211838O} by more than an order of magnitude \citep{2019MNRAS.486.3636P, 2022ApJ...927...59L, 2023MNRAS.519..821O, 2025ApJ...992..206C, 2025PASA...42..133S}. Assuming negligible scattering contribution from any intervening galaxy halos (See Section 5.2 in Paper~I), enforcing this threshold gives us a high level of confidence that the observed scattering timescales are dominated by contributions from the FRB host galaxy or immediate environment. After this cut, we retain all of the depolarizing FRBs in this work and eight of the depolarizing FRBs in the literature. Converting the scattering timescale to the FRB rest frame to account for cosmological time dilation, assuming $\tau \propto \nu^{-4}$, we apply the equation
\begin{equation}
\tau_\mathrm{host} = \tau(1+z)^3\,, \label{eq:tau_host}
\end{equation}
\citep{Macquart_2013_ApJ}. Thus, the intrinsic scattering timescale in the rest-frame of the FRB host galaxy is larger than the value observed. Note that if the scattering index diverges from the assumed $\nu^{-4}$, it can change the significance of any observed correlation. In the middle panel of Figure \ref{fig:sigmarm_correlation}, we plot $\tau_\mathrm{host}$ versus $\sigma_\mathrm{RM,host}$ of depolarizing FRBs. Performing the same statistical tests on $\mathrm{log}_{10}(\sigma_\mathrm{RM,host})-\mathrm{log}_{10}(\tau_\mathrm{host})$, we find test statistics of $0.828$ and $0.681$ with p-values $5 \times 10^{-4}$ ($\sim 3.5 \sigma$) and $1 \times 10^{-2}$ ($\sim 2.6 \sigma$) for the Pearson R and Spearman $\rho$ tests, respectively. To be conservative, we label this correlation as marginally significant.

In the right panel of Figure \ref{fig:sigmarm_correlation}, we plot $\tau_\mathrm{host}$ versus $|\mathrm{RM}_\mathrm{host}|$ for all depolarizing FRBs with measured scattering timescales. Pearson R and Spearman $\rho$ tests conducted on $\mathrm{log}_{10}(|\mathrm{RM}_\mathrm{host}|)-\mathrm{log}_{10}(\tau_\mathrm{host})$ find a statistically significant correlation with test statistics of $0.761$ and $0.747$ with p-values equal to $3 \times 10^{-3}$ ($\sim 3.0 \sigma$) in both cases. We consider this correlation statistically significant.

We caution that the correlations presented above may be sensitive to the inclusion of FRB~20190520B, which has $\sigma_\mathrm{RM,host}$ and $\tau_\mathrm{host}$ that are an order of magnitude or more larger than the rest of the sample. Removing FRB~20190520B and repeating the Pearson R and Spearman $\rho$ tests, we find p-values $1 \times 10^{-3}$ ($\sim 3.2 \sigma$) and $7 \times 10^{-4}$ ($\sim 3.4 \sigma$) for the $\mathrm{log}_{10}(\sigma_\mathrm{RM,host})-\mathrm{log}_{10}(|\mathrm{RM}_\mathrm{host}|)$ correlation, $3 \times 10^{-2}$ ($\sim 2.2 \sigma$) and $4 \times 10^{-2}$ ($\sim 2.1 \sigma$) for the $\mathrm{log}_{10}(\sigma_\mathrm{RM,host})-\mathrm{log}_{10}(\tau_\mathrm{host})$ correlation, and $2 \times 10^{-2}$ ($\sim 2.3 \sigma$) for both for the $\mathrm{log}_{10}(|\mathrm{RM}_\mathrm{host}|)-\mathrm{log}_{10}(\tau_\mathrm{host})$ correlation. 

\begin{figure}
    \centering
    \includegraphics[width=0.47\textwidth]{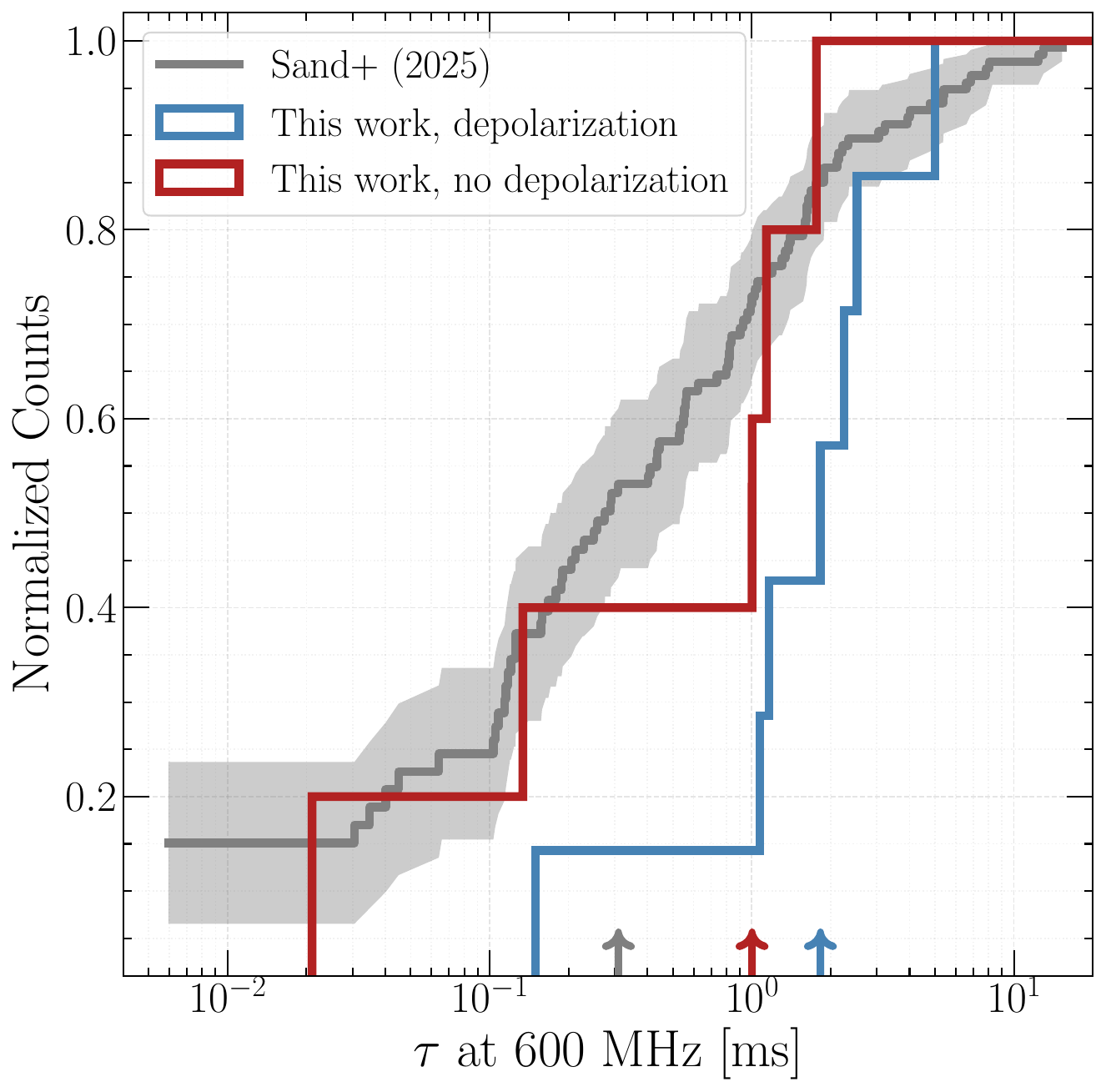}
    \caption{Cumulative distribution functions of the scattering timescale (observer frame; referenced to $600$~MHz) of the seven depolarizing (blue) and five non-depolarizing (red) co-detected FRBs. The median of each distribution is marked by an arrow in their respective colors. Overplotted in gray is the empirical cumulative distribution function of scattering timescales for the $137$ FRBs from the first CHIME/FRB baseband catalog \citep{2025ApJ...979..160S}; here, $\tau$ upper limits have been incorporated as censored data and the shaded area is the 95\% confidence interval.} 
    \label{fig:scat_hist}
\end{figure}

The typical burst durations and scattering timescales of the co-detected FRBs are smaller than most DSA-110-detected FRBs, but larger than those of the CHIME/FRB population; this is likely due to a confluence of selection effects driven by the $0.262$~ms time resolution of the DSA-110's FRB search at higher frequencies and CHIME/FRB's decreased sensitivity to highly scattered FRBs \citep{2026arXiv260626334M}. In Figure \ref{fig:scat_hist}, we plot the cumulative distribution function of $\tau$ (observer frame) for FRBs with and without depolarization; for comparison, we also show the distribution of $\tau$ measured in the first CHIME/FRB baseband catalog \citep{2025ApJ...979..160S}. Here, we have employed survival analysis that take censored data into account \citep[see, e.g.,][]{2012msma.book.....F}, via the {\tt NADA V1.6-1.2} package\footnote{\url{https://rdrr.io/cran/NADA/}} in {\tt R} \citep{helsel2005nondetects, NADApackage, Rsoftware}, to incorporate $\tau$ upper limits into the empirical cumulative distribution function. The mean scattering timescale, referenced to $600$~MHz, for the depolarizing FRBs is $\tau = 2\pm1$~ms, while the mean for our FRBs without depolarization is $\tau = 0.8\pm0.7$~ms. Building upon Paper I, we qualitatively find that co-detected FRBs undergoing depolarization may peak at larger scattering timescale than those with no measurable depolarization (see Figure \ref{fig:scat_hist}).

\section{Discussion} \label{sec:discussion}
We have explored spectropolarimetry of FRBs with low-frequency, broadband co-detections, finding that only one third (four sources) of the FRBs have constant $(L/I) (\nu)$, seven exhibit depolarization consistent with multi-path propagation through an inhomogeneous, magnetized plasma, and three sources show a decrease in $L/I$ towards higher frequencies. A suppression of PA variability is also seen in two FRBs that show clear PA variations at 1280$-$1530~MHz, yet their PA profiles appear more constant at 400$-$800~MHz, where scatter broadening is more prominent. Below, we interpret the physical origins of these spectropolarimetric features and discuss implications for population-level studies of FRB polarimetry.

\subsection{Evidence for a magnetospheric emission process} \label{sec:discussion_emission}
As discussed in Section \ref{subsec:model_3}, a decrease in the $L/I$ with increasing frequency is a commonly observed phenomenon in polarized radio pulsars that is typically attributed to the interaction of orthogonal polarization modes with birefringent plasma in the neutron star magnetosphere. Despite the specific mechanism by which the resulting linearly polarized spectra is produced still being debated \citep[e.g.,][]{1975ApJ...196...51R, 1986ApJ...302..138B, 1997ApJ...475..763M, 1998A&A...336..209V, 2005MNRAS.359..481K, 2008MNRAS.388..261J}, all of these models require the emission site of the pulses to be within the magnetosphere of the neutron star itself. While \cite{2024ApJ...968...50P} noted one FRB with a tentative decrease in its $L/I$ across 400$-$800~MHz, there have been no other observations of FRB linear polarization decreasing towards higher frequencies. In Section \ref{sec:results_li}, we present three sources (FRBs~20230307C/A, 20230913F/G, and 20240229E/A) that show a decreasing $L/I$ with increasing frequency, each depicting a $\sim 20-40\%$ lower $L/I$ at the bottom of the DSA-110 frequency band compared to the top of the CHIME/FRB band. The $L/I$ spectra of these FRBs are well fit by Equation \ref{eq:model_3}, which includes a power-law decrease of $L/I$ towards higher frequencies motivated by the shape of polarized pulsar spectra in which this magnetospheric birefringence effect is observed \citep{1973ApJ...179L...7M}. In a radius-to-frequency mapping scheme, it is also expected that the burst width should increase with decreasing frequency \citep[e.g.,][]{1997ApJ...475..763M}. However, due to the significant scatter broadening at 400$-$800~MHz, in conjunction with our lack of spectral coverage between 800$-$1280~MHz, it is challenging for us to unambiguously measure the burst width versus frequency of our FRB sample. If the $L/I$ decrease with increasing frequency is due to mixing of orthogonal propagation modes, we might see associated orthogonal PA jumps at higher frequencies \citep[though note that not all pulsars undergoing a decrease in linear polarization show these PA jumps, e.g.,][]{2008MNRAS.388..261J}. However, the PA profiles of these three FRBs are poorly sampled in time at 1280$-$1530~MHz ($\leq 8$ PA samples across the burst), and it has been shown that even modest scattering can wash out orthogonal PA jumps at $1.4$~GHz \citep{2009MNRAS.392L..60K}; thus, we cannot robustly test for the presence, or lack thereof, of orthogonal PA jumps in these FRBs. Regardless, the resemblance between the $(L/I) (\nu)$ of these FRB and those of pulsars points towards a potential magnetospheric emission process for these FRBs, but below we explore a few other possible propagation effects that can produce confounding complexity in the $L/I$ spectra.

Spectropolarimetry of star-forming galaxies has revealed regions with increased polarized fractions at lower frequencies \citep[e.g., NGC~6946 and M51;][]{1991A&A...251...15B, 1992A&A...265..417H}. These polarized intensity spectra can be produced by differential Faraday rotation in a helical magnetic field co-located with the radio-emitting plasma \citep{1998MNRAS.299..189S, 2014MNRAS.441.2049H}. However, this mechanism requires physically extended ($\sim$~kpc scale) emitting and Faraday rotating regions along the LOS. While this is reasonable for a synchrotron-emitting galaxy, it does not map on to FRB emission, which we expect to be much more compact \citep[e.g.][]{2025Natur.637...48N}. 

\cite{2014ApJS..212...15F} produced a broadband polarization catalog of extragalactic continuum radio sources and found $21$\% of sources exhibit what they label ``repolarization'' (i.e., decreasing total polarization fraction with increasing frequency). They argue that repolarization can arise from blending of unresolved components that have different spectral indices (e.g., the flat-spectrum core and steep-spectrum lobes of a radio galaxy). Analogously, in radio pulsars, it has been theorized that incoherently summing over temporally unresolved burst components can lead to lower $L/I$ at higher frequency under the radius-to-frequency framework, as higher frequency emission would originate from lower in the neutron star magnetosphere, where the increased density of field lines would produce more unresolved burst components \citep{2008MNRAS.388..261J}. While this interpretation is a geometric effect rather than a physical one, it is still suggestive of a magnetospheric emission process even in the absence of any orthogonal PA jumps. Our results for FRBs~20230307C/A, 20230913F/G, and 20240229E/A are also consistent with incoherent summing of unresolved burst components under radius-to-frequency mapping which, given our data, is not distinguishable from the interpretation of orthogonal polarization modes propagating through birefringent plasma, but both scenarios suggest that the FRBs may be produced in neutron star magnetospheres.

Radio emission propagating through plasma with elliptically or linearly polarized natural wave modes will undergo some level of Faraday conversion, causing frequency-dependent conversion between linear and circular polarizations \citep{1969SvA....13..396S, 1998PASA...15..211K}. This mechanism can arise under a variety of physical conditions \citep[e.g.,][]{2010ApJ...725.1600M, 2019ApJ...876...74G, 2019MNRAS.485L..78V, 2022ApJ...933L...6L} and, generally, the conversion between $L/I$ and $V/I$ becomes stronger at lower frequencies, but there is no uniform prescription for the spectral signature of this effect. Faraday conversion has been invoked to explain circular polarization of centimeter$-$millimeter wavelength emission from supermassive black holes, where the conversion can arise in the extreme, hot accretion flows around the black hole \citep[e.g., in Sgr A*;][]{2011MNRAS.416.2574H}, or in their jets \citep[][]{2003Ap&SS.288..143W, 2018ApJ...862...58M}. Faraday conversion is also speculated to attenuate the polarization levels of some pulsar systems that show substantial circular polarization; here, the conversion would originate in highly magnetized plasma in the neutron star magnetosphere \citep{2004A&A...421..681E, 2019MNRAS.483.2778I, 2024NatAs...8..606L} or in the circumsource environment of a pulsar binary system \citep{2023Natur.618..484L}. Without well-calibrated circular polarization data at 400$-$800~MHz, direct tests of Faraday conversion models are challenging. Under the conservative assumption that Faraday conversion is frequency dependent such that conversion between $L/I$ and $V/I$ is stronger at lower frequencies, the expected signature in the $L/I$ spectra would be a higher level of variation with decreasing frequency (e.g., $L/I$ decreases monotonically or oscillates more rapidly at low $\nu$). The observed $(L/I) (\nu)$ of FRBs~20230307C/A, 20230913F/G, and 20240229E/A all rise towards lower frequencies, with $L/I$ across 400$-$800~MHz consistent with constant linear polarization or a decrease that is well described by depolarization due to multipath propagation. Thus, our observations do not match the expectations from Faraday conversion, though our lack of well-characterized $V/I$ spectra at 400$-$800~MHz, and gap in observing frequencies between 800$-$1280~MHz, make it impossible to entirely rule out this scenario.

The gap in frequency coverage between $800$ and $1280$~MHz is the foremost impediment for this work insofar as it restricts us from irrefutably showing that the co-detected FRBs have the same emission components extending across 400$-$1530~MHz. While rare, \cite{2020ApJ...891L..38C} and \cite{2024ApJ...968...50P} showed that different components of the same burst can have distinct polarization levels, which could complicate the $(L/I)(\nu)$ modeling. This needs to be considered for those FRBs that do not show consistent $L/I$ values between the DSA-110 band and the top of the CHIME/FRB band: FRBs~20220506A/D, 20230307C/A, 20230913F/G, and 20240229E/A. FRBs 20220506A/D and 20240229E/A both show only a single component in each frequency band, lessening the concern of probing different burst components at each observatory. DSA-110 observes two distinct components in FRB~20230307C/A and the best-fit CHIME/FRB model contains two components that are blended due to scatter broadening (Paper I). This similar morphology between frequency bands gives us some confidence that we are probing the same emission components; we also compute our polarization properties averaged across all burst components, reducing the impact of any potential component-to-component $L/I$ variation. CHIME/FRB detects multi-component structure in FRB~20230913F/G while there is only one component visible in the DSA-110 data. For this FRB, we cannot ensure that differing polarization levels between components do not contaminate its $(L/I)(\nu)$. In general, we do not expect any unmodeled instrumental polarization effects to manifest in a frequency-dependent manner resembling depolarization due to multipath propagation or a power-law depolarization towards higher frequencies (see Section \ref{sec:obs} for details on our polarization data reduction). Thus, we interpret the polarized spectra of all three FRBs best fit by Equation \ref{eq:model_3} to be suggestive of a magnetospheric emission process, but we caution against drawing a stronger conclusion as we cannot unambiguously confirm that we are detecting the same emission components across our observing bands. 

These findings are concordant with some other features seen in FRBs that point towards a magnetospheric origin and/or a compact emission site: downward-drifting burst morphology is suggested to be a consequence of radius-to-frequency mapping \citep[e.g., as seen in many repeating FRBs;][]{2019ApJ...876L..23H, 2020ApJ...889..135L, 2022MNRAS.509.5679T}, short duration microstructure \citep[e.g.,][]{2021NatAs...5..594N, 2023NatAs...7.1486S, 2024ApJ...974..274F}, rapid PA swings \citep[e.g.,][]{2025Natur.637...43M}, and two-screen scattering constraints \citep[e.g.,][]{2025Natur.637...48N}. While downward-drifting morphology is primarily observed in repeating FRBs, short-duration microstructure is seen in both populations, and the three FRBs that we find show increasing $L/I$ with decreasing frequency are all apparently non-repeating, lending credence to the idea that some FRBs may share common emission sites despite their visually distinct burst morphologies.

\subsection{Prevalence of depolarizing environments} \label{sec:discussion_depol}
Measurements of spectral depolarization in FRB sources have thus far been exceedingly rare (11 in total), with most detections being associated with repeating FRBs that have long activity periods, allowing for follow-up observations at multiple frequency bands \citep{2022Sci...375.1266F, 2026ApJ..1000L..53P}. Among apparently non-repeating FRBs, there are only a few concrete depolarization measurements \citep{2025MNRAS.tmp.1894U, 2026arXiv260918086B, 2026arXiv260921177D}. Apparently non-repeating FRBs still show a typically high median $L/I\sim0.65$ in CHIME/FRB, which is consistent with the distribution of non-repeater $L/I$ seen by DSA-110 \citep{2024ApJ...968...50P, 2024ApJ...964..131S}, pointing to the lack of strong, population-level depolarization. These results have led to the preferred association of inhomogeneous, depolarizing environments with repeating FRBs embedded in extreme magnetoionic plasma \citep{2020ApJ...895....7Y, 2022ApJ...928L..16Y, 2026arXiv260317615Y}.

However, the bandwidth of apparently non-repeating FRBs is limited to the instantaneous frequency coverage of the telescope, thus restricting the lever arm in frequency and biasing us against the detection of depolarization in apparent non-repeaters. We measure depolarization in $7/12$ co-detected FRBs, significantly increasing the population of FRB sources in which depolarization has been detected thus far. Outside of FRBs~20121102A and 20190520B, which inhabit two of the most extreme magnetoionic FRB environments currently known \citep[][]{2018Natur.553..182M, 2022Sci...375.1266F, 2023Sci...380..599A}, there is considerable overlap in the $\sigma_\mathrm{RM,host}$ distributions of repeaters and the apparent non-repeaters, suggesting that the FRB population originates from a continuum of inhomogeneous environments. Given the high occurrence rate of depolarization in our sample, our results point to a significant fraction of FRBs being associated with interstellar or circumburst media that contain turbulent, ionized gas, which appear to originate from markedly different environments than the exceptional FRBs found in quiescent regions, such as FRBs~20200120E and 20240209A \citep{2022Natur.602..585K, 2025ApJ...979L..21S, 2025ApJ...979L..22E}. Pinning down the true fraction of repeating and non-repeating FRBs that undergo depolarization is challenging given the limited sample size and complex inter-facility selection effects inherent to the co-detected FRBs, but this question may be answered by larger polarimetric population studies in the future.

Depolarization due to multipath propagation is not commonly observed in radio pulsars. PSRs~J0742$-$2822 and J1721$-$3532 show depolarization with estimated $\sigma_\mathrm{RM} = 0.13~\mathrm{rad}~\mathrm{m}^{-2}$ \citep{2019PASA...36...25X} and $\sim 20~\mathrm{rad}~\mathrm{m}^{-2}$ \citep{2015A&A...576A..62N}, respectively, but these are likely due to intervening structure in the Gum nebula and in an \ion{H}{2} region. At $\lesssim 200$~MHz, \cite{2015A&A...576A..62N} see a drop in $L/I$ at the $\sim 10$\% level for only $3/16$ of their pulsars, which would result in order $\sigma_\mathrm{RM} \lesssim 0.1~\mathrm{rad}~\mathrm{m}^{-2}$. At 1000$-$1500~MHz, \cite{2023RAA....23j4002W} find only $3/682$ pulsars show a $\gtrsim$~few percent decrease in $L/I$ with decreasing frequency. The lack of depolarization in most polarized pulsars, in conjunction with the significant depolarization in PSRs~J0742$-$2822 and J1721$-$3532 likely being due to intervening Galactic structures, is consistent with the fact that most known pulsars are seen in the diffuse interstellar medium. Further, some repeating FRBs show years-long DM variation that are significantly larger than DM variability typically observed in radio pulsars due to their motion through the interstellar medium \citep[e.g.,][]{2026ApJ..1000L..53P, 2026arXiv260409098A}. Therefore, FRBs, on average, may originate from more inhomogeneous magnetoionic environments than those surrounding Galactic pulsars.

We also report the first statistically significant detections of correlations between $\sigma_\mathrm{RM,host}-|\mathrm{RM}_\mathrm{host}|$ and $|\mathrm{RM}_\mathrm{host}|-\tau_\mathrm{host}$ in depolarizing FRBs, and a marginally significant correlation between $\sigma_\mathrm{RM,host}-\tau_\mathrm{host}$. These results suggest that most of the Faraday rotation attributed to the FRB host galaxy is contributed by the same turbulent, ionized environment that causes depolarization of the FRB signal. The observed scattering may also arise from a screen co-located with this medium, though the scattering and depolarization do not necessarily need to originate in the same screen.\footnote{If the two did originate from the exact same screen, the $L/I$ is expected to show a temporal decline in the scattering tail, which we do not see clear evidence for in our data.} A larger sample size of depolarizing FRBs is needed to robustly test the statistical significance of the $\sigma_\mathrm{RM,host}-\tau_\mathrm{host}$ correlation.

\subsection{Implications of scatter-induced PA flattening for FRB population studies} \label{sec:discussion_pa}
Models in which FRB emission originates near the surface of a rapidly-rotating compact object \citep[``close-in'' models, e.g. from coherent-curvature radiation or magnetic reconnection;][]{2017MNRAS.468.2726K, 2018ApJ...868...31Y, 2020arXiv200505093L, 2021ApJ...922..166L} can produce PA variations on $\lesssim$~ms timescales; meanwhile, models in which the FRB emission site is at large distances from its central engine \citep[``far-out'' models, e.g. by synchrotron maser emission;][]{2019MNRAS.485.4091M, 2020ApJ...896..142B} predict predominantly constant PA profiles. A geometric model describing the projection of beamed emission from a dipolar, rapidly rotating neutron star magnetic field towards an observer \citep[i.e., the ``rotating vector model'';][]{1969ApL.....3..225R} has proved to be well-suited in explaining the rich phenomenology of PA variability observed in most integrated profiles of radio pulsars \citep[e.g.,][]{2023MNRAS.520.4801J, 2024MNRAS.530.4839J} and also in many single pulses \citep{2016ApJ...833...28M, 2023ApJ...952..151M}. However, the majority of FRBs display little to no measurable PA changes across their burst durations \citep{2024ApJ...968...50P, 2024ApJ...964..131S}. Such a high occurrence rate of constant PA profiles is challenging to reconcile with an emission mechanism operating near the surface of a rapidly-rotating object, indicating that instead the FRB source must be rotating much slower than its burst durations, or the emission is produced much further away from the central engine.

The above conclusion presupposes that there are no confounding biases that may lead to an over representation of flat PA profiles in these surveys. Such an effect has been noted for radio pulsars, where scatter broadening of the pulse leads to the mixing of signals from different pulse phases, averaging over the temporal evolution of the PA profile and suppressing PA variability \citep[e.g., ][]{2003A&A...410..253L, 2023RAA....23j4002W, 2025arXiv250614519J}. PA flattening is less commonly seen in FRBs \citep{2025ApJ...989L..48C}, likely owing to the limited instantaneous bandwidth for most observations. We show that $2/12$ co-detected FRBs exhibit PA flattening between $1530$~MHz and $400$~MHz. While component-to-component PA variability could influence our finding (e.g., if higher frequency components have larger intrinsic PA variability), we do not deem this likely as none of these 12 co-detected FRBs show evidence for frequency-time drifting in either observing band (Paper~I); we instead argue in favor of scattering-induced PA flattening. While FRBs~20220310A/F and 20230814C/B are both visibly scatter broadened at 400$-$800~MHz, they have low to moderate scattering timescales compared to the rest of the co-detected FRBs. This may highlight that most FRBs in our sample already have their PA variability (if there is any intrinsically) flattened at 1280$-$1530~MHz.

While our sample size of $2/12$ is not large enough to place strong constraints on the rate at which intrinsically variable PA profiles are suppressed at low frequencies, multiple detections of PA flattening in our limited sample does suggest that the effect is an important systematic bias that must be addressed in polarimetric FRB studies. Importantly, many of the DSA-110 polarized FRB sample that do not show significant scatter broadening above $1280$~MHz, presented here as co-detected FRBs or previously by \cite{2024ApJ...964..131S}, still have flat PA profiles. This is in agreement with observations at $\gtrsim 1400$~MHz of some repeating FRBs, in which many bursts show flat PA profiles despite the lack of scatter broadening \citep[e.g.,][]{2018Natur.553..182M, 2021NatAs...5..594N, 2021MNRAS.508.5354H, 2025ApJ...988..175L, 2026arXiv260320663W}. Observations of the repeating FRB~20180301A at L-band show some bursts having flat PAs while others have variability on the order of $10~\mathrm{deg}$ \citep{2020Natur.586..693L}, indicating that even a single source can produce a diversity of PA profiles. Together, these results imply that many FRBs do indeed have intrinsically variable or flat PA profiles, but that a fraction of bursts showing flat PA profiles are contaminants from PA flattening due to scatter broadening, which is more prominent at lower observing frequencies. Future FRB polarization studies should be mindful of, and attempt to account for, this systematic bias by analyzing PA variability in FRBs whose burst profiles are minimally modulated by scattering.

\section*{Conclusions} \label{sec:conclusions}
We report on the spectropolarimetric properties of twelve FRBs co-detected by CHIME/FRB and DSA-110. The joint spectral coverage of these two telescopes, spanning almost two octaves (400$-$1530~MHz), enables tests of subtle propagation effects imprinted in the polarized spectra of these FRBs. We test the data against three physically motivated models of the frequency evolution of the linear polarization fraction, $L/I$: (i) a constant $L/I$ versus frequency, (ii) spectral depolarization in which the $L/I$ decreases exponentially at lower frequencies, and (iii) model (ii) multiplied by a power-law $L/I$ decrease at high frequencies from propagation through a birefringent magnetosphere. We also analyze the evolution of polarization angle, PA, with frequency, focusing on how scattering of the FRB emission may affect the observed PA profile at different frequencies. Our key conclusions from this work are:
\begin{enumerate}
    \item \textbf{Magnetospheric FRB emission process.} In three sources (FRBs~20230307C/A, 20230913F/G, and 20240229E/A), we discover a clear decrease in $L/I$ with increasing frequency. This high-frequency depolarization has previously been observed in broadband radio pulsars, where it is explained as: the mixing of orthogonal polarization modes as they propagate through birefringent plasma or incoherent summing of unresolved burst components; both scenarios are indicators of magnetospheric emission. Thus, our results are suggestive of a magnetospheric emission process for FRBs~20230307C/A, 20230913F/G, and 20240229E/A, with the caveat that we are indeed observing the same emission components between the DSA-110 and CHIME/FRB frequency bands.
    \item \textbf{Turbulent FRB environments.} We measure frequency-dependent depolarization due to multipath propagation of the FRB signal through inhomogeneous magnetoionic plasma in $7/12$ of the co-detected FRBs. This significantly increases the population of depolarized FRB sources currently known, particularly among apparently non-repeating FRBs. Our results challenge the view that depolarization occurs primarily in a subset of repeating FRBs; instead, they suggest that both repeating and apparently non-repeating FRBs$-$unlike, for example, most pulsars$-$can be commonly found in turbulent magnetoionic environments. For the first time, we demonstrate a significant correlation between the depolarization and amplitude of Faraday rotation of FRBs; we also find a marginally significant correlation between the depolarization and scattering timescale in the same sample. The Faraday rotation and depolarization likely arise from the same inhomogeneous FRB environment, and that the observed scattering may also originate from a screen co-located with this environment.
    \item \textbf{Systematic bias in PA studies.} We find flattening in the PA profiles of two FRBs from 1280$-$1530~MHz, where they show PA variation of $\gtrsim 40~\mathrm{deg}$, to 400$-$800~MHz, where their PA profiles are consistent with being constant. These results provide only the second and third such cases in which PA profile flattening has been measured in FRB data. Assuming that we measure the same burst components across our frequency range, this suppression of PA variability is most likely induced by scatter broadening of the FRB at lower frequencies, as has been observed previously in radio pulsars. Though, we note that there are some FRBs which may have intrinsically constant PA profiles \citep[e.g., FRB~20121102A shows constant PAs even at high frequencies;][]{2018Natur.553..182M}. Hence, FRB polarization studies, particularly at low frequencies, are biased towards detecting preferentially constant PA profiles at a population level. We argue that a sample of FRB sources observed at high frequencies, or one at low frequencies that is controlled for scattering, are necessary to determine the fraction of FRBs that have intrinsically constant or variable PA profiles.
\end{enumerate}

This work acts as a blueprint for the range of science that will be possible with broadband FRB surveys on upcoming radio telescopes, namely the Canadian Hydrogen Observatory and Radio transient Detector \citep[CHORD;][]{CHORD} and the full Deep Synoptic Array \citep[DSA;][]{DSA2000}. Both of these instruments will have contiguous bandwidths centered at low frequencies (300$-$1500~MHz for CHORD and 700$-$2000~MHz for DSA) that are ideal for constraining the spectropolarimetric propagation effects highlighted in our work. The increased sensitivity of CHORD and DSA will also drastically increase the population of detected FRBs, as they are projected to discover on the order of $10^4$~FRBs in their first few years of operations. These instruments will empower analyses similar to this work to be conducted on a much larger scale, thus constraining the origins of FRB emission and the properties of their immediate environments on a population level.

\section*{Acknowledgments}
We acknowledge that CHIME is located on the traditional, ancestral, and unceded territory of the Syilx/Okanagan people. We are grateful to the staff of the Dominion Radio Astrophysical Observatory, which is operated by the National Research Council of Canada. CHIME is funded by a grant from the Canada Foundation for Innovation (CFI) 2012 Leading Edge Fund (Project 31170) and by contributions from the provinces of British Columbia, Québec and Ontario. The CHIME/FRB Project is funded by a grant from the CFI 2015 Innovation Fund (Project 33213) and by contributions from the provinces of British Columbia and Québec, and by the Dunlap Institute for Astronomy and Astrophysics at the University of Toronto. Additional support was provided by the Canadian Institute for Advanced Research (CIFAR), McGill University and the Trottier Space Institute at McGill thanks to the Trottier Family Foundation, and the University of British Columbia. The baseband recording system for CHIME/FRB is funded in part by a CFI John R. Evans Leaders Fund award to IHS. This research was enabled in part by software provided by the Digital Research Alliance of Canada (\href{https://alliancecan.ca/}{alliance.can.ca}). 

The authors thank staff members of the Owens Valley Radio Observatory and the Caltech radio group, whose tireless efforts were instrumental to the success of the DSA-110. The OVRO is located on the ancestral homelands of the Big Pine Paiute Tribe of the Owens Valley; we recognize and acknowledge the historical and cultural significance of these lands to members of the Tribe. The DSA-110 is supported by the National Science Foundation Mid-Scale Innovations Program in Astronomical Sciences (MSIP) under grant AST-1836018.

The AstroFlash research group at McGill University, University of Amsterdam, ASTRON, and JIVE is supported by: a Canada Excellence Research Chair in Transient Astrophysics (CERC-2022-00009); an Advanced Grant from the European Research Council (ERC) under the European Union's Horizon 2020 research and innovation programme (`EuroFlash'; Grant agreement No. 101098079); and an NWO-Vici grant (`AstroFlash'; VI.C.192.045).
FRB research at UBC is supported by an NSERC Discovery Grant and by the Canadian Institute for Advanced Research. 

A.~P.~is a Trottier Space Institute postdoctoral fellow and a Fonds de recherche du Québec - Nature et technologies (FRQNT) postdoctoral fellow.
K.~N.~acknowledges support by NASA through the NASA Hubble Fellowship grant \# HST-HF2-51582.001-A awarded by the Space Telescope Science Institute, which is operated by the Association of Universities for Research in Astronomy, Incorporated, under NASA contract NAS5-26555.
A.~P.~C.~is a Canadian SKA Scientist and is funded by the Government of Canada / est financé par le gouvernement du Canada.
V.~M.~K.~holds the Lorne Trottier Chair in Astrophysics \& Cosmology, a Distinguished James McGill Professorship, and receives support from an NSERC Discovery grant (RGPIN 228738-13).
M.~N.~is a Fonds de Recherche du Québec - Nature et Technologies (FRQNT) postdoctoral fellow.
A.~B.~P.~acknowledges support by NASA through the NASA Hubble Fellowship grant \mbox{HST-HF2-51584.001-A} awarded by the Space Telescope Science Institute, which is operated by the Association of Universities for Research in Astronomy, Inc., under NASA contract \mbox{NAS5-26555}. A.~B.~P.~also acknowledges prior support from a Banting Fellowship, a McGill Space Institute~(MSI) Fellowship, and a Fonds de Recherche du Québec -- Nature et Technologies~(FRQNT) Postdoctoral Fellowship.
M.~W.~S.~acknowledges support from the French government under the France 2030 investment plan, as part of the Initiative d’Excellence d’Aix-Marseille Université - AMIDEX (AMX-23-CEI- 088).
P.~S.~acknowledges the support of an NSERC Discovery Grant (RGPIN-2024-06266).

\facilities{CHIME and DSA-110.}

\software{{\tt Astropy} \citep{2013A&A...558A..33A, 2018AJ....156..123A, 2022ApJ...935..167A},
{\tt Matplotlib} \citep{Hunter:2007}, 
{\tt NADA} \citep{helsel2005nondetects, NADApackage, Rsoftware},
{\tt NumPy} \citep{harris2020array},
{\tt PyGEDM} \citep{2021PASA...38...38P},
{\tt RM-CLEAN} \citep{2009A&A...503..409H},
{\tt RM-synthesis} \citep{2005A&A...441.1217B}, 
{\tt RM-tools} \citep{2020ascl.soft05003P, 2026arXiv260120092V}, and
{\tt SciPy} \citep{2020SciPy-NMeth}.}

\bibliography{ref}{}
\bibliographystyle{aasjournalv7}

\end{document}